\documentclass[aps,onecolumn,showpacs,amssymb,superscriptaddress,notitlepage]{revtex4-1}

\usepackage{graphicx}
\usepackage{bm}
\usepackage{verbatim}
\usepackage[section]{placeins}
\usepackage[normalsize,bf]{subfigure}
\usepackage{amsmath}
\usepackage{todonotes}
\usepackage{array}

\begin{document}

\newcommand{\kmax}{k_{\mathrm{max}}}
\newcommand{\dd}{\mathrm{d}}


\title{\large{Stochastic epidemic dynamics on extremely heterogeneous networks}}

\author{C\'{e}sar Parra-Rojas}
\email{cesar.parrarojas@postgrad.manchester.ac.uk}
\affiliation{Theoretical Physics Division, School of Physics and Astronomy, 
The University of Manchester, Manchester M13 9PL, UK}
\author{Thomas House}
\email{thomas.house@manchester.ac.uk}
\affiliation{School of Mathematics, The University of Manchester, Manchester M13 9PL, UK}
\author{Alan~J. McKane}
\email{alan.mckane@manchester.ac.uk}
\affiliation{Theoretical Physics Division, School of Physics and Astronomy, 
The University of Manchester, Manchester M13 9PL, UK}

\begin{abstract}
Networks of contacts capable of spreading infectious diseases are often observed to be highly heterogeneous, with the majority of individuals having fewer contacts than the mean, and a significant minority having relatively very many contacts. We derive a two-dimensional diffusion model for the full temporal behavior of the stochastic susceptible-infectious-recovered (SIR) model on such a network, by making use of a time-scale separation in the deterministic limit of the dynamics. This low-dimensional process is an accurate approximation to the full model in the limit of large populations, even for cases when the time-scale separation is not too pronounced, provided the maximum degree is not of the order of the population size.
\end{abstract}


\maketitle

\section{Introduction}

Mathematical models are used throughout infectious disease epidemiology to
understand the dynamical processes that shape patterns of
disease~\cite{Rock:2014,Heesterbeek:2015}. While early models did not include
complex population structure, modeling approaches now frequently let the
epidemic spread take place on a network, which enables greater realism than a
model in which all individuals mix homogeneously. It does, however, pose many
technical problems for model analysis, particularly the question of how
heterogeneity in the number of links each individual participates in---their
degree---influences the
epidemic~\cite{Danon:2011,PastorSatorras:2015,Pellis:2015}.

Similarly, even though some of the earliest mathematical models of infectious
diseases were stochastic, accounting for the chance nature of transmission of
infection~\cite{Bailey:1957}, much of the applied modeling that followed was
deterministic and based on non-linear differential
equations~\cite{Anderson:1991}. More recent applied work has, however,
recognized the importance of using stochastic epidemic models~\cite{King:2015}
and also of the development of associated methodology~\cite{Britton:2015}. 

The difficulty in mathematically analyzing models that include both stochastic
elements and network structure can be a reason for not including these factors,
but we prefer to include them, and subsequently to systematically reduce the
complexity of the resulting model. This is the approach we adopt in this paper;
the reduction process being made possible because of the existence of a
separation of time-scales: many variables ``decaying'' away rapidly, leaving a
few slow variables, which govern the medium- to long-term dynamics.

Compartmental models of epidemics typically assume that the majority of the
population starts susceptible to infection, with a small number infectious, who
then spread infection to others before recovering. A key distinction is between
susceptible-infectious-susceptible (SIS) dynamics in which recovery returns an
individual to the susceptible compartment and susceptible-infectious-recovered
(SIR) dynamics in which recovery removes an individual from a further role in
the next epidemic~\cite{PastorSatorras:2015}, with the former being used to
model, e.g.,\ sexually transmitted infections other than HIV and the latter, e.g.,\
pandemic influenza~\cite{Anderson:1991,Keeling:2007,Heesterbeek:2015}. 

In the theoretical analysis of epidemic models, a crucial quantity corresponds to
the basic reproductive ratio $R_0$---verbally described as the expected number
of secondary cases infected per primary case early in the
epidemic~\cite{Diekmann:1990}. Depending on the value of $R_0$, either
$\mathcal{O}(N)$ individuals will experience infection in a population of size
$N$---the model is then said to be supercritical---or $\mathcal{O}(1)$ individuals
will experience infection in a population of size $N$---the model is then
subcritical---thus defining an epidemic threshold.

In reality, the contact network on which the disease spreads will often change over
the course of the epidemic; while approaches exist in which the dynamics of
both the disease and the network are considered~\cite{Miller:2012}, it is more
common to consider two limiting cases. The first of these is a \emph{static}
approach (also called `quenched') in which the network is assumed to evolve
much more slowly than the disease and which is typically approached
analytically through the use of pair approximation and related
techniques~\cite{Miller:2012,Eames:2002}. The second is a \emph{dynamic} limit
(also called `annealed' or `discrete heterogeneous') in which the network is
assumed to evolve much more quickly than the epidemic, and can be described by
an effective network characterized by its degree distribution, in which all
individuals sharing the same degree are considered equivalent. This case can be
analyzed through use of a set of degree-indexed differential equations (often
called the `heterogeneous mean field' approach) provided the maximum degree is
not too large~\cite{May:1988,Ferreira:2012,house2014non}.

When the distribution of degrees in the population is highly variable---a
situation that appears to be supported
empirically~\cite{LeighBrown:2011,House:2015}---it was recognized that the
epidemic may not exhibit straightforward critical behavior. This happens
because as the population size $N$ becomes large, extremely small levels of
infectiousness can lead to large epidemic sizes~\cite{Pastor-Satorras:2001} or
more accurately speaking the critical level of infectiousness can depend very
sensitively on the largest degree, $K$, in the
network~\cite{May:2001,Pastor-Satorras:2002}. The behavior of highly
heterogeneous network epidemics near criticality continues to generate interest
in both the physics and mathematics
literature~\cite{Berger:2005,Durrett:2010,Ferreira:2012,PastorSatorras:2015}.

In this paper we investigate the stochastic behavior of heterogeneous network
epidemics over time. We study a network in the dynamic limit, characterized by
a power-law degree distribution such that the probability of an individual
having degree $k$ is given by $d_k\propto k^{\alpha}$, with $-3<\alpha<-2$,
although the method of analysis is applicable to other distributions.  This
type of network has the property that the basic reproductive ratio, which is
found to be proportional to the second moment of the degree distribution
$\langle k^2\rangle$, diverges in the limit of infinite populations, leading to
the absence of an epidemic threshold. This is evidently not the case when the
population---and thus the degree cutoff---is finite, but the second moment of
the distribution can still be extremely large for sufficiently large $N$, and
heterogeneity can play an important role in the dynamics of the system.

For the case of large but finite population size, we derive a two-dimensional
stochastic differential equation (SDE) approximation to full SIR dynamics,
which reduces to an analytically solvable one-dimensional system early in the
epidemic. We perform simulations using a power-law degree distribution with a
maximum degree cutoff $K$, which show that our approach provides a good
approximation provided $K$ is not too close to the population size. There have,
to our knowledge, been no directly comparable studies of this kind. Perhaps
those closest have been, first the study of finite-size scaling of the SIS
model near the critical point to produce a phenomenological one-dimensional
SDE~\cite{Hong:2007}, second, the derivation of a one-dimensional SDE for the
SIS model from first principles, but using an adiabatic approximation that
would only be expected to hold near the critical point~\cite{Boguna:2009}, and
finally the study of a four-dimensional SDE for the SIR model on a static
network that is much more complex than ours and less amenable to
analysis~\cite{Graham:2014}.

The outline of the paper is as follows. In Sec.~\ref{sec:formulation} we
introduce the model, and formulate it first at the level of individuals (the
microscale) and then at the level of populations, but retaining a stochastic
element (the mesoscale). We begin to explore the reduction of the
mesoscopic-level SDEs in Sec.~\ref{sec:det} and complete the process in
Sec.~\ref{sec:fastvarelim}. In Sec.~\ref{sec:CIR} we perform a further
reduction, which allows us to make contact with a model that has already been
discussed in the literature. In Sec.~\ref{sec:size} these reduced models are used to find the distribution of epidemic sizes, and we conclude with a discussion of our results in Sec.~\ref{sec:discussion}. There are two appendices
where technical details relevant to Secs.~\ref{sec:det} and
\ref{sec:fastvarelim} have been relegated.

\section{Formulation of the Model}

\label{sec:formulation}

In this section we formulate the model first at the microscale, that is, in
terms of individuals, and from this derive a mesoscopic version of the model
where the variables are continuous and represent the fractions of types of
individuals who are infected or who are susceptible to infection. The general
procedure to do this is reviewed in Ref.~\cite{mckane_14} and has been
previously applied to models of epidemics on
networks~\cite{rozhnova_11,rozhnova_12}, albeit in situations where the nodes
of the network contained many individuals, rather than just one, as in the
present context. 

As discussed in the Introduction, individuals are located at the nodes of a
network through which they are exposed to infection from individuals at
neighboring nodes. Since we work in the dynamic limit, individuals do not alter
their number of contacts when acquiring the infection or recovering from it,
and the network plays no role other than encoding the information about how
many connections to other individuals a given individual possesses. In practice, this means that the network is not explicitly generated, and it is only characterized through its degree distribution.  

An individual is labeled according to whether it is (i) susceptible, infected or
recovered, and (ii) the degree of the node where it is located. Thus, the
variables at the microscale are $S_k$, $I_k$, and $R_k$, $k=1,\ldots,K$. We will
assume that the population is closed, i.e., that
$S_k(t)+I_k(t)+R_k(t)=N_k$, with total population $N=\sum_{k=1}^K N_k$, at all
times, so that we can remove one of the variables: $R_k(t)=N_k-S_k(t)-I_k(t)$.

Two types of events can occur: the infection of a susceptible individual,
whenever an individual of type $I$ comes in contact with one of type
$S$---contacts between individuals $i$ and $j$ having degrees $k_i$ and $k_j$,
respectively, are governed by a Poisson process with rate $\beta k_ik_j$; and
the recovery of an infectious individual with rate $\gamma$. In this paper we
will be interested in the properties of an epidemic that takes place over a
much shorter timeframe than demographic processes and hence will ignore the
birth and death of individuals, leaving the only interactions between the
individuals to be:
\begin{enumerate}
\item Infection of an individual at a node of degree $k$. The transition rate
	for this process is given by
\begin{align}
T_1(S_k-1,I_k+1\vert S_k, I_k) &= \beta k \frac{S_k}{N} \sum^K_{l=1}lI_l,
\label{eqn:rate_1}
\end{align}
where the summation is over infected individuals labeled according to the
degree of the node on which they are located. The only arguments of $T_1$ which
are shown explicitly are those involving individuals on nodes of degree $k$,
since only these change in this process; those on the right represent the
initial state, and those on the left represent the final state.
\item Recovery of an infected individual at a node of degree $k$. This proceeds
	at a rate given by $\gamma$, and so the transition rate for this process is
	given by
\begin{align}
T_2(S_k,I_k-1\vert S_k,I_k) &= \gamma I_k.
\label{eqn:rate_2}
\end{align}
\end{enumerate}

The dynamics of the process can be described by writing down an equation for
the probability that the system is in a state $\{ S_l, I_l\,| l=1,\ldots,K\}$
at time $t$, which will be denoted by $P(\{ S_l, I_l\}, t)$. This is the master
equation~\cite{vanKampen_07}, which equates $dP/dt$ to the net rate of increase
of $P$ due to the processes given by Eqs.~(\ref{eqn:rate_1}) and
(\ref{eqn:rate_2}), and is given by
\begin{align}
\frac{dP(\{ S_l,I_l\}, t)}{dt} &= \sum^K_{k=1} \left[ T_1(S_k,I_k\vert S_k+1, I_k-1)P(S_k+1,I_k-1, t)+T_2(S_k,I_k\vert S_k, I_k+1)P(S_k,I_k+1, t) \right] 
\nonumber \\
& \quad - \sum^K_{k=1} \left[T_1(S_k-1,I_k+1\vert S_k, I_k)+T_2(S_k,I_k-1\vert S_k, I_k)\right]P(S_k,I_k, t).
\label{eqn:master}
\end{align}
Once again, the dependence of the probability distribution, $P$, on the state
variables $\{ S_l, I_l\,| l \neq k\}$ has been suppressed in the sum over $k$
on the right-hand side of this equation.

We will refer to the discrete model defined by
Eqs.~\eqref{eqn:rate_1}--\eqref{eqn:master} as the full microscopic model,
since we shortly derive a mesoscopic model from it, and subsequently further
reduce this model. Once an initial state has been specified, the model is
completely determined, and in principle $P(\{ S_l,I_l\}, t)$ can be found for
any state $\{ S_l, I_l\,| l=1,\ldots,K\}$ at any time $t$. This behavior can be
explored numerically \emph{via} Monte Carlo simulations (e.g.\ using
Gillespie's method~\cite{gillespie_76,gillespie_92} or other
approaches~\cite{house2014non}); however, our main interest in this paper is in
approximating the master equation to obtain models that are more amenable to
analysis.

A mesoscopic description of the process involves going over to new variables
$s_k = S_k/N$ and $i_k = I_k/N$, $k=1,\ldots,K$, which are assumed continuous
for large $N$, but retaining the stochastic nature of the system by keeping $N$
large but finite. A macroscopic description, on the other hand, would involve
taking the limit $N \to \infty$, which renders the process deterministic, and
which eliminates all effects of the original discrete nature of the system. 

The construction of the mesoscopic model involves expanding the master equation,
Eq.~\eqref{eqn:master}, in powers of $1/N$ and truncating this expansion at
order $\mathcal{O}\left(N^{-2}\right)$. This is discussed in many places in the
literature, but here we follow Ref.~\cite{mckane_14}, which gives explicit
forms for the functions $A^{(k)}_{\mu}$ and $B^{(k)}_{\mu \nu}$, $\mu, \nu
=1,2$, which define the mesoscopic model. Carrying out this procedure, the
master equation becomes a Fokker-Planck equation~\cite{Gardiner_09} of the form
\begin{align}
\frac{\partial P\left(\bm{x}, t\right)}{\partial t} &= - \frac{1}{N} \sum^K_{k=1}\sum\limits_{\mu=1}^2\frac{\partial}{\partial x^{(k)}_\mu}\left[A^{(k)}_\mu(\bm{x})P\left(\bm{x}, t\right)\right]+\frac{1}{2N^2}\sum^K_{k=1}\sum\limits_{\mu,\nu = 1}^2\frac{\partial ^2}{\partial x^{(k)}_\mu \partial x^{(k)}_\nu}\left[B^{(k)}_{\mu \nu}(\bm{x}) P\left(\bm{x}, t\right)\right],
\label{eqn:FPE}
\end{align}
where $\bm{x}^{(k)}=(s_k, i_k)$; $x^{(k)}_1 = s_k$, $x^{(k)}_2 = i_k$. We have
also introduced $\bm{x} = (\bm{x}^{(1)},\ldots,\bm{x}^{(K)})$. The explicit
form of the functions $A^{(k)}_{\mu}$ and $B^{(k)}_{\mu \nu}$
are~\cite{mckane_14}
\begin{align}
A^{(k)}_1 &= - \mathcal{T}_1^{(k)}; \ \ A^{(k)}_2 = \mathcal{T}_1^{(k)} - \mathcal{T}_2^{(k)}; \ \ B^{(k)}_{1 1} = \mathcal{T}_1^{(k)}; \ \ B^{(k)}_{1 2} = B^{(k)}_{2 1} = - \mathcal{T}_1^{(k)}; \ \ B^{(k)}_{2 2} = \mathcal{T}_1^{(k)} + \mathcal{T}_2^{(k)},
\label{eqn:A_and_B}
\end{align}
where
\begin{align}
\mathcal{T}_1^{(k)} &\equiv \beta ks_k\sum^K_{l = 1} l i_l, \ \ \ \mathcal{T}_2^{(k)} \equiv \gamma i_k,
\label{eqn:rescaled_rates}
\end{align}
are the original transition rates (scaled by $N$) given in
Eqs.~(\ref{eqn:rate_1}) and (\ref{eqn:rate_2}), but written in terms of the
continuous state variables $s_k$ and $i_k$. 

The final stage of the construction of the mesoscopic form of the model is to
adopt a coarser time scale by introducing a new time variable $\tau = t/N$.
From Eq.~(\ref{eqn:FPE}) it can be seen that without this change of variable,
the macroscopic ($N \to \infty$) limit eliminates the right-hand side of the
equation, giving a trivial macroscopic limit. With the change of variable, the
first term on the right-hand side survives (but without the factor $N^{-1}$),
giving a Liouville-like equation for $P$, which implies a non-trivial
deterministic dynamics.

A clearer way to see this, and in fact a more intuitive formulation of the
mesoscopic dynamics to that of the Fokker-Planck equation, is to use the SDE
which is equivalent to Eq.~(\ref{eqn:FPE}). We utilize the general result that
a Fokker-Planck equation of the form given by Eq.~(\ref{eqn:FPE}) is equivalent
to the SDE $dx^{(k)}_\mu/d\tau = A^{(k)}_\mu(\bm{x}) +
\eta^{(k)}_\mu(\tau)/\sqrt{N}$, where the $\eta_{\mu}^{(k)}$ are Gaussian
random variables with zero mean and a correlation matrix given by the matrix
$\bm{B}^{(k)}$. In terms of the variables $s_k$ and $i_k$, the SDEs take the form
\begin{align}
\frac{ds_k}{d\tau} &= -\beta ks_k\sum^K_{l=1} li_l + \frac{\eta_1^{(k)}(\tau)}{\sqrt{N}},\label{eqn:sk}\\
\frac{di_k}{d\tau} &= \beta ks_k\sum^K_{l=1} li_l-\gamma i_k + \frac{\eta_2^{(k)}(\tau)}{\sqrt{N}},\label{eqn:ik}
\end{align}
where 
\begin{align}
\left\langle \eta_\mu^{(k)}(\tau) \eta_\nu^{(l)}(\tau ')\right\rangle &= B^{(k)}_{\mu \nu}(\bm{x}) \delta_{k l} \delta(\tau - \tau '). 
\label{eqn:noise_corr_1}
\end{align}
Here the noise is to be interpreted in the sense of It\={o}~\cite{Gardiner_09}.
It is clear from Eqs.~(\ref{eqn:sk}) and (\ref{eqn:ik}) that the limit $N \to
\infty$ eliminates the noise terms leaving deterministic equations, and we
recover Eq.~(2.6) from Ref.~\cite{house2014non}.

It is convenient to make the noise in the SDEs explicitly multiplicative. Since
the rates $\mathcal{T}^{(k)}_\mu$ are non-negative we may introduce new
functions $\sigma^{(k)}_\mu(\bm{x}) \equiv
\sqrt{\mathcal{T}^{(k)}_\mu(\bm{x})}$, $\mu=1,2$, and then transform to new
noise variables
\begin{align}
\begin{split}
\eta_1^{(k)}(\tau) &= \sigma^{(k)}_1(\bm{x})\rho_1^{(k)}(\tau),\\
\eta_2^{(k)}(\tau) &= - \sigma^{(k)}_1(\bm{x})\rho_1^{(k)}(\tau)+\sigma^{(k)}_2(\bm{x})\rho_2^{(k)}(\tau).
\label{eqn:noise_transform}
\end{split}
\end{align}
It is straightforward to check that $\langle \rho_\mu^{(k)}(\tau)
\rho_\nu^{(l)}(\tau ')\rangle = \delta_{\mu \nu}\delta_{kl}\delta(\tau - \tau
')$, and therefore that all the state dependence has been transferred from the
correlator to the SDEs. Equations \eqref{eqn:sk}--\eqref{eqn:ik} correspond to
the full mesoscopic model.

In order to describe the evolution of the entire population, we need to
consider the SDEs for all $k =1,\ldots,K$; that is, we have to work with $2 K$
equations. In this paper, we will be interested in networks where individuals
can pick degrees from a truncated Zipf distribution of the form
\begin{align}
d_k &= \kappa_{\alpha} k^\alpha,\qquad \kappa_{\alpha}=\sum_{k=1}^{K} k^\alpha,
\end{align}
and we let $1\leq K\leq N-1$. Since we are interested in the case when $K$ is
large, the number of equations making up the full mesoscopic model are so large
that it can make any direct treatment impractical. Therefore, in the following
sections we will be concerned with reducing the number of equations, and we
start by looking at the deterministic limit of the system. This will also have
a crucial role to play in the reduction of the stochastic model, described in
Sec.~\ref{sec:fastvarelim}.

\section{Deterministic dynamics and reduction of the equations}\label{sec:det}

The deterministic limit that controls the dynamics of the macroscopic version
of the model is found by taking the $N \to \infty$ limit of
Eqs.~\eqref{eqn:sk}--\eqref{eqn:ik}, and so eliminating the noise terms. For
notational convenience we reorder the entries of the state vector, so that
$\bm{x}(\tau)=(s_1(\tau),\ldots,s_K(\tau),i_1(\tau),\ldots,i_K(\tau))$, and write Eqs.~\eqref{eqn:sk}--\eqref{eqn:ik} in the form $d{\bm{x}}/d\tau =
\bm{A}(\bm{x})$. To find the fixed points of the dynamics we set the time
derivatives on the left-hand side of these equations to zero; these fixed
points will be denoted by an asterisk and obey $\bm{A}(\bm{x}^*)=\bm{0}$. It is
immediately clear that there is a set of stable fixed points with $i_k^* = 0$
and $s_k^*$ undetermined, for all $k$, corresponding to a disease-free state.

For a given initial condition, the fixed point is uniquely determined, however
it may not be stable. To investigate the stability of the fixed point
$(s^*_1,\ldots,{s}^*_{K},0,\ldots,0)$ we perform a linear stability analysis.
The $2K \times 2K$ stability matrix $\bm {J}^{\{ 2K\}}$ is defined so that near
$\bm{x}^*$, the dynamics are $d{\bm{x}}/d\tau = \bm{J}^{\{ 2K\}} (\bm{x}-\bm{x}^*) +
\mathcal{O}(\bm{x}-\bm{x}^*)^2$, and is found to have the form
\begin{align}
\label{eqn:J_2K_defn}
\bm{J}^{\{ 2K\}} &=\left(\begin{array}{rr}
0 \ & \ - \bm{\mathcal{J}} \\ \\
0 \ & \ \bm{\mathcal{J}} - \gamma \bm I 
\end{array}\right),
\end{align}
where $\bm{\mathcal{J}}$ is a $K \times K$ matrix with entries $\mathcal{J}_{k l} =
\beta s_k^* k l$ and $\bm I$ is the $K \times K$ unit matrix. The eigenvalues and
eigenvectors of $\bm{J}^{\{ 2K\}}$ are straightforward to determine. Any vector
that has the last $K$ entries equal to zero is an eigenvector with eigenvalue
zero, reflecting the degeneracies of the system. Similarly a vector with the
first $K$ entries set equal to zero and the last $K$ entries equal to $\psi_k$,
$k=1,\ldots,K$, is an eigenvector with eigenvalue $-\gamma$ if $\sum^K_{l=1}
\mathcal{J}_{k l}\psi_l = 0$, that is, if $\sum^{K}_{l=1} l\psi_l = 0$. These
then yield another degenerate set of $(K-1)$ eigenvalues $-\gamma$. The final
eigenvalue can be found by equating the trace of $\bm{J}^{\{ 2K\}}$ to the sum of
its eigenvalues. In summary, the $2K$ eigenvalues are 
\begin{align}
\Lambda^{\{ k\}} &= 0, \ k=1,\ldots,K; \ \ 
\Lambda^{\{ K+l\}} = - \gamma, \ l=1,\ldots,K-1; \ \
\Lambda^{\{ 2K\}} = -\gamma + \beta \sum_{k=1}^{K} k^2 s^*_k,
\label{eqn:eva_full}
\end{align}
while a suitable set of corresponding eigenvectors are listed in
Appendix~\ref{sec:eve}. From Eq.~(\ref{eqn:eva_full}) we see that the condition
for the fixed points to be (marginally) stable is
\begin{align}
\sum_{k=1}^{K} k^2 s^*_k & < \frac{\gamma}{\beta}.
\label{stability_cond_full}
\end{align}

As mentioned in the Introduction, there are two possible outcomes for the
deterministic dynamics depending on the parameter values: either a negligible
or a significant number of individuals in the population will get infected
during the course of the epidemic. The quantity that determines which of these
cases we are in is often referred to as the basic reproductive ratio; the final
size of the epidemic, in turn, can be obtained from the $s$ coordinates of the
fixed point, $s_1^*,\ldots,s^*_K$. In order to determine these in the general
heterogeneous case, we look first at the homogeneous case, where $K=1$. If we
set an initial condition with only a few infective individuals in an almost
completely susceptible population, we will have $i(\tau=0)=i_0\ll 1$ and
$s(\tau=0)=s_0\sim 1$, where $s=s_1$ and $i=i_1$; with this, the $N \to \infty$
limit of Eq.~(\ref{eqn:ik}) takes the form $di/d\tau = (\beta - \gamma)i$,
early in the epidemic. From this we obtain 
\begin{align}
i(\tau) &= i_0 e^{\gamma(R_0-1)\tau}\label{eqn:basic_homo},
\end{align}
where $R_0=\beta/\gamma$ corresponds to the basic reproductive ratio. This
defines an epidemic threshold, such that for $R_0>1$ the number of infectives
grows exponentially in the early stages of the epidemic, while for $R_0<1$ the
epidemic does not take off and its final state corresponds to
$(s^*,i^*)=(s_0,i_0)\sim(1,0)$.

For the case when $R_0>1$, we can find the $s$ coordinate of the fixed point by
dividing Eq.~\eqref{eqn:ik} by Eq.~\eqref{eqn:sk} with $K=1$; this yields
\begin{align}
\frac{di}{ds} &= -\frac{\beta s-\gamma}{\beta s}.\label{eqn:size_deriv}
\end{align}
If we set again an initial condition $i_0\ll 1$, with $s_0\sim 1$, we can
integrate the equation above to find
\begin{align}
\frac{\gamma}{\beta}\ln s^* - s^* + 1 &= 0,\label{eqn:size_deriv_2}
\end{align}
where we have used that the final state of the epidemic corresponds to $i^*=0$.
Therefore, to a very good approximation, we can find $s$ coordinate of the
fixed point as the smallest solution to
\begin{align}
s^* &= e^{-R_0 (1-s^*)};\label{eqn:size_homo}
\end{align}
note that $s^*=1$ is always a solution, and the only one for $R_0<1$. The total
epidemic size will be given by $r^*=1-s^*$, which corresponds to the fraction
of recovered individuals at the end of the outbreak. This is due to the fact
that the initial number of recovered individuals is zero, and that every
individual that gets infected during the course of the epidemic will eventually
recover and remain permanently recovered.

Now, a significant simplification can be made in the deterministic limit of the
heterogeneous case, which reduces the $2K$ dynamical equations to $2$ equations
(see Refs.~\cite{Volz:2008,House:2010,Miller:2011}, which consider the static case).
This is achieved by use of the Ansatz $s_k(\tau) = d_k \theta(\tau)^k$; we also
define $\lambda(\tau)$ such that $\lambda(\tau)\equiv \sum_{k}ki_k(\tau)$. With
these two variables, Eqs.~\eqref{eqn:sk}--\eqref{eqn:ik}, in the case when
$N\to \infty$, can be rewritten independently of $k$ as
\begin{align}
\frac{d \theta}{d\tau} &= -\beta \theta \lambda, \label{eqn:theta_det}\\
\frac{d\lambda}{d\tau} &= \beta \lambda \phi(\theta)-\gamma \lambda, \label{eqn:lambda_det}
\end{align}
where we have introduced $\phi(x) = \sum^K_{k=1} k^2 d_k x^k$. While the new
variables $\theta$ and $\lambda$ are justified mathematically through the fact
that they simplify the equations, they also have a physical interpretation. For
the case of $\theta$, this is related to the recently-introduced concept of a
\textit{test node}~\cite{Miller:2012,Miller:2013}, which plays an analogous
role in epidemic dynamics to the concept of a test charge in electrostatics. If
we introduce a single initially susceptible individual with one link into the
system, and assume that the population is so large that this individual has
negligible influence on the epidemic dynamics, then $\theta(\tau)$ is the
probability that such a `test node' has avoided infection at time $\tau$. The
quantity $\lambda$ is related to the \textit{force of infection} that is defined in epidemiology as the per-capita rate at which susceptible individuals become infective~\cite{Keeling:2007}. Our $\lambda(\tau)$ is the network generalization of this, being instead the per-\textit{link} rate at which susceptible individuals become infective at time $\tau$. We therefore argue that where a non-network SIR epidemic can have its dynamics represented in phase space with coordinates $(s(\tau),i(\tau))$, for the network model the physically relevant coordinates, designed to account for link heterogeneity, are $(\theta(\tau),\lambda(\tau))$.

If we also introduce the probability generating function of the degree
distribution $G(x) = \sum^K_{k=1} d_k x^k$, then $\phi(x) = x^2 G^{\prime
\prime}(x)+x G^{\prime}(x)$. In this reduced model, there is a line of stable
fixed points, uniquely determined by the initial conditions, which satisfy
$\lambda^*=0$. The stability matrix for these fixed points is the $2 \times 2$
matrix
\begin{align}
\label{eqn:J_2_defn}
\bm{J}^{\{ 2\}} &=\left(\begin{array}{rr}
0 \ & \ - \beta \theta^* \\ \\
0 \ & \ \beta \phi(\theta^*) - \gamma 
\end{array}\right),
\end{align}
which has a similar form to the stability matrix for the full system given by
Eq.~(\ref{eqn:J_2K_defn}). Once again, an eigenvector exists that has the last
entry equal to zero. This corresponds to an eigenvalue zero, reflecting the
existence of a line of fixed points. The trace of $\bm{J}^{\{ 2\}}$ then gives the
second eigenvalue, and the two eigenvalues are thus
\begin{align}
\Lambda^{\{ 1\}} &= 0 \ ; \ \ \ \
\Lambda^{\{ 2\}} = -\gamma + \beta \phi(\theta^*) = -\gamma + \beta \sum_{k=1}^{K} k^2 s^*_k,
\label{eqn:eva_total}
\end{align}
with the eigenvectors given in Appendix~\ref{sec:eve}. The stability condition
is now
\begin{align}
\phi(\theta^*)<\gamma/\beta.
\label{eqn:stability_cond_total}
\end{align}

Now we can find analogous expressions to those in Eqs.~\eqref{eqn:basic_homo}
and \eqref{eqn:size_homo} for the heterogeneous case. We start by noting that,
in terms of $\theta$, the total number of susceptible individuals will be given
by
\begin{align}
\sum_{k=1}^K s_k &= \sum_{k=1}^K d_k \theta^k = G(\theta).
\end{align}
It is clear that if $\theta<1$, then $G(\theta)< G(1) = 1$; conversely, if
$\theta>1$ then $G(\theta)> G(1) = 1$, and so $G(\theta)=1\Leftrightarrow
\theta=1$. This means that a completely susceptible population will correspond
to $\theta$ being equal to unity. A small initial number of infectives will, in
turn, correspond to $\lambda \ll 1$. In addition from
Eq.~(\ref{eqn:lambda_det}), $d\lambda/d\tau = \gamma(\rho - 1)\lambda$, where
$\rho = \beta \phi(1)/\gamma = \beta \langle k^2\rangle/\gamma$, at early
times. Therefore
\begin{align}
\lambda(\tau) &= \lambda_0 e^{\gamma(\rho-1)\tau}.
\label{eqn:lambda_early}
\end{align}
From Eq.~(\ref{eqn:lambda_early}) it is clear that there is a critical value of
$\rho$ equal to $1$; if $\rho$ is larger than $1$ the infection grows, and if
it is below $1$ it does not. Thus, $\rho$ is the heterogeneous basic
reproductive ratio, generalizing $R_0 = \beta/\gamma$ in the homogeneous case.
Central to our investigation is the fact that this quantity does not have a
deterministic threshold for networks characterized by power-law degree
distributions in the case of interest, i.e., when $-3<\alpha<-2$.

In order to find the heterogeneous final size, we proceed again as in the
homogeneous case; dividing Eq.~\eqref{eqn:lambda_det} by
Eq.~\eqref{eqn:theta_det}, we find
\begin{align}
\frac{d\lambda}{d\theta}&= -\frac{\beta \phi(\theta)-\gamma}{\beta \theta}.
\end{align}
From this, the heterogeneous analog of Eq.~\eqref{eqn:size_deriv_2} is given by
\begin{align}
\frac{\gamma}{\beta}\ln \theta^* - \theta^* G^\prime (\theta^*) + \langle k\rangle =0,
\end{align}
where we again set an initial condition with a very small number of infectives,
$\lambda_0\ll 1$, in an almost completely susceptible population, $\theta_0
\sim 1$, and the final state is given by $\lambda^*=0$. The $\theta$ component
of the fixed point will then correspond to the smallest solution to
\begin{align}
\theta^* &= \exp \left\lbrace -\frac{\rho}{\langle k^2\rangle} \left[ \langle k\rangle - \theta^* G^\prime(\theta^*)\right]\right\rbrace, \label{eqn:theta_fix}
\end{align}
with $\rho$ the heterogeneous basic reproductive ratio defined above. The
epidemic size, by definition, will be given by $r^*=1-G(\theta^*)$.
Furthermore, the $s_k$ components of the fixed point correspond to $s^*_k=d_k
{\theta^*}^k$.

Our interest in this paper is in the (finite $N$) stochastic dynamics of the
model, where a complete reduction of the kind that led to
Eqs.~(\ref{eqn:theta_det}) and (\ref{eqn:lambda_det}) is not possible. In order
to see this, we proceed in the same manner as in the deterministic case: we
start from Eq.~\eqref{eqn:sk}, this time without taking the $N\to\infty$ limit,
and use the Ansatz $s_k(\tau)=d_k\theta(\tau)^k$; summing over $k$ and dividing
by $G^\prime (\theta)$, we arrive at
\begin{align}
\frac{d\theta}{d\tau} &= -\beta \theta \lambda + \frac{1}{\sqrt{N}}\xi_1(\tau),\label{eqn:theta_stoch_deriv}
\end{align}
where we have introduced $\xi_1(\tau)=\left[G^\prime (\theta)\right]^{-1}\sum_k
\sigma^{(k)}_1 \rho^{(k)}_1(\tau)$. Since the noises $\rho^{(k)}_1(\tau)$ have
zero mean and are independent of each other, we can rewrite $\xi_1(\tau)$ as
\begin{align}
\xi_1(\tau) &= \bar{\sigma}_1 \zeta_1(\tau),
\label{eqn:xi_to_zeta}
\end{align}
where $\zeta_1(\tau)$ has zero mean and unit variance, and $\bar{\sigma}_1^2 =
\left[G^\prime (\theta)\right]^{-2}\sum_k
\left[\sigma^{(k)}_1\right]^2=\left[G^\prime (\theta)\right]^{-1}\beta \lambda
\theta$. With this, we arrive at a reduced equation for $\theta$ given by
\begin{align}
\frac{d\theta}{d\tau} &= -\beta \theta \lambda + \frac{1}{\sqrt{N}}\bar{\sigma}_1\zeta_1(\tau).\label{eqn:theta_stoch}
\end{align}

As in the deterministic case, then, we can express the $K$ equations for the
$s_k$ variables, described by Eq.~\eqref{eqn:sk}, entirely in terms of $\theta$
and $\lambda$ \emph{via} Eq.~\eqref{eqn:theta_stoch}. However, we will see that the
same does not hold for the dynamics of the $i_k$ variables, as we have
anticipated.

In a similar fashion as with the $s_k$ variables, we take the equation for the
$i_k$ variables, Eq.~\eqref{eqn:ik}, multiply it by $k$ and sum over $k$, to
obtain
\begin{align}
\frac{d\lambda}{d\tau} &= \beta \lambda \phi(\theta)-\gamma \lambda - \frac{1}{\sqrt{N}} \sum_k k \sigma^{(k)}_1 \rho^{(k)}_1(\tau) + \frac{1}{\sqrt{N}}\sum_k k \sigma^{(k)}_2 \rho^{(k)}_2(\tau).\label{eqn:lambda_stoch_deriv}
\end{align}

Let us now denote the noise terms appearing in the equation for $\lambda$ above
as $\xi_2(\tau)=-\sum_k k \sigma^{(k)}_1 \rho^{(k)}_1(\tau)$ and
$\xi_3(\tau)=\sum_k k \sigma^{(k)}_2 \rho^{(k)}_2(\tau)$. These two noise
terms, like $\xi_1(\tau)$, have zero mean; also, it can be readily verified
that the correlations between all three of them are given by $\langle
\xi_\mu(\tau)\xi_\nu(\tau')\rangle=\delta(\tau-\tau')\bar{B}_{\mu \nu}$, where
now $\mu,\nu=1,2,3$, with
\begin{align}
\bar{B}_{11} &= \bar{\sigma}_1^2,\qquad \bar{B}_{12} = -\frac{\phi(\theta)}{\theta}\bar{\sigma}_1^2,\qquad \bar{B}_{22}=\beta \lambda \left[\phi(\theta)+\psi(\theta)\right],\qquad \bar{B}_{33}=\gamma \sum_k k^2 i_k,
\end{align}
and $\bar{B}_{13}=\bar{B}_{23}=0$, where $\psi(x) = \sum^{K}_{k=1} \left(k^3-k^2\right) d_k x^k
= x^3 G^{\prime \prime \prime}(x)+2 x^2 G^{\prime \prime}(x)$. 

We can now, as in Sec.~\ref{sec:formulation}, express $\xi_1(\tau)$,
$\xi_2(\tau)$ and $\xi_3(\tau)$ in terms of three independent
normally-distributed random variables, $\zeta_1(\tau)$, $\zeta_2(\tau)$, and
$\zeta_3(\tau)$, with zero mean and $\langle
\zeta_\mu(\tau)\zeta_\nu(\tau')\rangle=\delta(\tau-\tau')\delta_{\mu \nu}$,
$\mu,\nu=1,2,3$. The definition of $\zeta_1(\tau)$ in terms of $\xi_1(\tau)$
has already been given by Eq.~(\ref{eqn:xi_to_zeta}); the other two are:
\begin{align}
\xi_2(\tau) &= - \frac{\phi(\theta)}{\theta}\,\bar{\sigma}_1 \zeta_1(\tau) + 
\bar{\sigma}_2 \zeta_2(\tau), \ \ \ \xi_3(\tau) = \bar{\sigma}_3 \zeta_3(\tau),
\label{eqn:xi_to_zeta_2}
\end{align}
where $\bar{\sigma}_2\equiv \sqrt{\bar{B}_{22} - \bar{\sigma}^2_1}$ and
$\bar{\sigma}_3\equiv \sqrt{\bar{B}_{33}}$. Substituting all this back into
Eq.~\eqref{eqn:lambda_stoch_deriv} yields
\begin{align}
\frac{d\lambda}{d\tau} &= \lambda (\beta \phi(\theta)-\gamma) +\frac{1}{\sqrt{N}}\left[-\frac{\phi(\theta)}{\theta}\bar{\sigma}_1 \zeta_1(\tau) +\bar{\sigma}_2\zeta_2(\tau) + \bar{\sigma}_3 \zeta_3(\tau)\right]. \label{eqn:lambda_stoch}
\end{align}

The main feature we note from this is that the two-dimensional system of
equations cannot be closed. This is so because $\bar{\sigma}_3$ is still
dependent on $i_k$ for all values of $k$, and so it is not possible to achieve
a complete reduction of the stochastic system. Therefore, when we take the
intrinsic noise present in the system into account, a partially reduced model
of $(K+1)$ equations---Eq.~\eqref{eqn:theta_stoch} for $\theta$, with
$\lambda=\sum_k k i_k$, and Eq.~\eqref{eqn:ik} for $i_k$---is the best we can
achieve.

Due to this, the exploration of the epidemic dynamics for extremely
heterogeneous cases, in which $K$ can take very large values, is rendered
impractical: even after reducing the number of equations, we will still be left
with a very-high-dimensional system. However, progress can be made by observing
that, under certain conditions, the deterministic limit of the model presents a
separation of time-scales, as we shall see in the following section.

We will end this section by carrying out this partial reduction in the
deterministic case, as a preparation for the stochastic treatment. This partial
reduction involves applying the above reduction to the $s_k$ variables only, by
writing $s_k(\tau) = d_k \theta(\tau)^k$, but keeping the $K$ $i_k$ variables.
There are now $(K+1)$ dynamical equations, which take the form
\begin{align}
\frac{d \theta}{d\tau} &= -\beta \theta \sum^K_{l=1} l i_l, \label{eqn:semi_1}\\
\frac{di_k}{d\tau} &= \beta kd_k \theta^k \sum^K_{l=1} li_l-\gamma i_k.\label{eqn:semi_2}
\end{align}
There is again a line of fixed points corresponding to disease-free states,
$i_k=0$, $k=1,\ldots,K$, but with $\theta$ undetermined. The $(K+1) \times
(K+1)$ stability matrix is now
\begin{align}
\label{eqn:J_K_plus_one_defn}
\bm{J}^{\{ K+1\}} &=\left(\begin{array}{rr}
0 \ & \ - \bm{j}^{\top} \\ \\
0 \ & \ \bm{\mathcal{J}} - \gamma \bm I 
\end{array}\right),
\end{align}
where $\bm{j}$ is a $K$-dimensional vector whose $l$-th entry is $\beta
\theta^* l$, $\bm{\mathcal{J}}$ is a $K \times K$ matrix with entries
$\mathcal{J}_{k l} = \beta d_k (\theta^*)^k k l$, and $\bm I$ is the $K \times K$
unit matrix. The eigenvalues and eigenvectors can be found in the same way as
before, the eigenvalues having the form
\begin{align}
\begin{split}
\Lambda^{\{ 1\}} &= 0\ ; \ \  
\Lambda^{\{ l+1\}} = - \gamma, \ l=1,\ldots,K-1; \ \
\Lambda^{\{ K+1\}} = -\gamma+\beta \phi(\theta^*),\label{eqn:eva_semi}
\end{split}
\end{align}
with the eigenvectors given in Appendix~\ref{sec:eve}. The stability condition is
once again given by Eq.~(\ref{eqn:stability_cond_total}).

\section{Slow dynamics in the partially reduced model}\label{sec:fastvarelim}

As we have discussed, the deterministic limit of
Eqs.~\eqref{eqn:sk}--\eqref{eqn:ik} can be greatly simplified by performing a
change of variables and describing the evolution of the system in terms of
$\theta$ and $\lambda$. The inclusion of demographic noise, however, hinders
such a reduction of dimensionality. It is not possible to obtain a closed
two-dimensional system of equations, thus enormously complicating the analysis
of the stochastic epidemic, compared to its deterministic limit.

\begin{figure}
\centering
\includegraphics[scale=0.42]{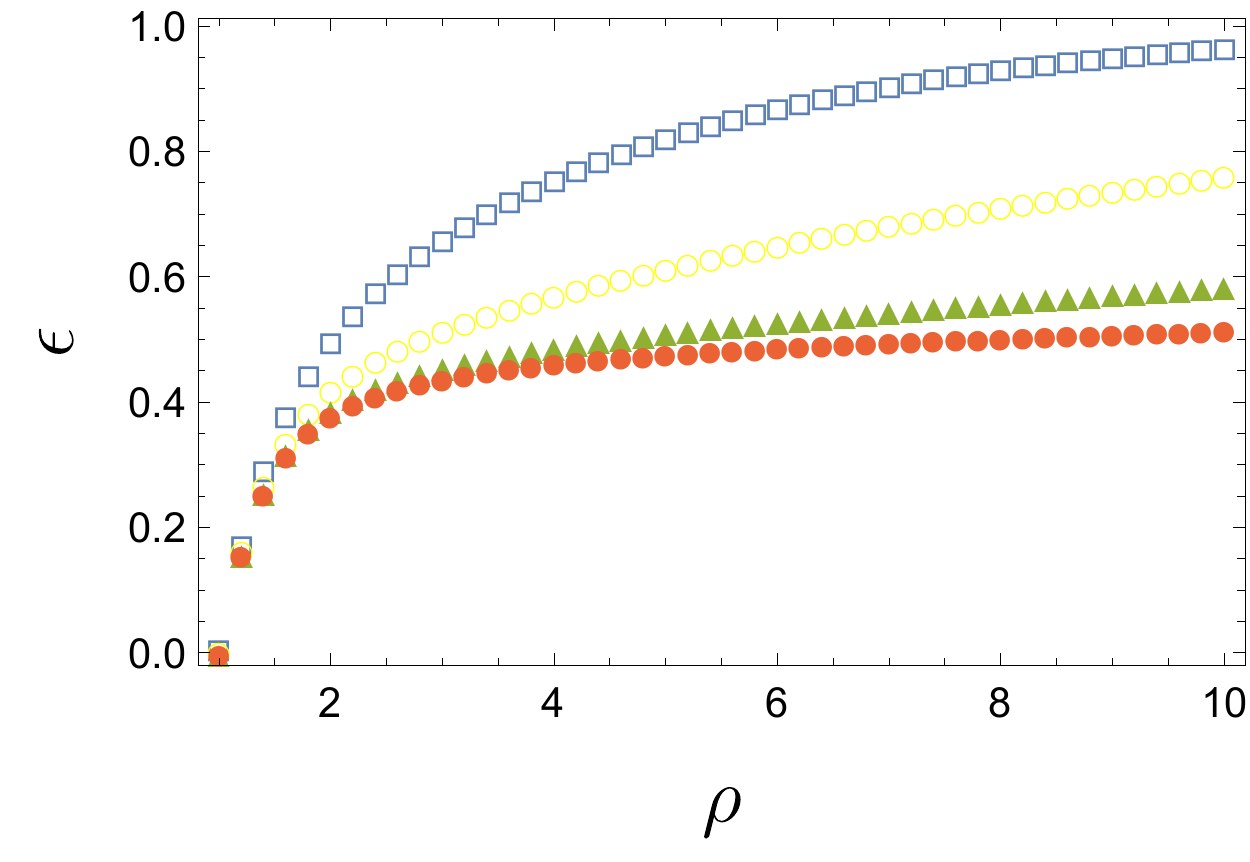}
\caption{Relative magnitudes of $\Lambda^{\{K+1\}}$ and $\Lambda^{\{2,\ldots,
K\}}$, $\epsilon = \vert \rho \phi(\theta^*)/\langle k^2\rangle - 1\vert$, as a
function of $\theta^*$ for ${\alpha=-2.5}$ and $K=10$ (blue squares), $K=10^2$ (empty yellow circles), $K=10^4$ (green triangles) and $K=10^6$ (filled red circles).}
\label{fig:L_R0}
\end{figure}

Here, we attempt to overcome this issue by exploiting the properties of the
deterministic limit of the model, and by finding a way to close the equations
for $\theta$ and $\lambda$ in the finite-$N$ case, while still retaining the
essential features of the full model. In order to do this, we start from
Eqs.~\eqref{eqn:semi_1}--(\ref{eqn:semi_2}) for the partially reduced $(K+1)$-dimensional
deterministic system for $\theta$ and $i_k$. Examination of the eigenvalues
appearing in Eq.~\eqref{eqn:eva_semi}, shows that $\vert\Lambda^{\{K+1\}}\vert
< \vert\Lambda^{\{2,\ldots, K\}}\vert$. If the ratio $\epsilon \equiv \vert
\Lambda^{\{K+1\}}/\Lambda^{\{2,\ldots, K\}} \vert$ is small, then there is the
potential to carry out a reduction of the system. In such case, the
deterministic dynamics would quickly evolve along the directions corresponding
to the set of most negative eigenvalues towards a lower-dimensional region of
the state-space, in which the system as a whole evolves in a much slower
time-scale, and which contains the fixed point $(\theta^*,0,\ldots,0)$. As
illustrated in Fig.~\ref{fig:L_R0}, showing the value of $\epsilon$ for different choices of $K$, we would expect this
time-scale separation to occur for combinations of parameters for which the
final size of the epidemic is small, i.e., for large $\theta^*$.

Since there is one most negative eigenvalue, with a degeneracy equal to
$(K-1)$, after sufficient time has passed the evolution of the system will then
be constrained to a two-dimensional surface, which we will refer to as the slow
subspace. With the aim of reducing the dimensionality of the problem, then, we
ignore the fast initial behavior of the system and focus the analysis on the
slow surface. We do so by imposing the condition that no dynamics exists along
the $(K-1)$ fast directions, i.e.,
\begin{align}
\bm{u}^{\{j\}} \cdot \bm{A}(\bm{x}) &= 0,\qquad \qquad j=2, \ldots, K, \label{eqn:kill_fast}
\end{align}
where $\bm{x}=(\theta, i_1, \ldots, i_K)$, and $\bm{u}^{\{j\}}$ is the $j$-th
left-eigenvector of the Jacobian $\bm{J}^{\{K+1\}}$ evaluated at the fixed point
$(\theta^*,0,\ldots,0)$, with $\theta^*$ given by Eq.~\eqref{eqn:theta_fix}.
This procedure goes under the name of adiabatic elimination, or fast-variable
elimination~\cite{haken_83}, and it is a common practice in the simplification
of deterministic non-linear dynamical systems---e.g., systems of
oscillating chemical reactions~\cite{tyson_82,zhabotinsky_87}.

For ease of application of the method, the left- and right-eigenvectors of the
Jacobian are normalized in such a way that
\begin{align}
\bm{u}^{\{i\}} \cdot \bm{v}^{\{j\}} &= \delta_{ij};\label{eqn:normal}
\end{align}
these are given, respectively, by Eqs.~\eqref{eqn:eve_semi} and
\eqref{eqn:leve_semi}. Imposing the condition given by
Eq.~\eqref{eqn:kill_fast}, one finds that $A_{k+1} = Y_k \sum^K_{l=1} l
A_{l+1}$, where $k=2,\ldots,K$ and $Y_k$ is given by Eq.~(\ref{XandY}). In
terms of the $\theta$ and $i_k$ variables, this reads
\begin{align}
i_k = \frac{k d_k}{\gamma \phi(\theta^*)}\,\left[\gamma  {\theta^*}^k-\beta  \left(\phi(\theta){\theta^*}^k-\phi(\theta^*)\theta^k\right)\right]\,\sum^K_{l=1} l i_l, \ \ k=2,\ldots,K,
\label{eqn:slow_ik_intermediate}
\end{align}
and this gives the behavior of $i_k$ which limits the dynamics of the
deterministic system to a slow subspace. A more convenient form can be found by
multiplying Eq.~(\ref{eqn:slow_ik_intermediate}) by $k$ and summing from $k=2$
to $k=K$. This allows us to relate $\sum^K_{l=1} l i_l$ to $i_1$ to find the
following $(K-1)$ expressions for $i_k$:
\begin{align}
i_k&=\frac{i_1}{d_1} \frac{1}{\gamma \theta^* -\beta \left[\theta^* \phi(\theta)-\theta \phi(\theta^*)\right]} k d_k \left[\gamma  {\theta^*}^k-\beta  \left(\phi(\theta){\theta^*}^k-\phi(\theta^*)\theta^k\right)\right], \ \ k=2,\ldots,K.\label{eqn:slow_ik}
\end{align}

It should be stressed that the $(K-1)$ equations (\ref{eqn:slow_ik}), which we
may write in the form $i_k(\tau) = F_k(\theta(\tau), i_1(\tau))$,
$k=2,\ldots,K$, should be interpreted as the time-evolution that the $i_k$ need
to follow, so that $d\bm{x}/d\tau = 0$ in the directions $\bm{v}^{\{ k \} }$,
$k=2,\ldots,K$. This follows from the deterministic equations of motion
$d\bm{x}/d\tau = \bm{A}(\bm{x})$ and from Eq.~(\ref{eqn:kill_fast}). Thus the
only temporal evolution is in the $\bm{v}^{\{ 1 \} }$ and $\bm{v}^{\{ K+1 \} }$
directions. 

\begin{figure}
\centering
\includegraphics[scale=.42]{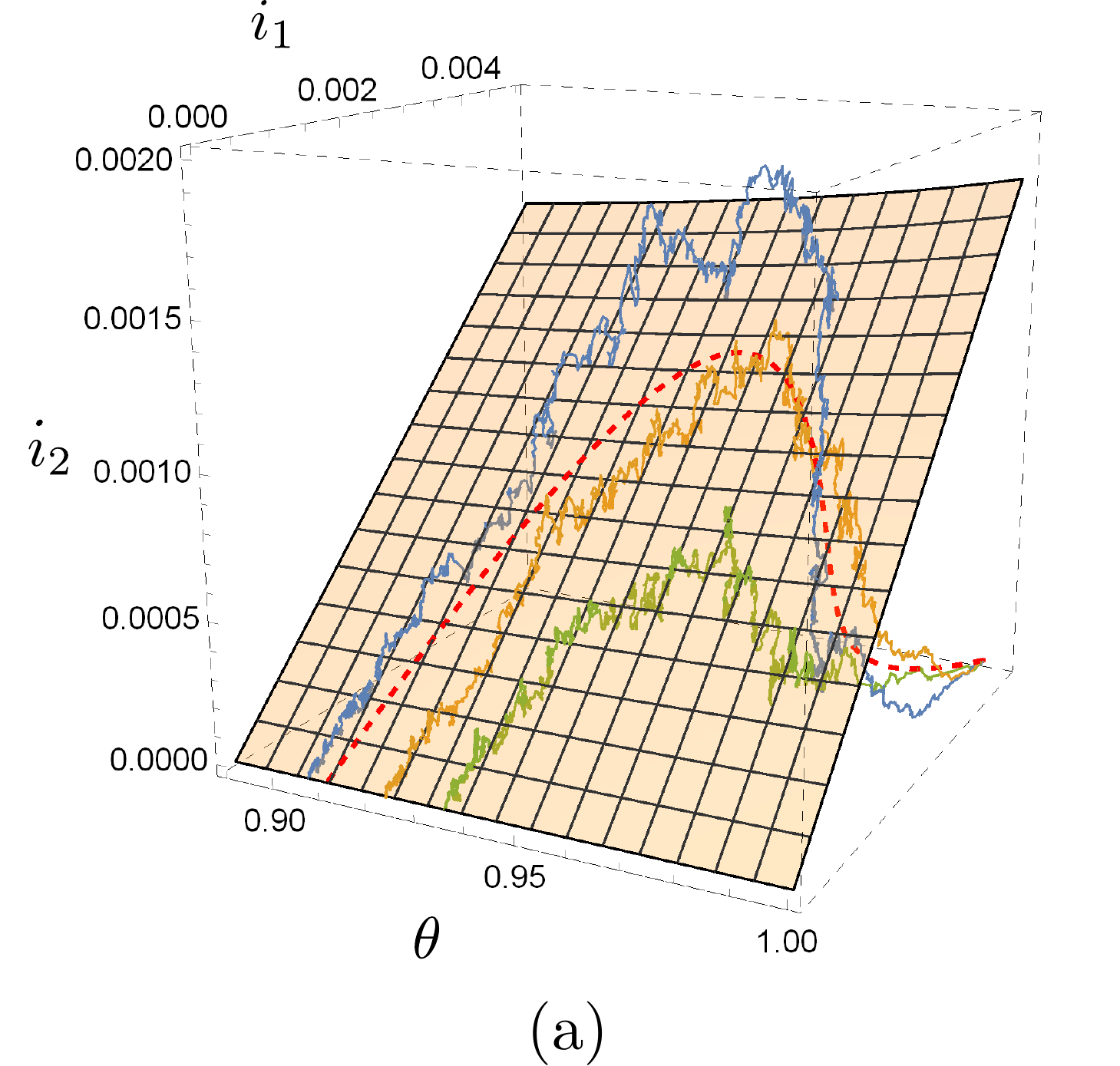}\quad \includegraphics[scale=.42]{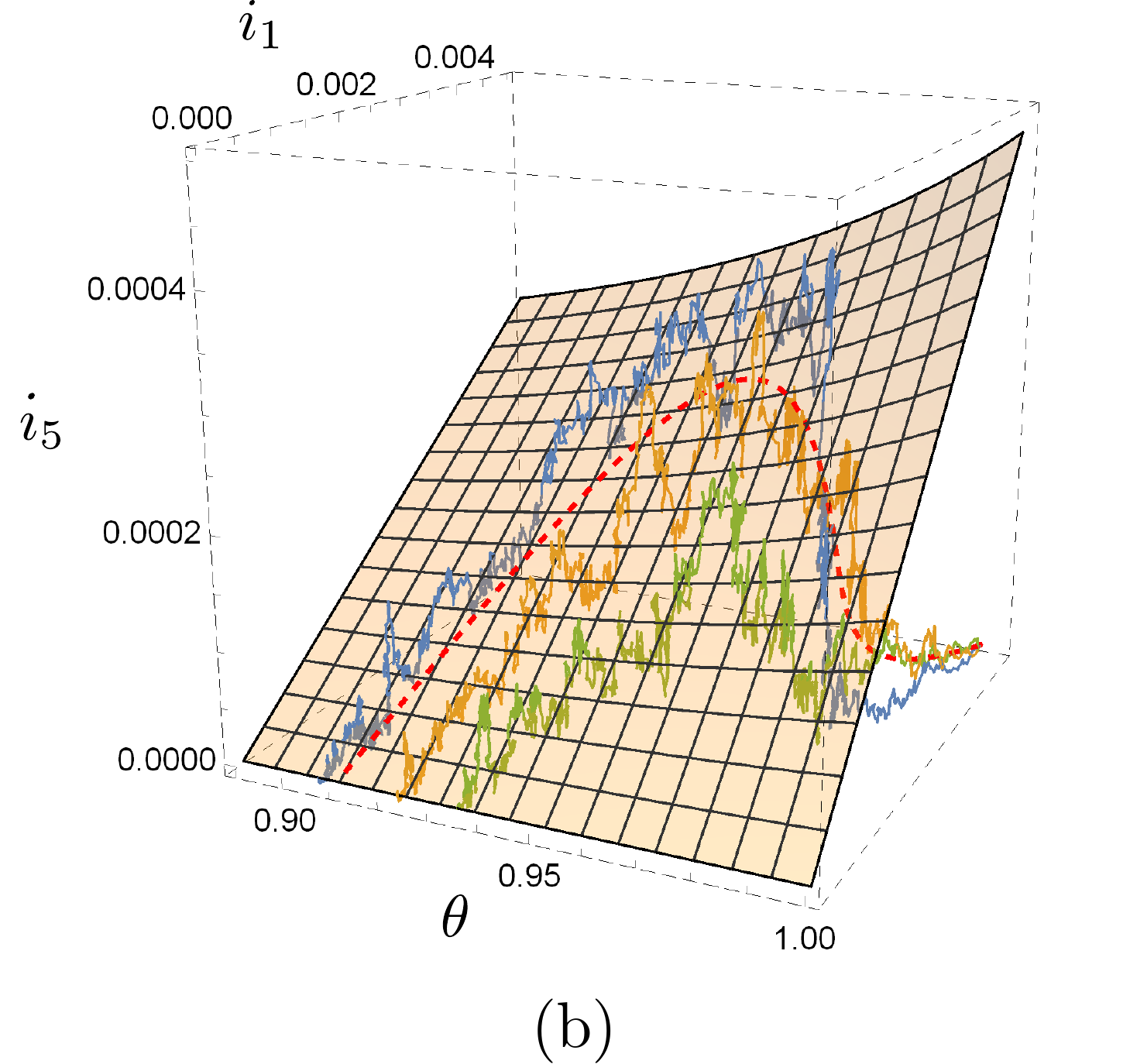}
\caption{Projection of the full epidemic dynamics onto \textbf{(a)} $(\theta,i_1,i_2)$, and \textbf{(b)} $(\theta,i_1,i_5)$, for $K=10$, $\alpha=-2.5$, $\rho=1.2$, $N=10^5$, where $s_k=d_k \theta^k$ and $i_k$ are the fractions of susceptible and infected individuals of degree $k$, respectively. Red, dashed line: deterministic limit, Eqs.~\eqref{eqn:semi_1}--\eqref{eqn:semi_2}; blue, yellow, green lines: stochastic differential equations \eqref{eqn:theta_stoch} and \eqref{eqn:ik}. The initial condition corresponds to $i_{1,0}=0.005$, $i_{k,0}=0$ for $k=2,\ldots,K$, and $G(\theta_0)=1-i_{1,0}$. In both cases, the system is pushed back from the initial condition (lower right in both panels) towards the slow subspace described by Eq.~\eqref{eqn:slow_ik}, on which it seems to stay during the rest of the course of the epidemic. Note the different scales in the $z$ axes of both panels, so the fluctuations away from the slow subspace are small in both cases; the fluctuations on the slow subspace can, however, be large, as seen from the departure of the stochastic trajectories from the deterministic dynamics.}
\label{fig:Slow_km10}
\end{figure}

The functional forms $i_k(\tau) = F_k(\theta(\tau), i_1(\tau))$, with $F_k$
defined by Eq.~(\ref{eqn:slow_ik}), formally only hold sufficiently near the
fixed point that linearization is accurate. However in
Fig.~\ref{fig:Slow_km10} we show three-dimensional projections of the
deterministic dynamics from Eqs.~\eqref{eqn:semi_1}--\eqref{eqn:semi_2} for an
epidemic on a network characterized by a power-law degree distribution with
$\alpha=-2.5$ and maximum degree $K=10$, with heterogeneous basic reproductive
ratio $\rho=1.2$, which gives $\epsilon\approx 0.184$. It appears that the
solid red line, which represents the deterministic dynamics, lies in the slow
subspace for most of the time, not just at late times when the system
approaches the fixed point. That this is the case can be seen from a
consideration of the Jacobian of the system at an arbitrary time, and not just
at the fixed point:
\begin{align}
\label{eqn:J_K_plus_one_arb_time}
\bm{J}^{\{ K+1\}} &=\left(\begin{array}{rr}
-\beta \sum_m m i_m \ & \ - \beta \theta l \\ \\
\beta k^2 d_k \theta^{k-1} \sum_m m i_m \ & \ \beta k l d_k \theta^k - \gamma \delta_{k l} 
\end{array}\right).
\end{align}

It is straightforward to check that $(0,\psi_1,\ldots,\psi_K)^{\top}$ is a
right-eigenvector of the matrix in Eq.~(\ref{eqn:J_K_plus_one_arb_time}) with
eigenvalue $-\gamma$, if $\sum^K_{l=1} l \psi_l = 0$. Therefore, remarkably,
the eigenvectors $\bm{v}^{\{k\}}$ of the Jacobian at the fixed point, defined
in Eq.~(\ref{eqn:eve_semi}), are eigenvectors of the Jacobian at all times. The
existence of this negative eigenvalue with degeneracy $(K-1)$ means that the
system remains in the plane described by the vectors $\bm{v}^{\{ 1 \} }$ and
$\bm{v}^{\{ K+1 \} }$, since any small deviations out of the plane will tend to
be pushed back again. Normally, when carrying out a fast-variable elimination,
one would find that the system possesses a line of fixed points which is
quickly approached at early times, dominated by the deterministic dynamics, and
along which the slow, stochastic dynamics takes place. In this case, however,
the line of fixed points corresponds to an absorbing boundary of the system:
the evolution of the epidemic stops altogether once it reaches the disease-free
state. Therefore, it is important that the projection onto the slow variables
constitutes a good approximation to the dynamics also far from the line of
fixed points.

We parameterize the slow subspace using the coordinates $z_1 \equiv \bm{u}^{\{
1 \} }\cdot\bm{x}$ and $z_2 \equiv \bm{u}^{\{ K+1 \} }\cdot\bm{x}$. Using
Eq.~(\ref{eqn:leve_semi}) and $\bm x = (\theta,i_1,\ldots,i_K)$, we obtain
\begin{align}
z_1 &= \theta + \frac{\beta \theta^*}{-\gamma+\beta \phi(\theta^*)}\sum^K_{k=1} k i_k, \label{eqn:slow_z1}\\
z_2 &= \frac{K d_{K}{\theta^*}^{K}}{\phi(\theta^*)} \sum^K_{k=1} k i_k.  \label{eqn:slow_z2}
\end{align}

A linear transformation of Eqs.~(\ref{eqn:slow_z1}) and (\ref{eqn:slow_z2})
shows that an equivalent set of coordinates are $\theta$ and $\sum_k k i_k$,
previously called $\lambda$. So, in summary, in Sec.~\ref{sec:det} we showed
the system of $2K$ equations leads to a closed set of equations in $\theta$ and
$\lambda$. Here, through a consideration of the slow modes of the deterministic equations, we have shown that not only can we describe the system in terms of $\theta$ and $\lambda$, but that it remains in the vicinity of a subspace that can be parameterized by these variables.

We would like to carry this reasoning over to the stochastic case, to find a
reduced set of SDEs, but which are closed, i.e.,~which can be expressed entirely
in terms of the slow variables $z_1$ and $z_2$ (or $\theta$ and $\lambda$). Before starting the actual treatment of the model, we visualize its dynamics to find whether the system evolves towards a lower-dimensional subspace, as it does in the deterministic case. Figure~\ref{fig:Slow_km10} also shows three-dimensional projections of stochastic trajectories from Eqs.~\eqref{eqn:theta_stoch} and \eqref{eqn:ik}, for a total population of $N=10^5$. The behavior is found indeed to be similar to that
of the deterministic case: the dynamics seems to quickly reach, and then fluctuate around, the deterministic slow subspace, with the early behavior dominated by the deterministic dynamics. Therefore, we expect that we should be able to effectively reduce the stochastic system from $(K+1)$ dimensions to two dimensions, by neglecting the fast initial behavior of the epidemic. 

We will do this by projecting the stochastic differential equations,
Eqs.~\eqref{eqn:theta_stoch} and \eqref{eqn:ik}, onto the slow degrees of
freedom of the deterministic limit of the model; that is, we will neglect the
fluctuations away from the slow subspace, along the $(K-1)$ fast directions
corresponding to the most negative eigenvalue. This can be formulated
mathematically  through the application of a projection operator that only
picks the components of $\bm{x}$, $\bm{A}(\bm{x})$, and $\bm \eta(\tau)$ along
the slow eigenvectors of the deterministic limit, $\bm{v}^{\{1\}}$ and
$\bm{v}^{\{K+1\}}~$\cite{constable2014fast}:
\begin{align}
\bm{P} &= \bm{v}^{\{1\}} [\bm{u}^{\{1\}}]^\top + \bm{v}^{\{K+1\}}[\bm{u}^{\{K+1\}}]^\top.
\label{eqn:projector}
\end{align}
The details of the derivation are given in Appendix~\ref{sec:projection}, where
it is shown that the SDEs given by Eqs.~(\ref{eqn:theta_stoch}) and
(\ref{eqn:lambda_stoch}) of Sec.~\ref{sec:det} are recovered, except that now
$\bar{\sigma}_3\equiv \sqrt{\bar{B}_{33}}$ is determined in terms of $\theta$
and $\lambda$. This closed system of stochastic differential equations for
$\theta$ and $\lambda$, which we call the \textit{reduced mesoscopic model},
constitutes the main result of this paper and so we reproduce it here:
\begin{align}
\frac{d\theta}{d\tau} &= -\beta \theta \lambda + \frac{1}{\sqrt{N}}\bar{\sigma}_1 \zeta_1(\tau),\label{eqn:slow_theta}\\
\frac{d\lambda}{d\tau} &= \lambda (\beta \phi(\theta)-\gamma) +\frac{1}{\sqrt{N}}\left[-\frac{\phi(\theta)}{\theta}\bar{\sigma}_1 \zeta_1(\tau) +\bar{\sigma}_2\zeta_2(\tau) + \bar{\sigma}_3 \zeta_3(\tau)\right], \label{eqn:slow_lambda}
\end{align}
where $\bar{\sigma}_1^2 =\left[G^\prime (\theta)\right]^{-1}\beta \lambda
\theta$, $\bar{\sigma}_2^2 = \beta \lambda [\phi(\theta) + \psi(\theta)] -
\bar{\sigma}^2_1$, and 
\begin{align}
\bar{\sigma}_3^2 &= \lambda\left(\gamma \frac{\phi(\theta^*)+\psi(\theta^*)}{\phi(\theta^*)} + \frac{\beta}{\phi(\theta^*)} \left[\phi(\theta^*)\psi(\theta)-\phi(\theta)\psi(\theta^*)\right]\right).
\label{sigma_3_explicit}
\end{align}

\begin{figure}
\centering
\subfigure{\includegraphics[scale=0.42]{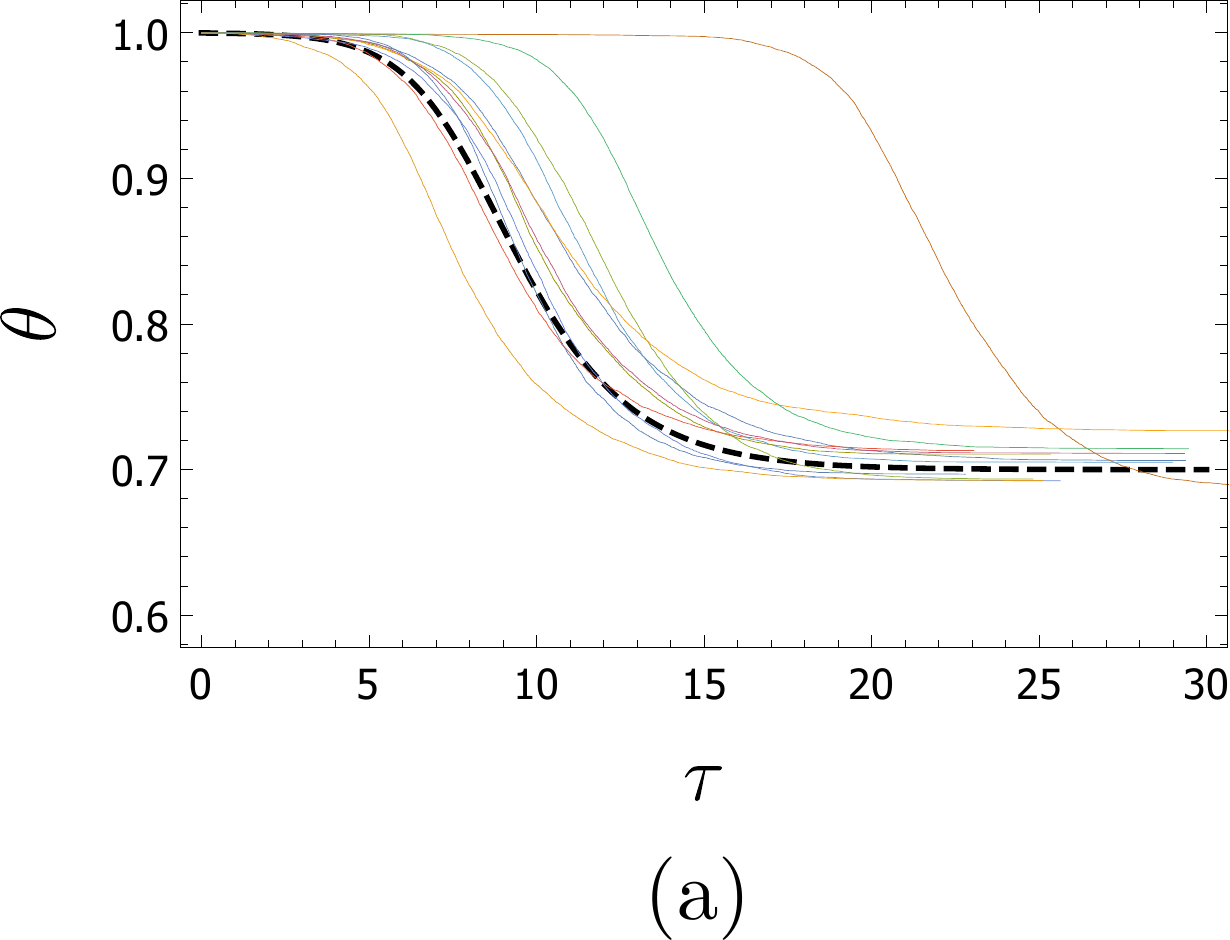}\quad \includegraphics[scale=0.42]{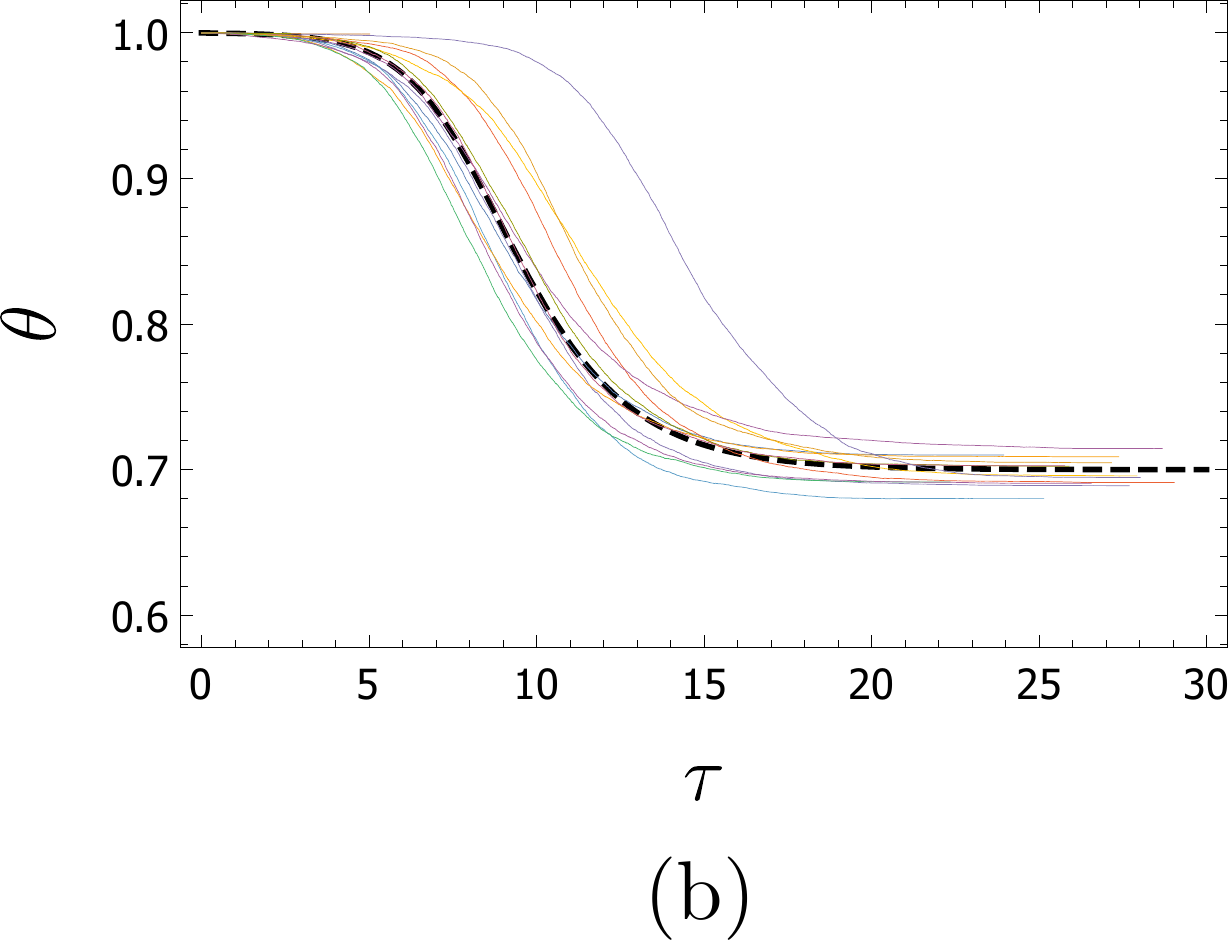}}\\
\ \\
\subfigure{\includegraphics[scale=0.42]{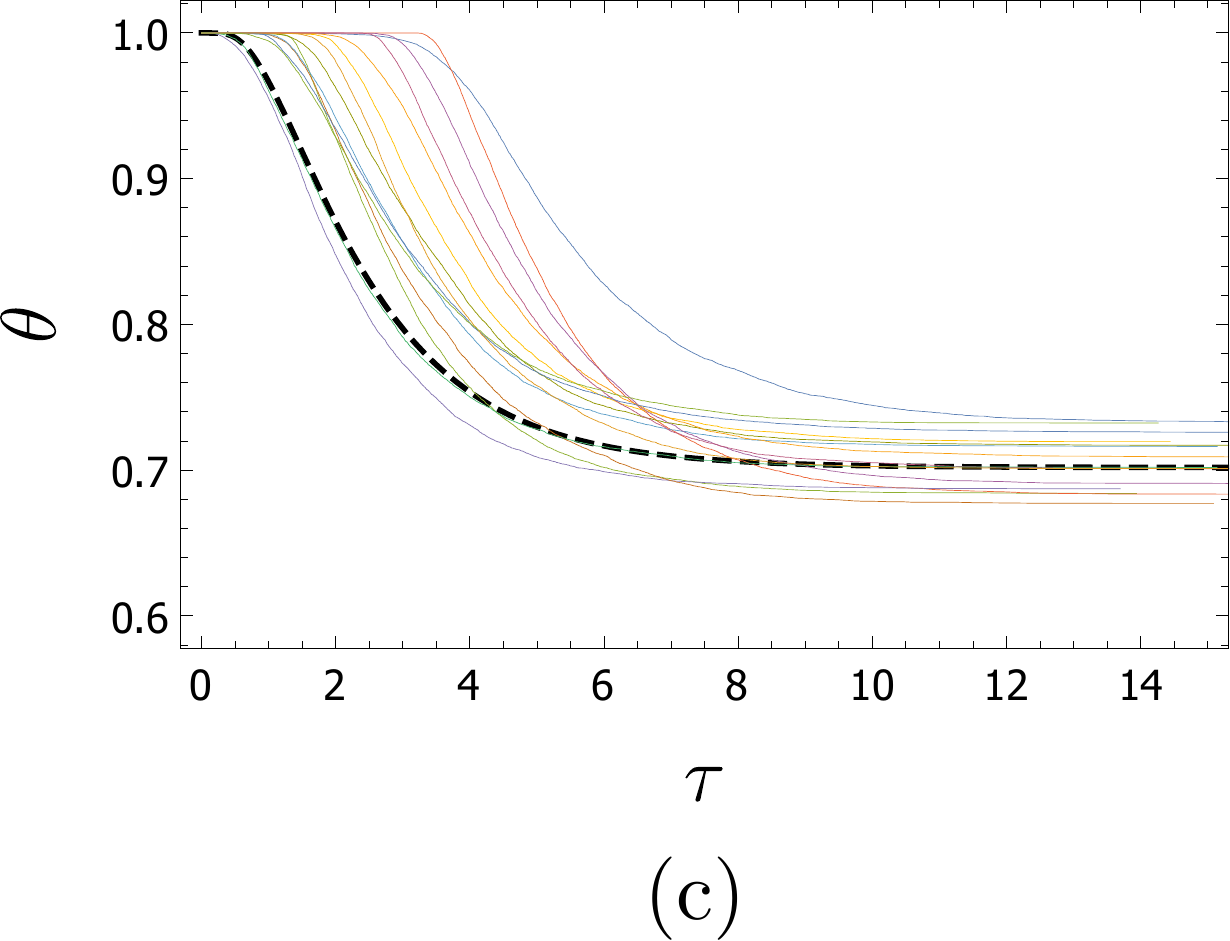}\quad \includegraphics[scale=0.42]{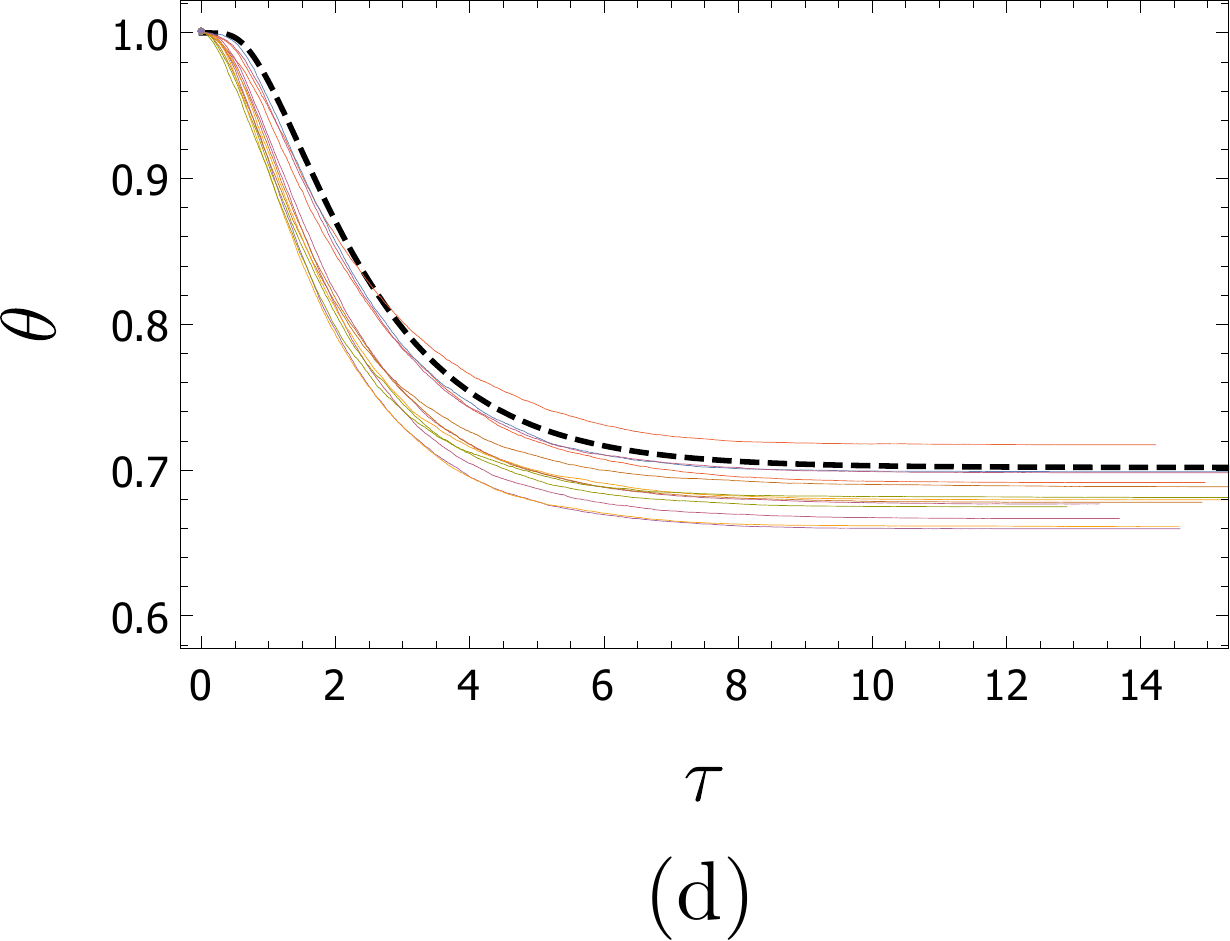}}\\
\ \\
\subfigure{\includegraphics[scale=0.42]{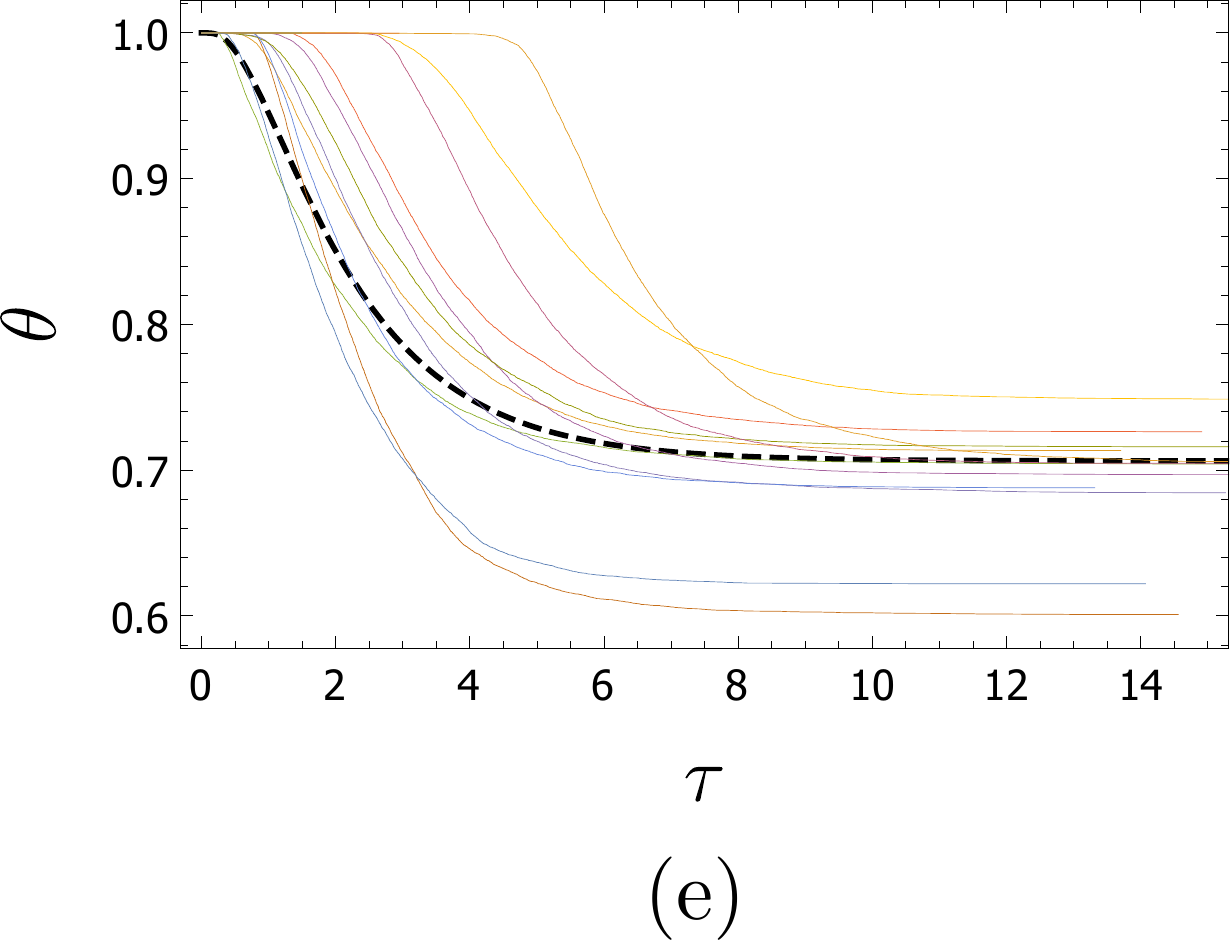}\quad \includegraphics[scale=0.42]{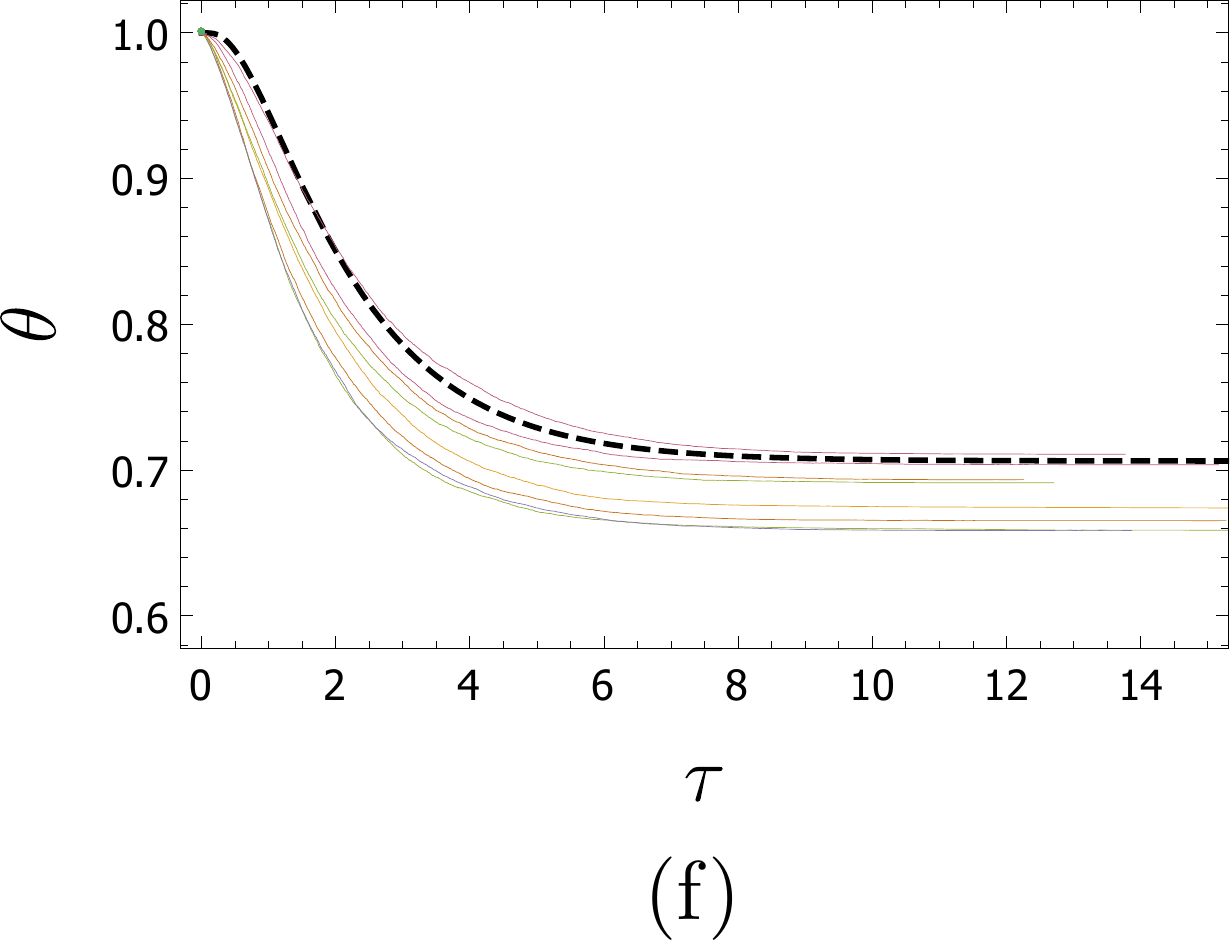}}\\
\caption{Time series of $\theta$ for $N=20000$, $\gamma=1$, and $\theta^*=0.7$, with \textbf{(a)--(b)} $K=10$, \textbf{(c)--(d)} $K=1000$, and \textbf{(e)--(f)} $K=19999$. Colored lines correspond to realizations of the full microscopic model (left panels) and the reduced mesoscopic model~\eqref{eqn:slow_theta}--\eqref{eqn:slow_lambda} (right panels) started with a total number of 5 infectives with degrees sampled from a truncated Zipf distribution with $\alpha=-2.5$. The black, dashed line corresponds to the deterministic solution.}
\label{fig:ts_full_red_3}
\end{figure}

\begin{figure}
\centering
\subfigure{\includegraphics[scale=0.42]{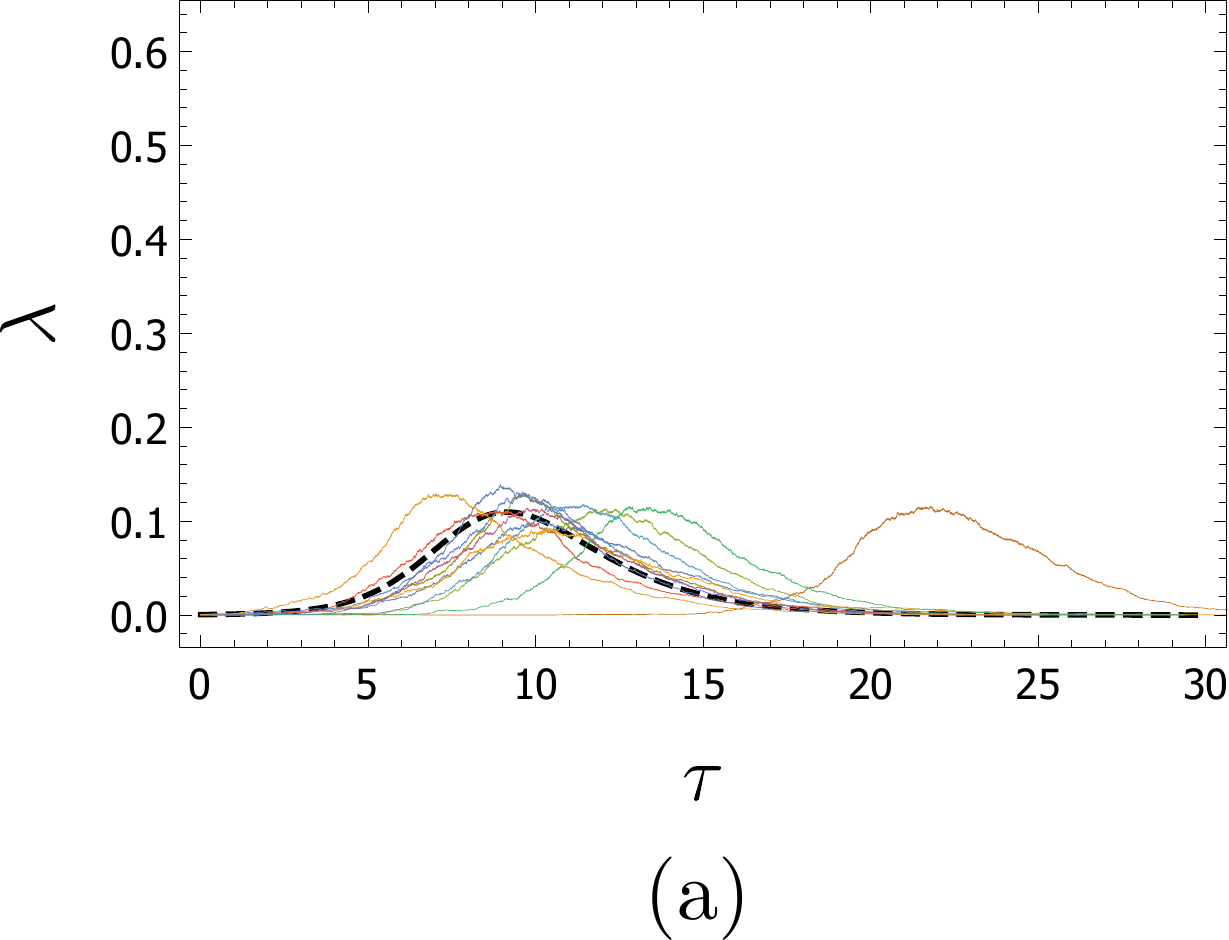}\quad \includegraphics[scale=0.42]{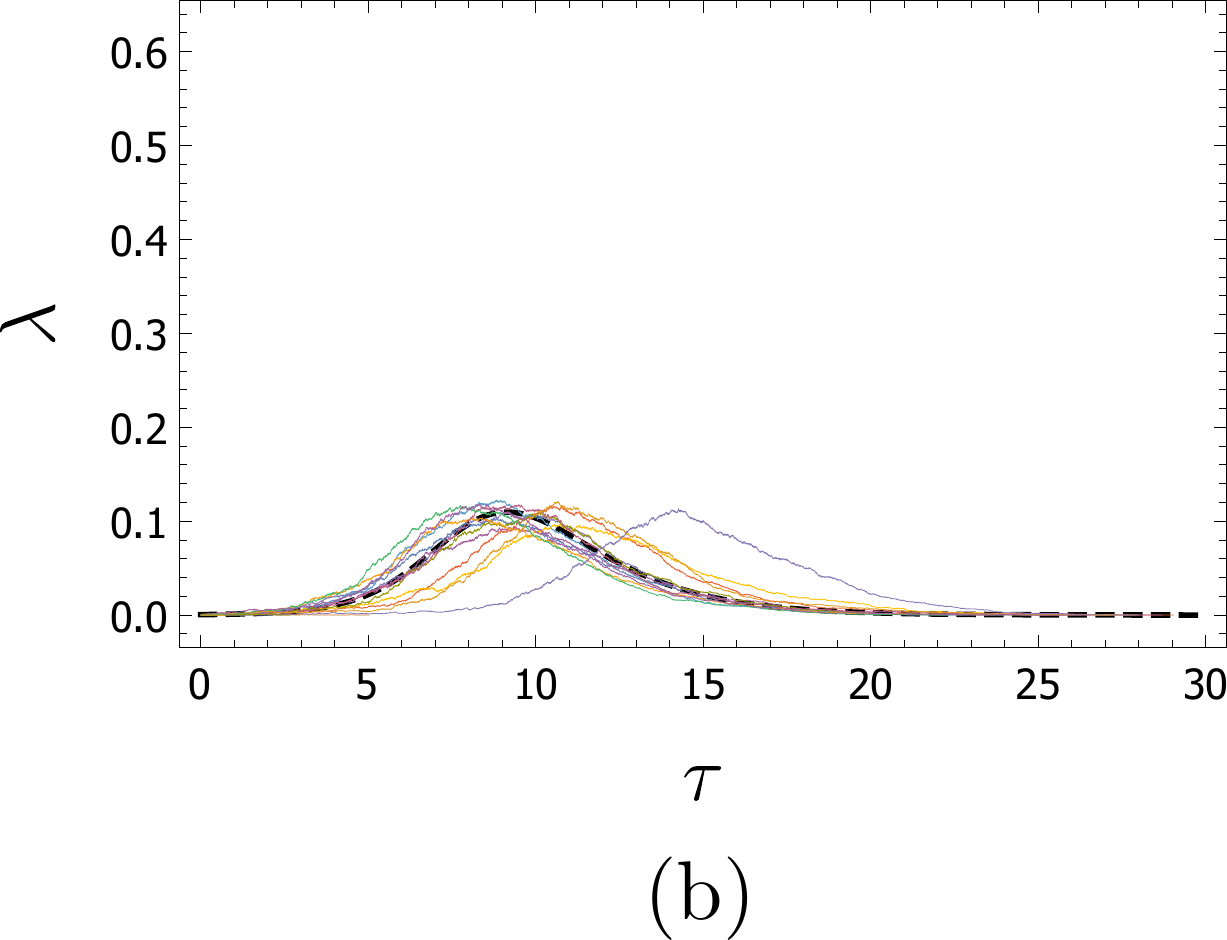}}\\
\ \\
\subfigure{\includegraphics[scale=0.42]{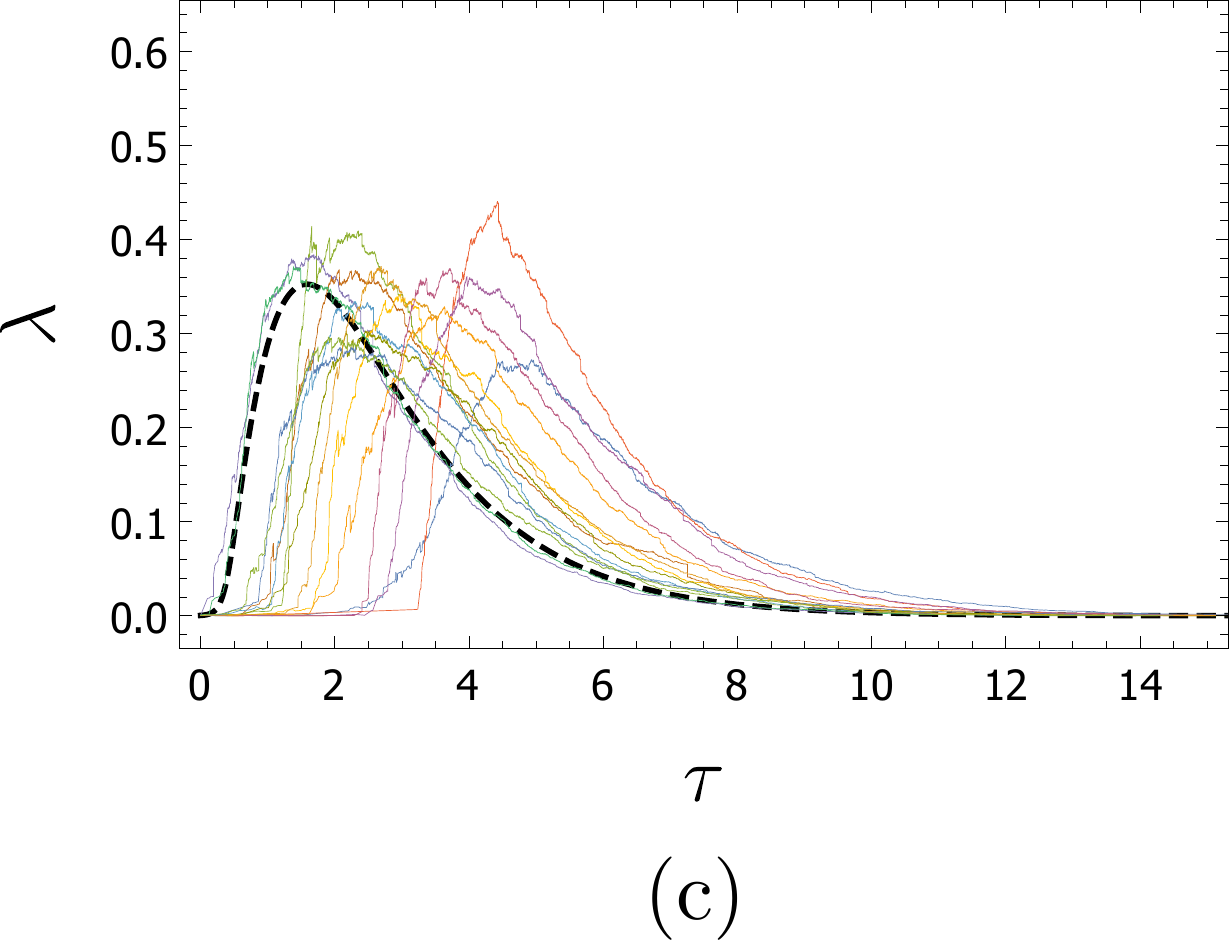}\quad \includegraphics[scale=0.42]{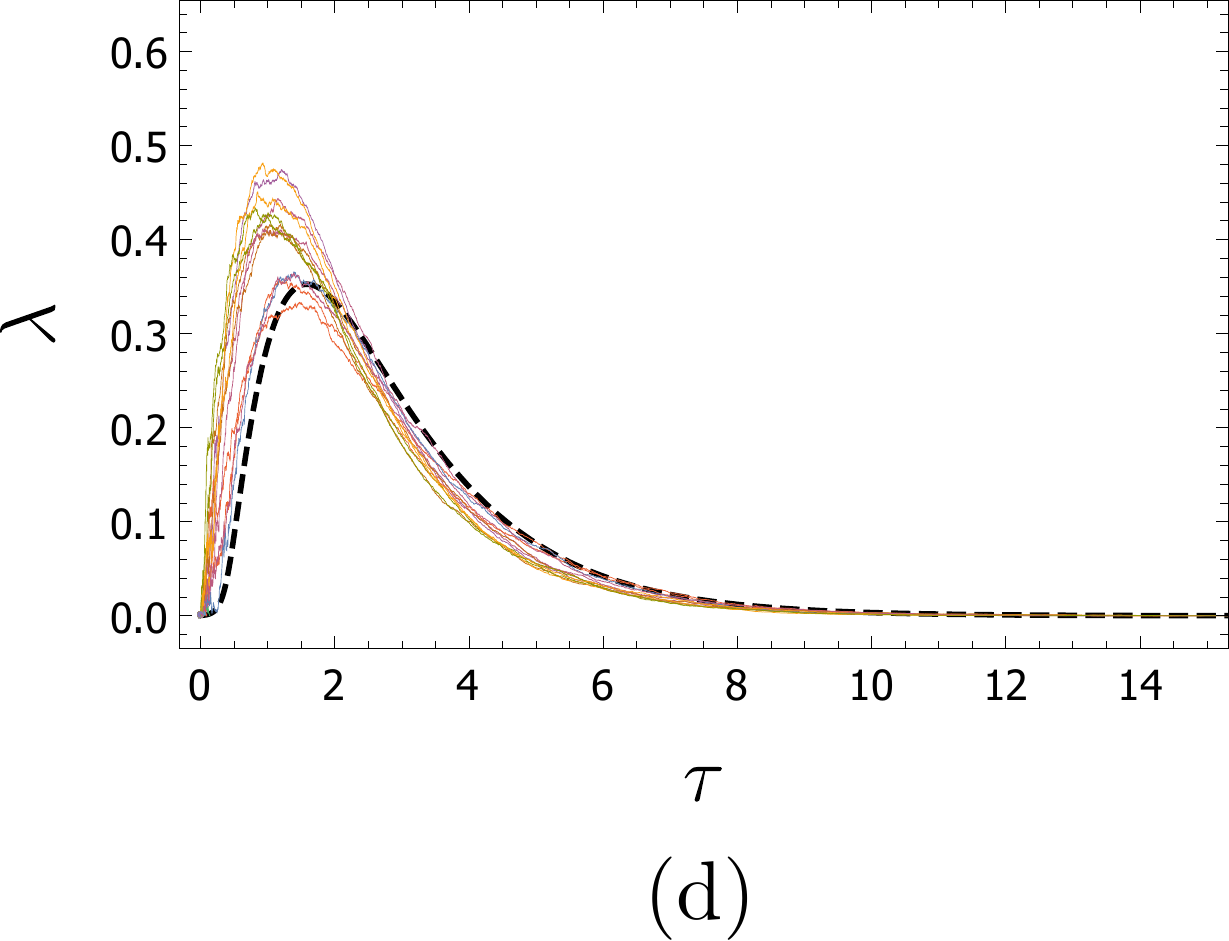}}\\
\ \\
\subfigure{\includegraphics[scale=0.42]{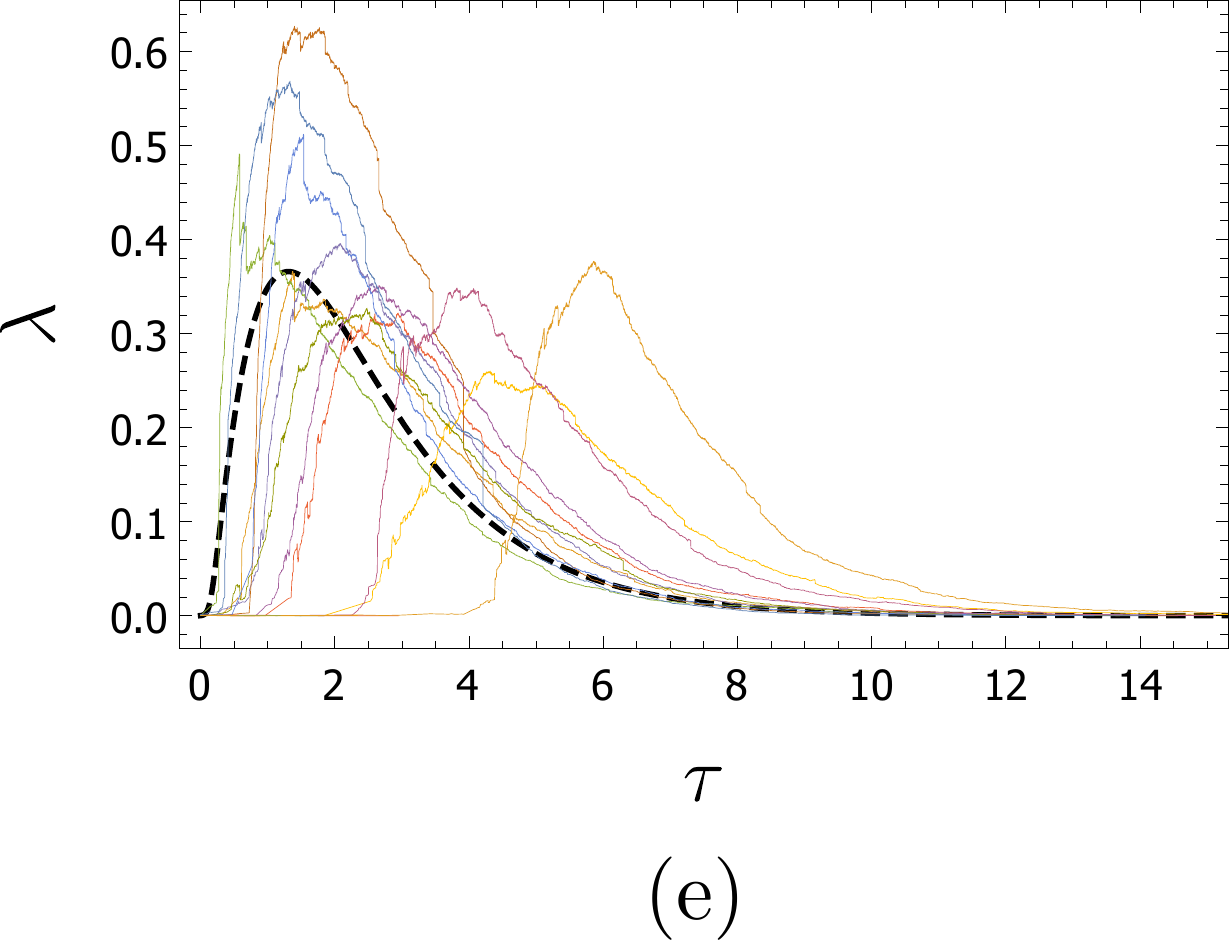}\quad \includegraphics[scale=0.42]{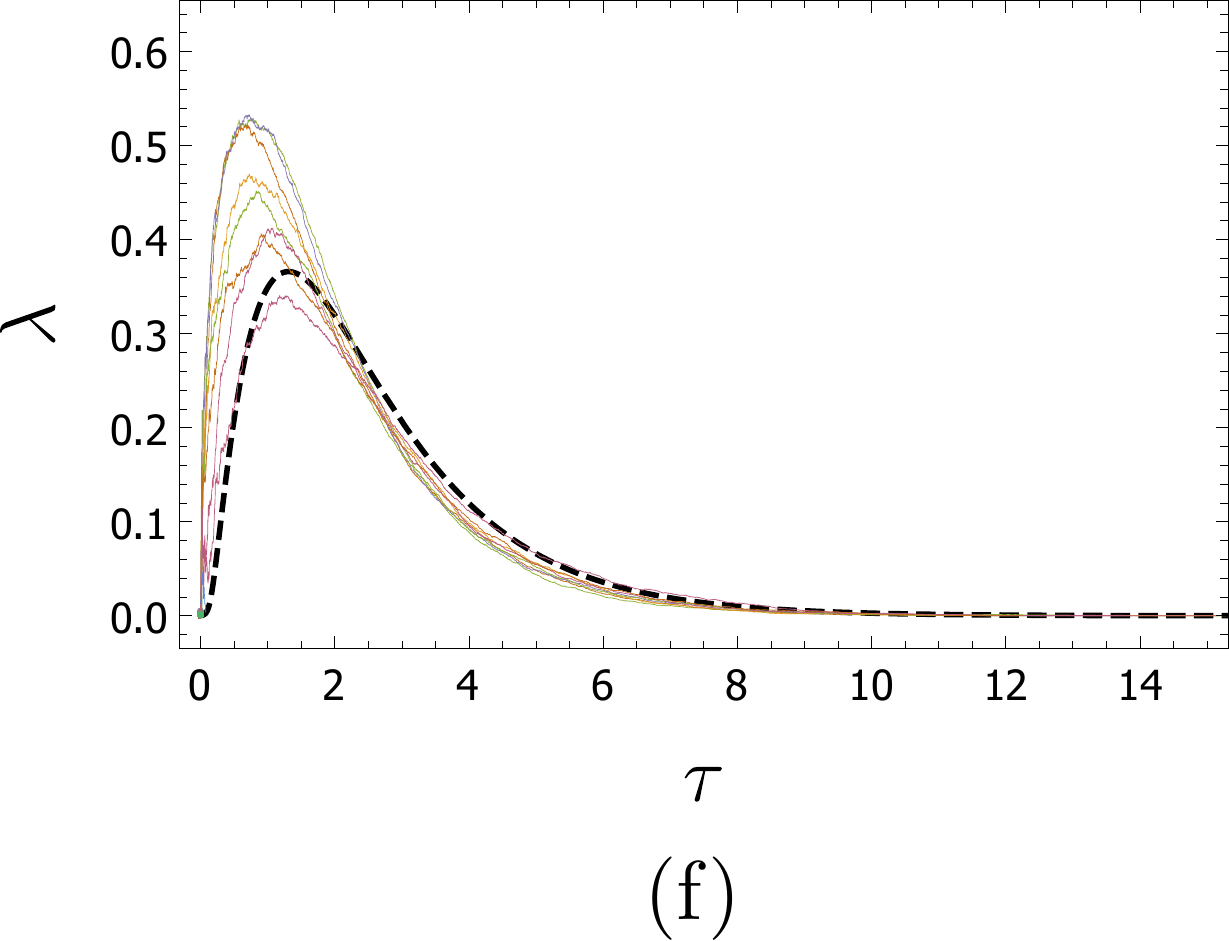}}\\
\caption{Time series of $\lambda$ for $N=20000$, $\gamma=1$, and $\theta^*=0.7$, with \textbf{(a)--(b)} $K=10$, \textbf{(c)--(d)} $K=1000$, and \textbf{(e)--(f)} $K=19999$. Colored lines correspond to realizations of the full microscopic model (left panels) and the reduced mesoscopic model~\eqref{eqn:slow_theta}--\eqref{eqn:slow_lambda} (right panels) started with a total number of 5 infectives with degrees sampled from a truncated Zipf distribution with $\alpha=-2.5$. The black, dashed line corresponds to the deterministic solution.}
\label{fig:ts_full_red_4}
\end{figure}

\begin{figure}
\centering
\subfigure{\includegraphics[scale=0.42]{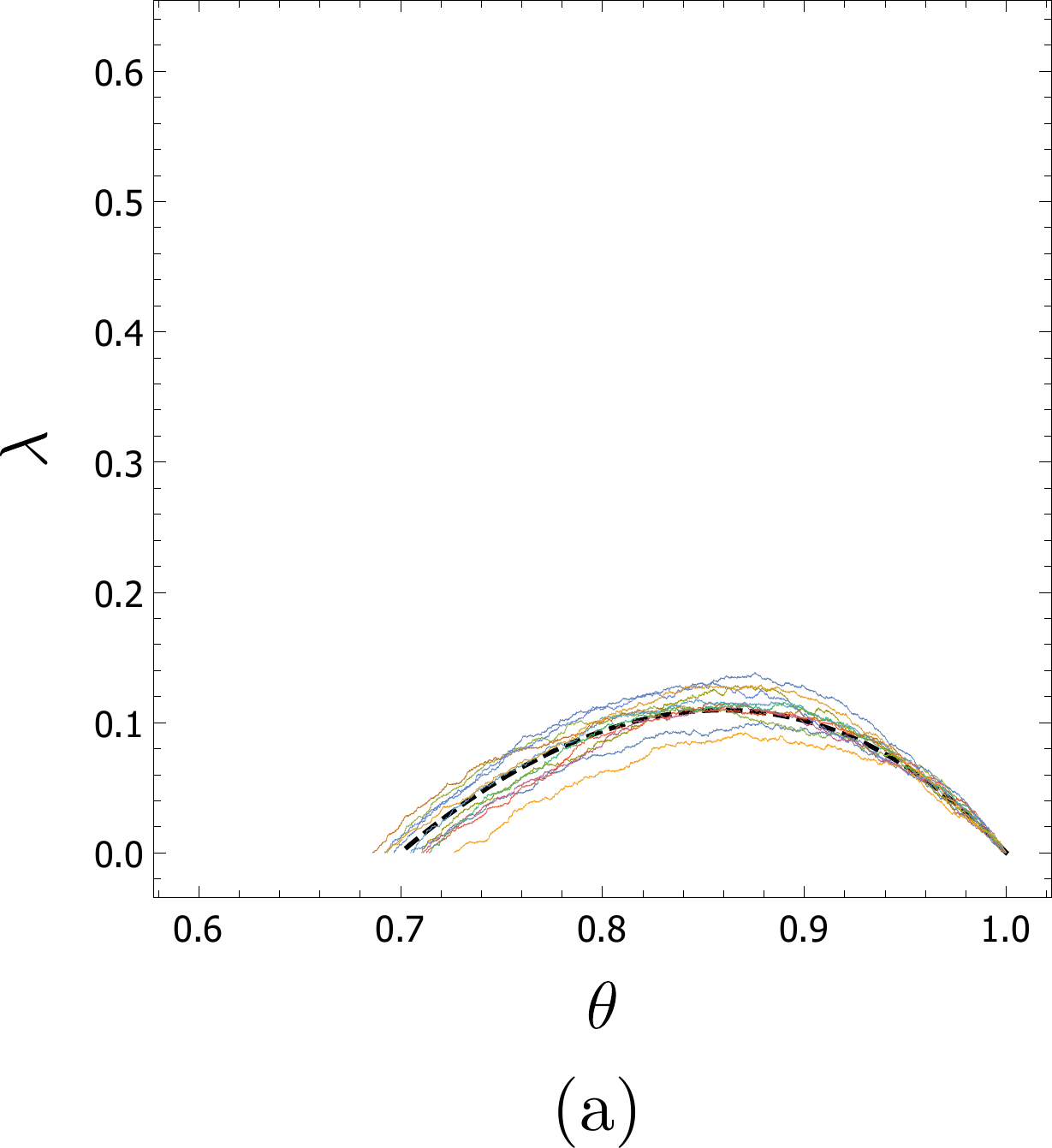}\quad \includegraphics[scale=0.42]{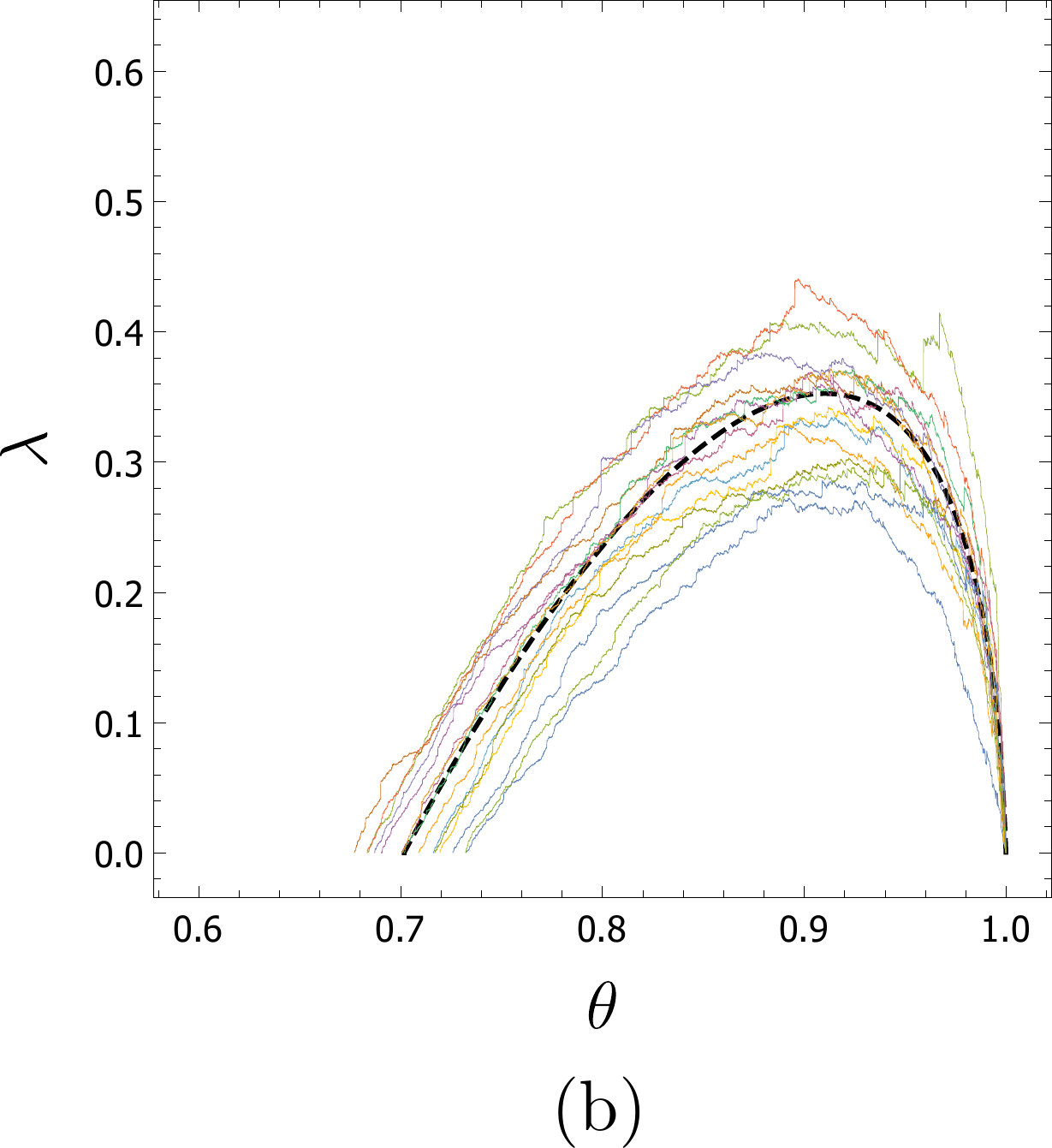}\quad \includegraphics[scale=0.42]{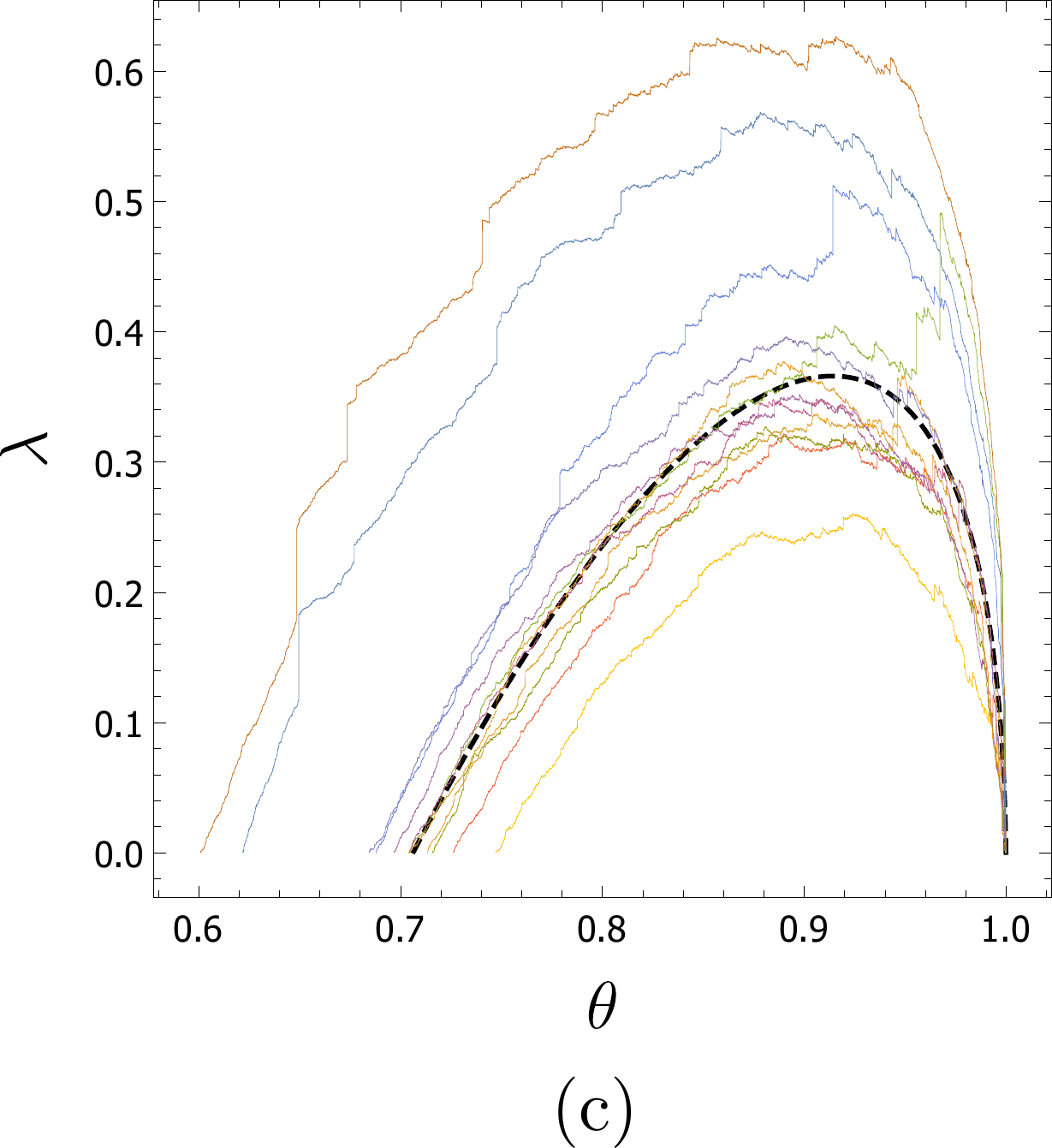}}\\
\ \\
\subfigure{\includegraphics[scale=0.42]{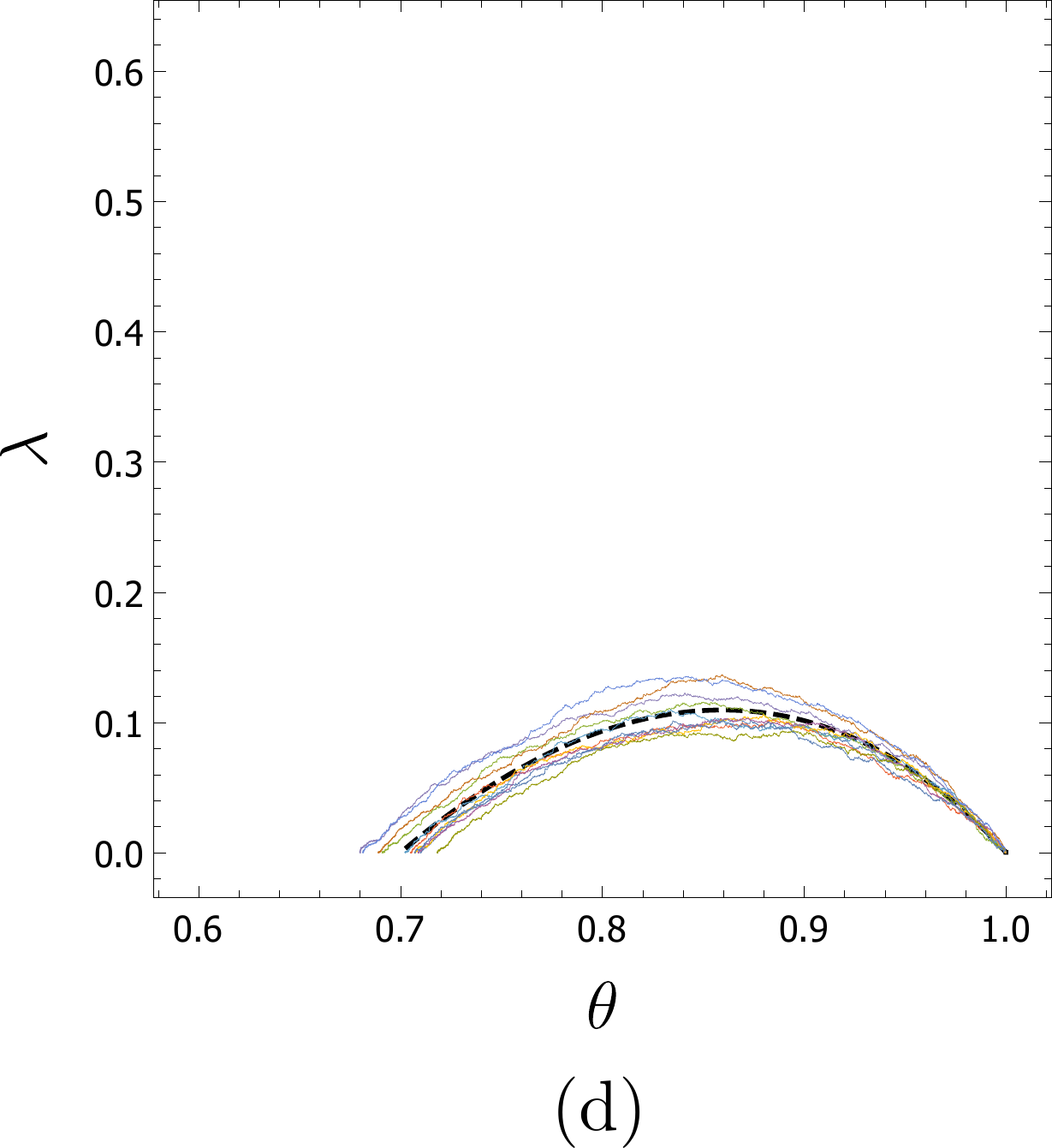}\quad \includegraphics[scale=0.42]{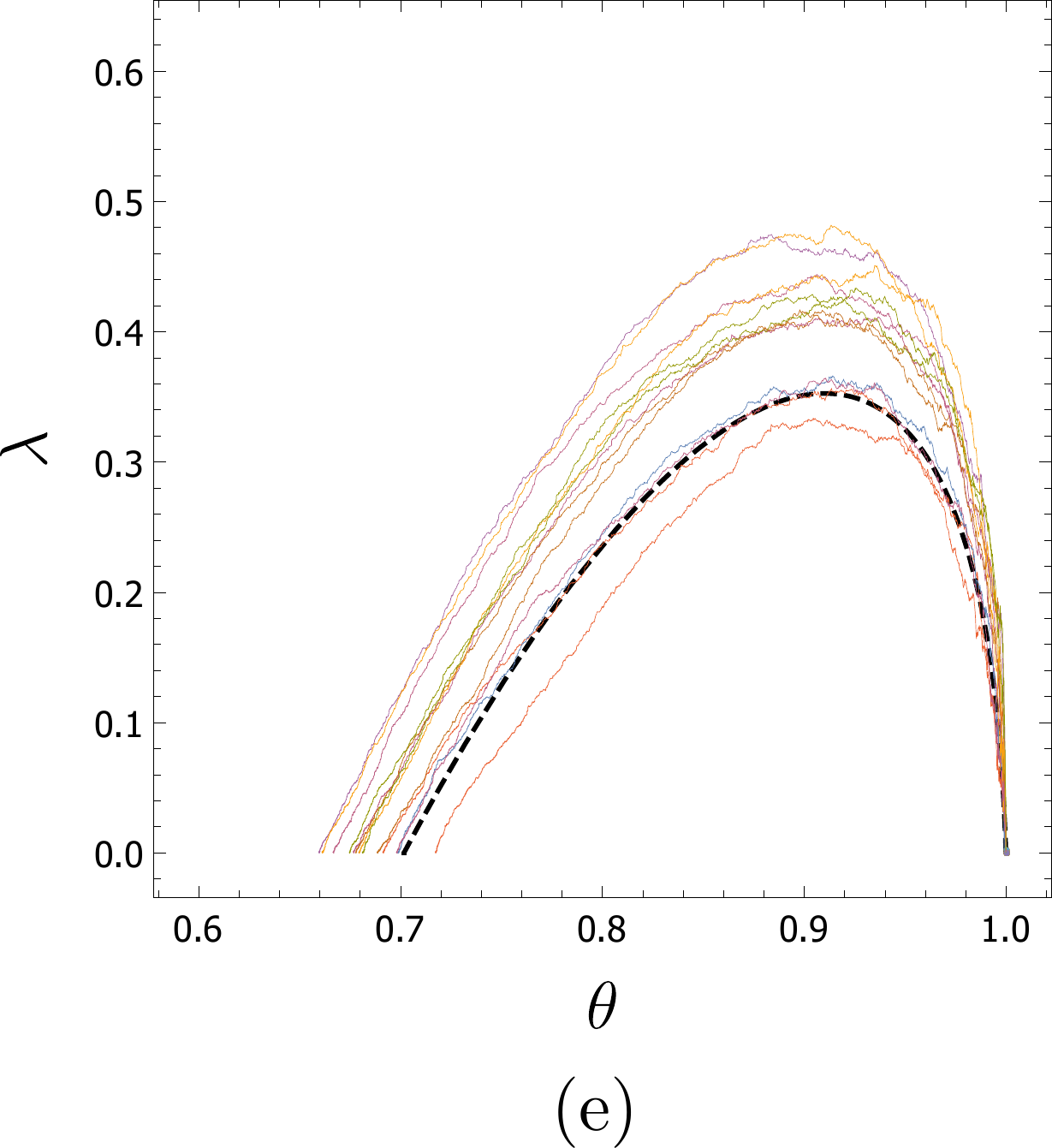}\quad \includegraphics[scale=0.42]{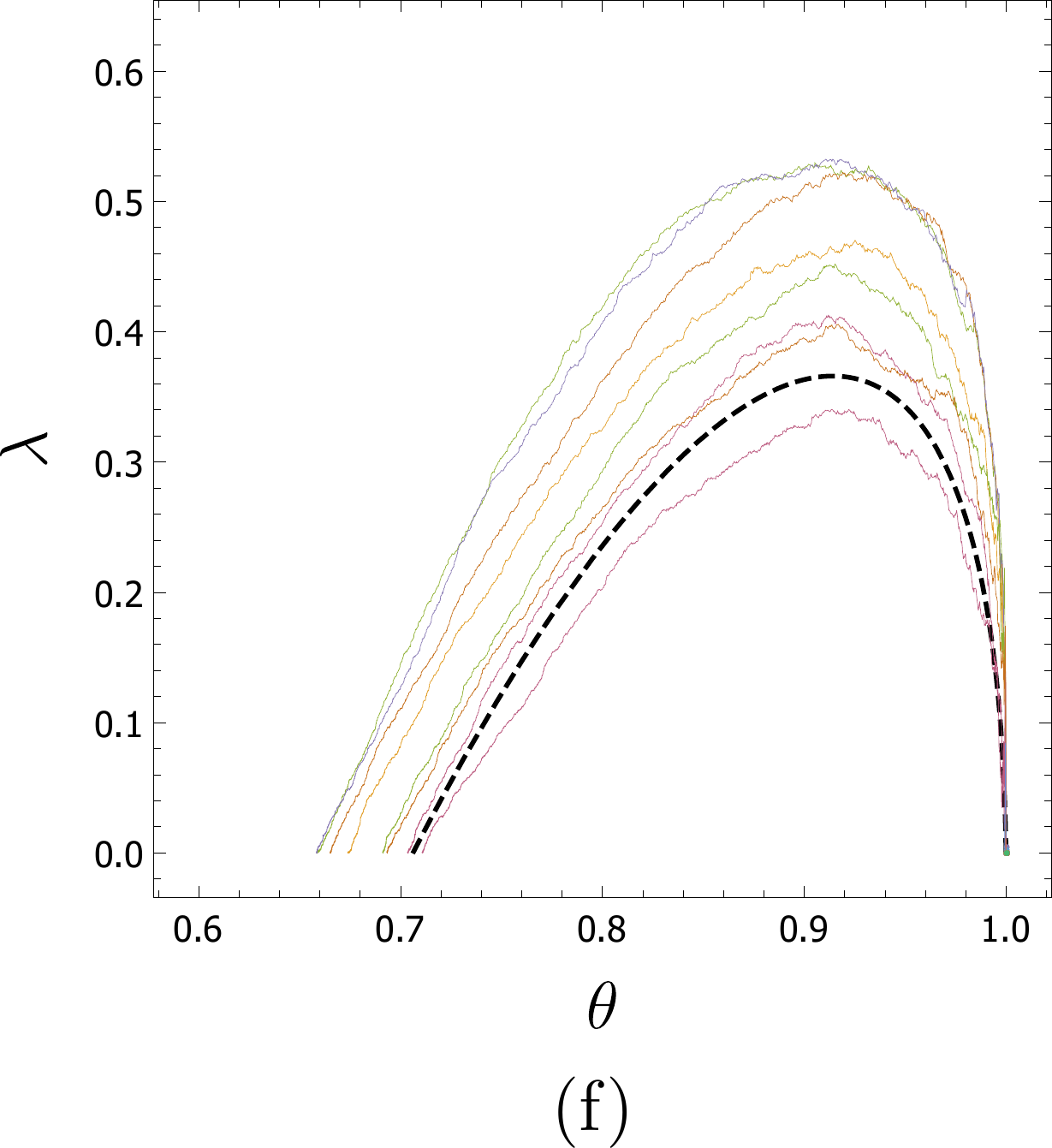}}
\caption{Phase diagrams for $N=20000$, $\gamma=1$, and $\theta^*=0.7$, for \textbf{(a)--(c)} the full microscopic model, and \textbf{(d)--(f)} the reduced mesoscopic model~\eqref{eqn:slow_theta}--\eqref{eqn:slow_lambda}. Colored lines correspond to stochastic realizations started with a total number of 5 infectives with degrees sampled from a truncated Zipf distribution with $\alpha=-2.5$ and $K=10$ (left), $K=1000$ (center), and $K=19999$ (right). The black, dashed line corresponds to the deterministic solution.}
\label{fig:phase_full_red_2}
\end{figure}

We now compare the dynamics of the full microscopic model, defined by
Eqs.~\eqref{eqn:rate_1}--\eqref{eqn:master}, to that of the reduced mesoscopic
model above. Figures~\ref{fig:ts_full_red_3} and \ref{fig:ts_full_red_4} show
the time series of the full microscopic model in terms of the variables
$\theta$ and $\lambda$, and that of the reduced mesoscopic model, for $N=20000$
and different values of $K$ having the same deterministic final size
$\theta^*$, obtained from Eq.~\eqref{eqn:theta_fix};
Figure~\ref{fig:phase_full_red_2}, in turn, compares the phase diagrams. We see
that, in the case of small $K$, the time series of both variables in the full
microscopic model are rather smooth, and are dominated by a horizontal spread
due to the fact that the epidemic takes off at different moments. As
heterogeneity increases, the dynamics of the system becomes noisier and
noisier, as seen in the phase diagrams; however, the time series of $\theta$
continues to be smooth. This observation will be useful in the following
section.

From the time series of the reduced mesoscopic model we see that, except for
small $K$, the temporal behavior of $\theta$ and $\lambda$ does not present the
horizontal spread observed in the full microscopic model. However, once the
epidemic has taken off, the dynamics of the full microscopic model is correctly
captured when $K$ is not so large; this can be seen from the phase diagrams. We
note that the relative magnitude of the eigenvalues for this choice of
parameters takes the values $\epsilon\approx 0.44\text{--}0.65$, and that the
reduced mesoscopic model fares well despite $\epsilon$ being rather large. We
find slightly worse agreement between the full and reduced models for smaller
values of $\theta^*$, corresponding to larger $\epsilon$. One further detail to
add is that in the case of the reduced model there are far fewer trajectories
in which the epidemic does take off, compared to the full microscopic model;
that is, the reduced model results in an over-representation of very early
extinctions.

Note that, for the value of $N$ employed in these figures, $K=19999$ is the
maximum possible degree cutoff. In this case, we would not expect the reduced
model to yield accurate results. This is due to the fact that not even the full
$2K$-dimensional mesoscopic model (Eqs.~\eqref{eqn:sk}--\eqref{eqn:ik})
provides a good approximation to the microscopic dynamics in this case, since
$K$ is of the order of $N$ and this would need to be explicitly taken into
account when performing the mesoscopic expansion leading to the Fokker-Planck
equation~\eqref{eqn:FPE}. The implementation of the reduced mesoscopic model, however, is much more practical than that of the full mesoscopic model: while we have found that simulations of the latter take a time that, as its number of equations, grows linearly with the degree of heterogeneity of the network, for the reduced version this time, besides being much shorter, appears to scale as $\sqrt{K}$.

We end this section with some further observations regarding the reduced
stochastic system. First, the existence of a $(K-1)$-fold degenerate eigenvalue
equal to $-\gamma$ throughout the time evolution also means that stochastic
trajectories, as well as deterministic ones, which venture out of the slow
subspace will tend to be pushed back into the plane. Thus we would expect the
reduced set of SDEs to be a good approximation throughout the motion, and not
just at late times. This may be the reason why the approximation works well
even if $\epsilon$ is not particularly small. Second, in retrospect, the
elimination of the fast variables in the stochastic system is not strictly
necessary, since the SDEs previously obtained in Sec.~\ref{sec:det} are
recovered, and $\bar{\sigma}_3$ could be found in terms of $\theta$ and
$\lambda$ using Eq.~(\ref{eqn:slow_ik}) obtained from the elimination of fast
variables in the deterministic system. One can compare the value of
$\bar{\sigma}_3$ in terms of the $i_k$ to the one in terms of $\lambda$ in the
deterministic limit and see that, indeed, the approximation is not very good
for moderate to large values of $\epsilon$. This does not seem to play a huge
role when considered as one of many terms in the stochastic evolution of
$\lambda$.

Finally, we can make a general comment about the nature of the fluctuations as
the epidemic starts to grow from its initial state. In this case the relevant eigenvalues will be
those of the Jacobian Eq.~(\ref{eqn:J_K_plus_one_arb_time}), but with $i_k =0$
and $\theta = 1$, and will therefore be given by those in
Eq.~(\ref{eqn:eva_semi}), but with $\theta^*$ set equal to $1$. Therefore, the
largest eigenvalue is now $(\rho - 1)\gamma$. In this case, if $\rho > 1$, so
that an epidemic starts to grow, the dynamics is pushed away from the line
$(\theta,0,\ldots,0)$ in its early stages, as can be seen from
Fig.~\ref{fig:Slow_km10}, where the trajectories leave the vicinity of the absorbing state after reaching the slow subspace.

\section{Further reduction}\label{sec:CIR}

In the previous section, we have derived a reduced two-dimensional system with
three white-noise processes, the reduced mesoscopic model
Eqs.~\eqref{eqn:slow_theta}--\eqref{eqn:slow_lambda}, which provides a
good approximation to the dynamics of the full microscopic model,
provided the maximum degree individuals can pick is not so large. This is true
even for relatively large values of $\epsilon$, corresponding to small
separation between the eigenvalues of the system. We have also noted that the
time series of $\theta$, obtained from the full microscopic model, presents
very smooth behavior over the range of parameters explored. Inspired by this
observation, we attempt to further simplify the model by noting that
Eqs.~\eqref{eqn:slow_theta}--\eqref{eqn:slow_lambda} have the form
\begin{align}
\frac{d\theta}{d\tau} &= -\beta \theta \lambda + \sqrt{\frac{\lambda}{N}}f_1(\theta)\zeta_1(\tau),\label{eqn:slow_theta_2}\\
\frac{d\lambda}{d\tau} &= \lambda \left(\beta \phi(\theta)-\gamma \right)- \sqrt{\frac{\lambda}{N}}f_2(\theta)\zeta_1(\tau) + \sqrt{\frac{\lambda}{N}}f_3(\theta)\zeta_2(\tau)+ \sqrt{\frac{\lambda}{N}}f_4(\theta)\zeta_3(\tau).\label{eqn:slow_lambda_2}
\end{align}

\begin{figure}
\centering
\subfigure{\includegraphics[scale=0.42]{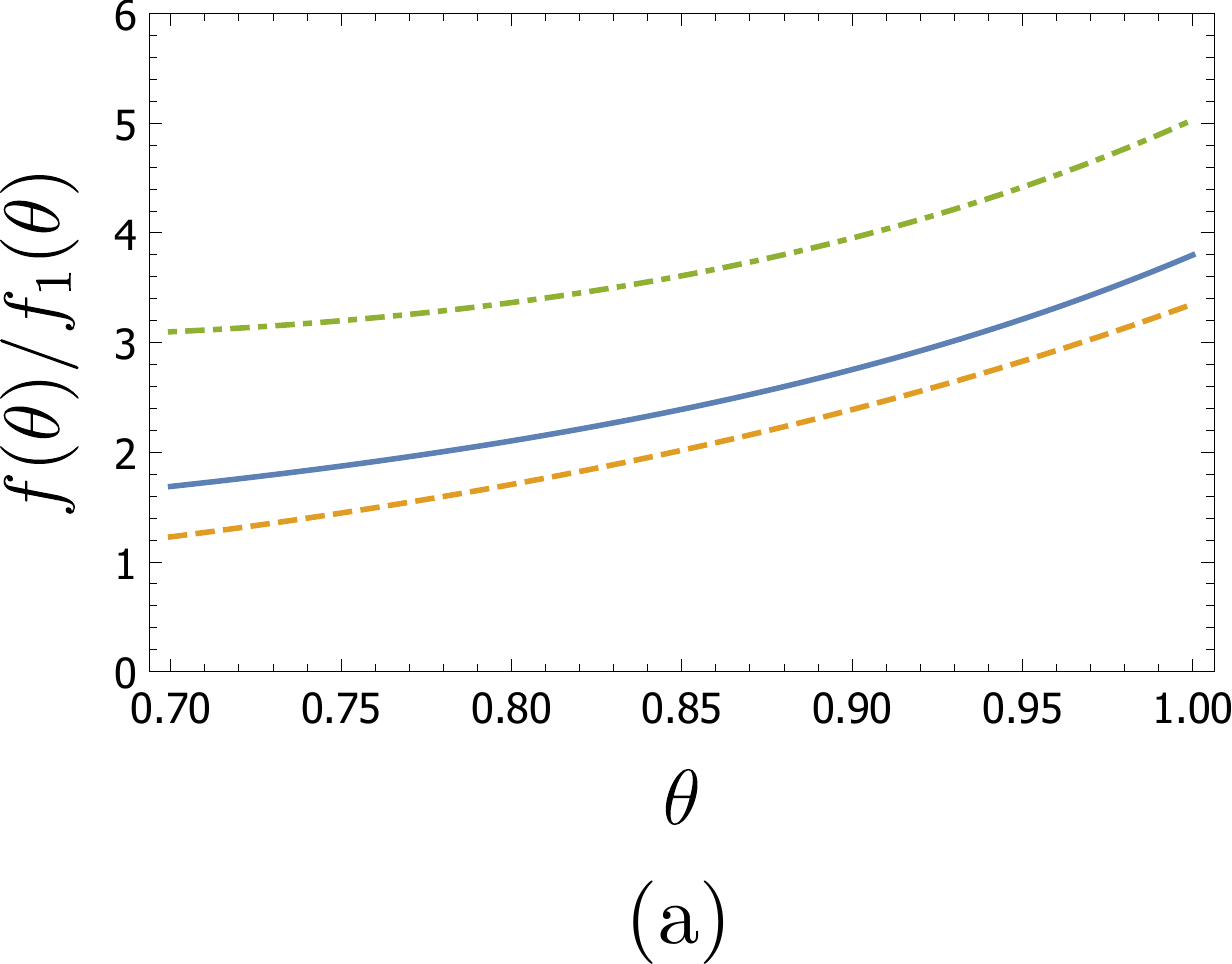}\quad \includegraphics[scale=0.42]{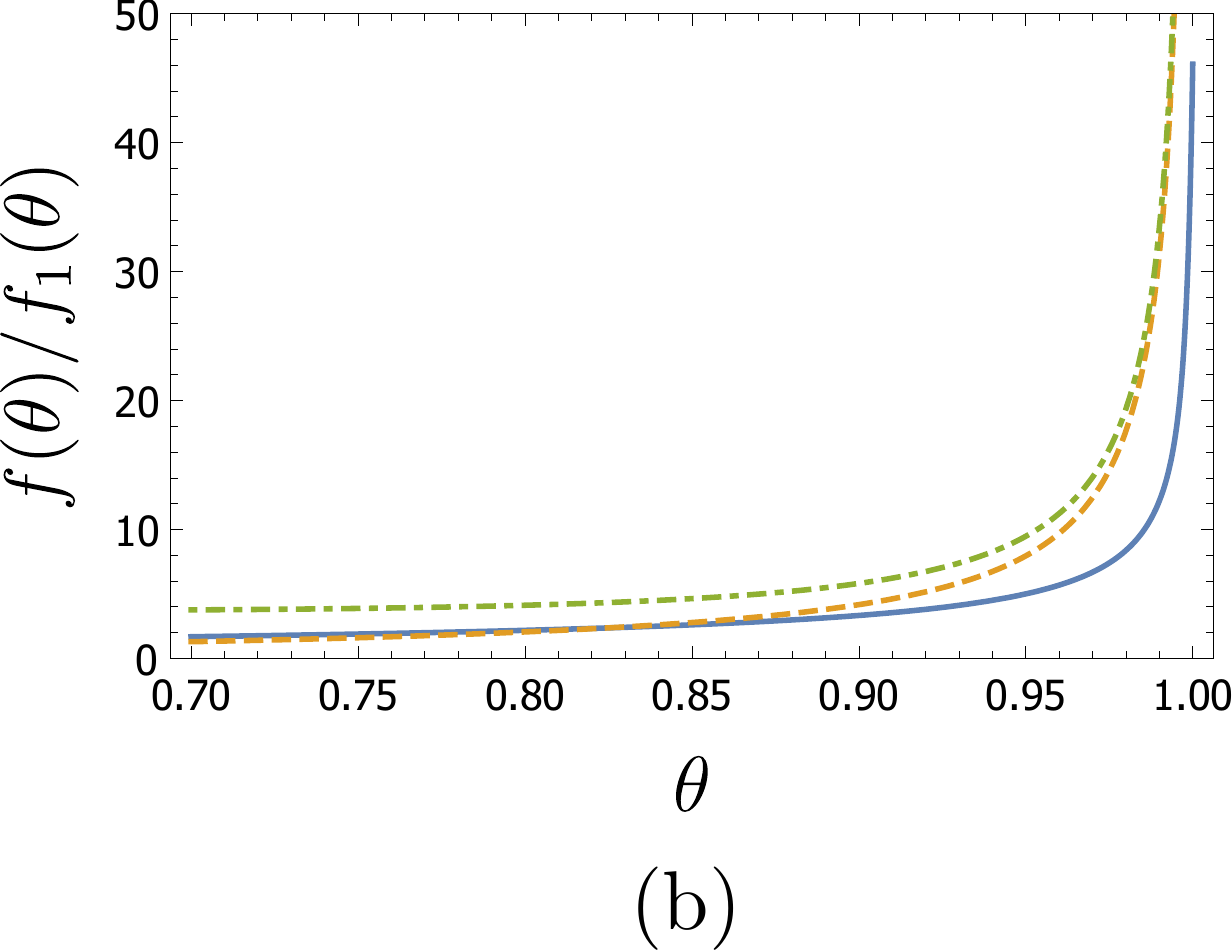}\quad \includegraphics[scale=0.42]{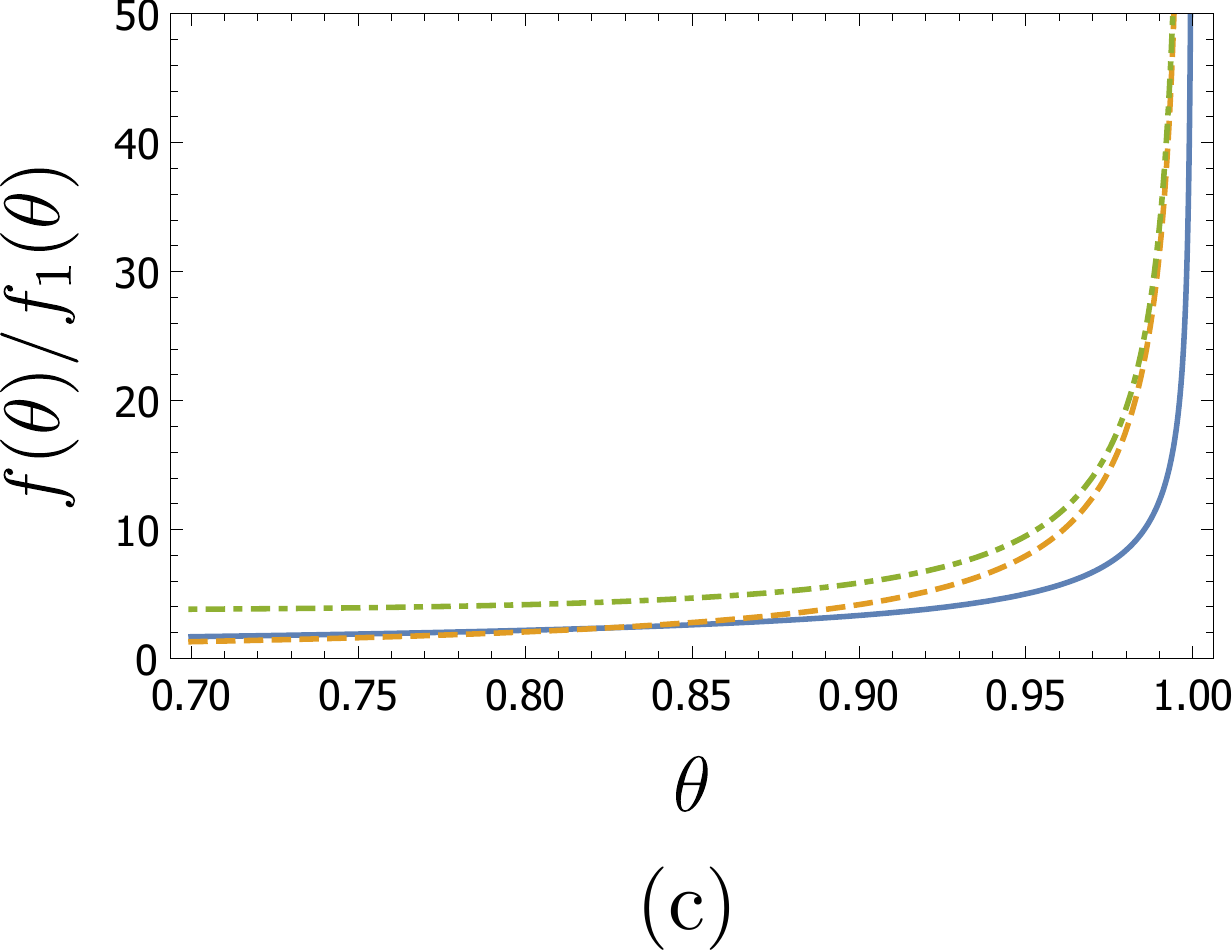}}\\
\ \\
\subfigure{\includegraphics[scale=0.42]{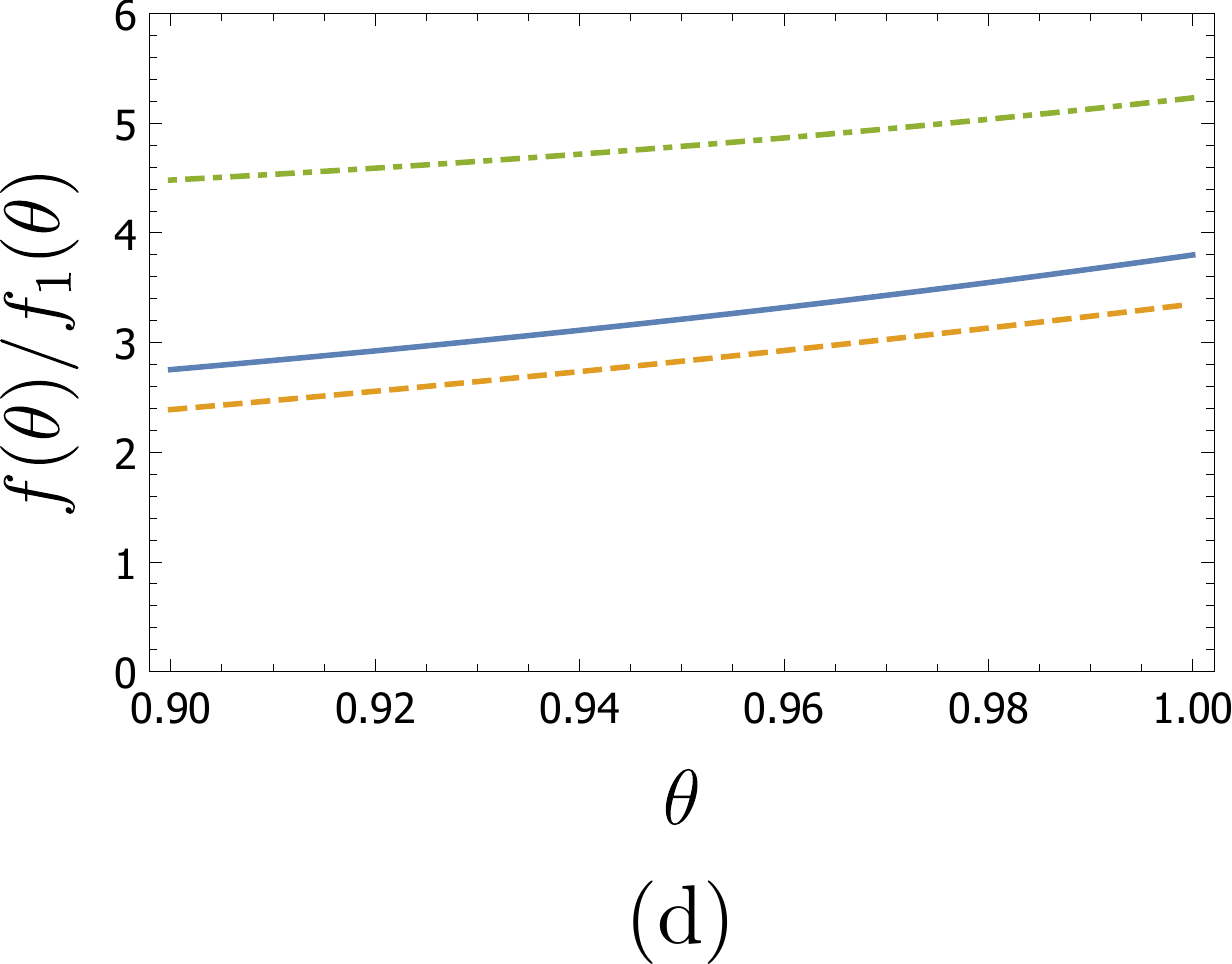}\quad \includegraphics[scale=0.42]{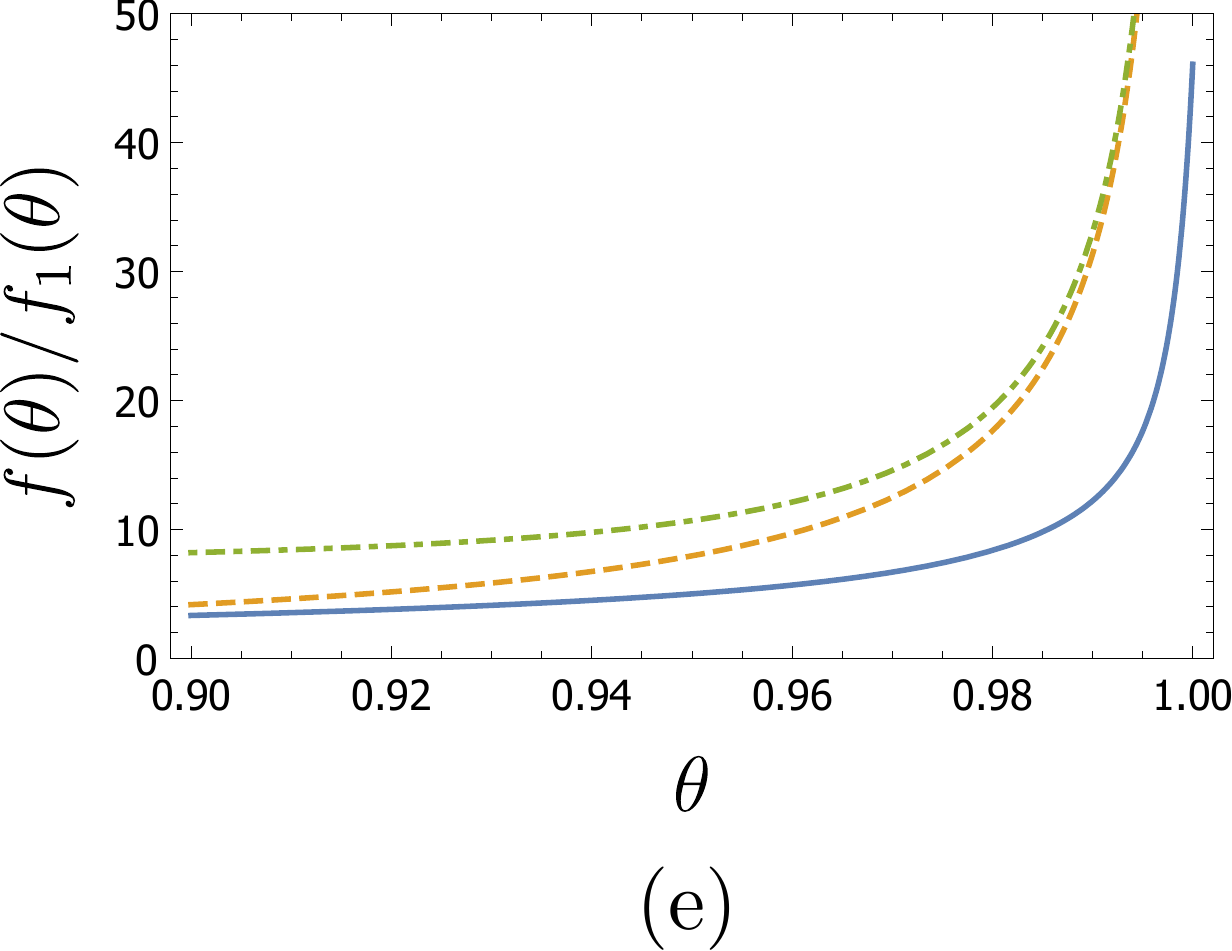}\quad \includegraphics[scale=0.42]{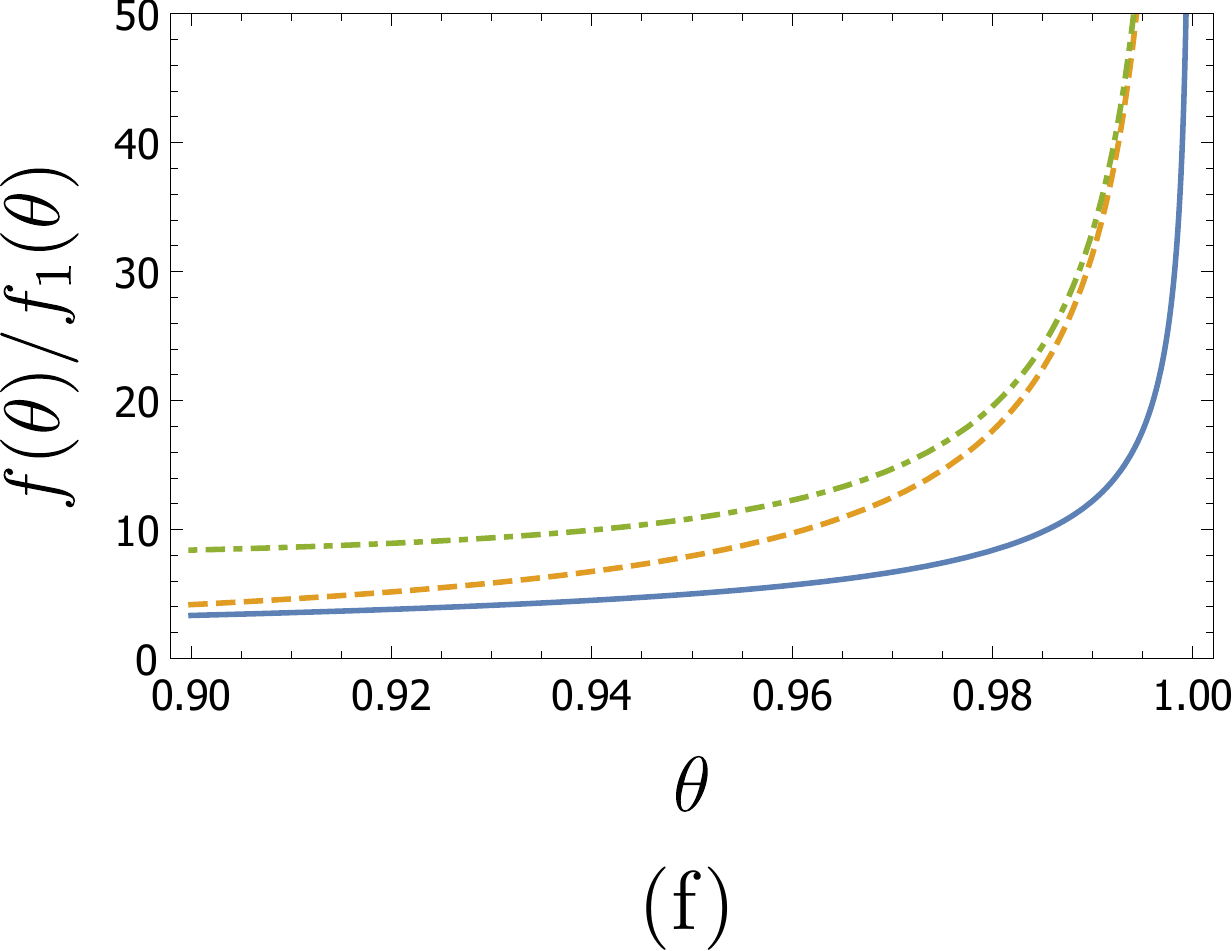}}
\caption{Relative intensities of the noise terms in Eqs.~\eqref{eqn:slow_theta_2}--\eqref{eqn:slow_lambda_2} for \textbf{(a)--(c)} $\theta^*=0.7$, and \textbf{(d)--(f)} $\theta^*=0.9$, with $K=10$ (left), $K=1000$ (center), $K=19999$ (right). Blue, solid line: $f=f_2$; yellow, dashed line: $f=f_3$; green, dot-dashed line: $f=f_4$.}
\label{fig:noises}
\end{figure}

Taking into account the fact that, in any given realization of the process, the
lowest value reached by $\theta$ will be roughly around $\theta^*$, we compare
the magnitudes of the noise intensities $f_1,\ldots, f_4$ between
$\theta=\theta^*$ and $\theta=1$ for fixed $\theta^*$ and different values of
$K$. We find that, in general, and especially during the early stages of the
epidemic when $\theta\approx 1$, $f_1$ is small compared to $f_2, f_3$ and
$f_4$---see Fig.~\ref{fig:noises}; the effect is more pronounced for more
heterogeneous distributions with smaller deterministic final sizes,
i.e., larger $\theta^*$. Therefore, the system may be described simply
by a deterministic ordinary differential equation (ODE) for $\theta$, and
Eqs~\eqref{eqn:slow_theta}--\eqref{eqn:slow_lambda} become
\begin{align}
	\frac{d \theta}{d \tau} &= -\beta \theta \lambda, \label{eqn:theta_simple}\\
\frac{d \lambda}{d \tau} &= \lambda \left(\beta \phi(\theta)-\gamma \right) + \frac{1}{\sqrt{N}}\bar{\sigma}\zeta(\tau), \label{eqn:lambda_simple}
\end{align}
with $\zeta(\tau)$ a Gaussian noise with zero mean and $\langle \zeta(\tau)
\zeta(\tau') \rangle=\delta(\tau-\tau')$, with the noise intensity,
$\bar{\sigma}$, given by
\begin{align}
\bar{\sigma} &= \left[\left(\frac{\phi(\theta)}{\theta}\right)^2 \bar{\sigma}_1^2 + \bar{\sigma}_2^2 + \bar{\sigma}_3^2\right]^{1/2}.
\end{align}
We call this system the \textit{semi-deterministic mesoscopic model}, which is
two-dimensional like the reduced mesoscopic model, but has only one white-noise
process in place of the latter's three.  Here we have used that $\langle
\zeta_\mu(\tau)\rangle=0$, and $\langle \zeta_\mu(\tau) \zeta_\nu(\tau')
\rangle=\delta_{\mu\nu}\delta(\tau-\tau')$, $\mu,\nu=1,2,3$. Now, we note that
Eq.~\eqref{eqn:lambda_simple} corresponds to a Cox-Ingersoll-Ross (CIR)
model~\cite{CIR},
\begin{align}
\frac{d\lambda}{d\tau} &= a(\theta(\tau))(b - \lambda) d \tau + \sigma(\theta(\tau)) \sqrt{\lambda}\zeta(\tau),\label{eqn:lambda_CIR}
\end{align}
where $\theta(\tau)$ is a time-dependent parameter that is the solution of the
deterministic differential equation (\ref{eqn:theta_simple}), $b=0$, and
\begin{align}
a(\theta(\tau)) &= \gamma - \beta \phi(\theta(\tau)),\\
\sigma (\theta(\tau)) &= \frac{1}{\sqrt{N}}\sqrt{\beta \left[\phi(\theta(t))\left(1-\frac{\psi(\theta^*)}{\phi(\theta^*)}\right)+2\psi(\theta(t))\right]+\gamma \frac{\phi(\theta^*)+\psi(\theta^*)}{\phi(\theta^*)}}.
\end{align}
For fixed $a$ and $\sigma$, the conditional probability distribution of
$\lambda(\tau+\Delta \tau)=\lambda$, given $\lambda(\tau)=\lambda'$, is known
in closed form, and corresponds to a non-central chi-square distribution given
by~\cite{CIR,Feller}
\begin{align}
P(\lambda, \tau+\Delta \tau \vert \lambda',\tau) &= c e^{-u-v}\left(\frac{v}{u}\right)^{q/2}I_q(2\sqrt{u v}),\label{eqn:CIR_pdf}
\end{align}
where
\begin{align}
\renewcommand{\arraystretch}{2}\begin{array}{rclcrcl}
c & \equiv & \dfrac{2 a}{\sigma^2\left(1-e^{-a\Delta \tau}\right)}, & & u &\equiv& \displaystyle c \lambda'\, e^{-a \Delta \tau},\\
v & \equiv & \displaystyle c \lambda, & & q &\equiv& \dfrac{2 a b}{\sigma^2} -1 = -1,
\end{array}
\end{align}
and where $I_q$ is the modified Bessel function of the first kind of order $q$.

\begin{figure}
\centering
\includegraphics[scale=0.42]{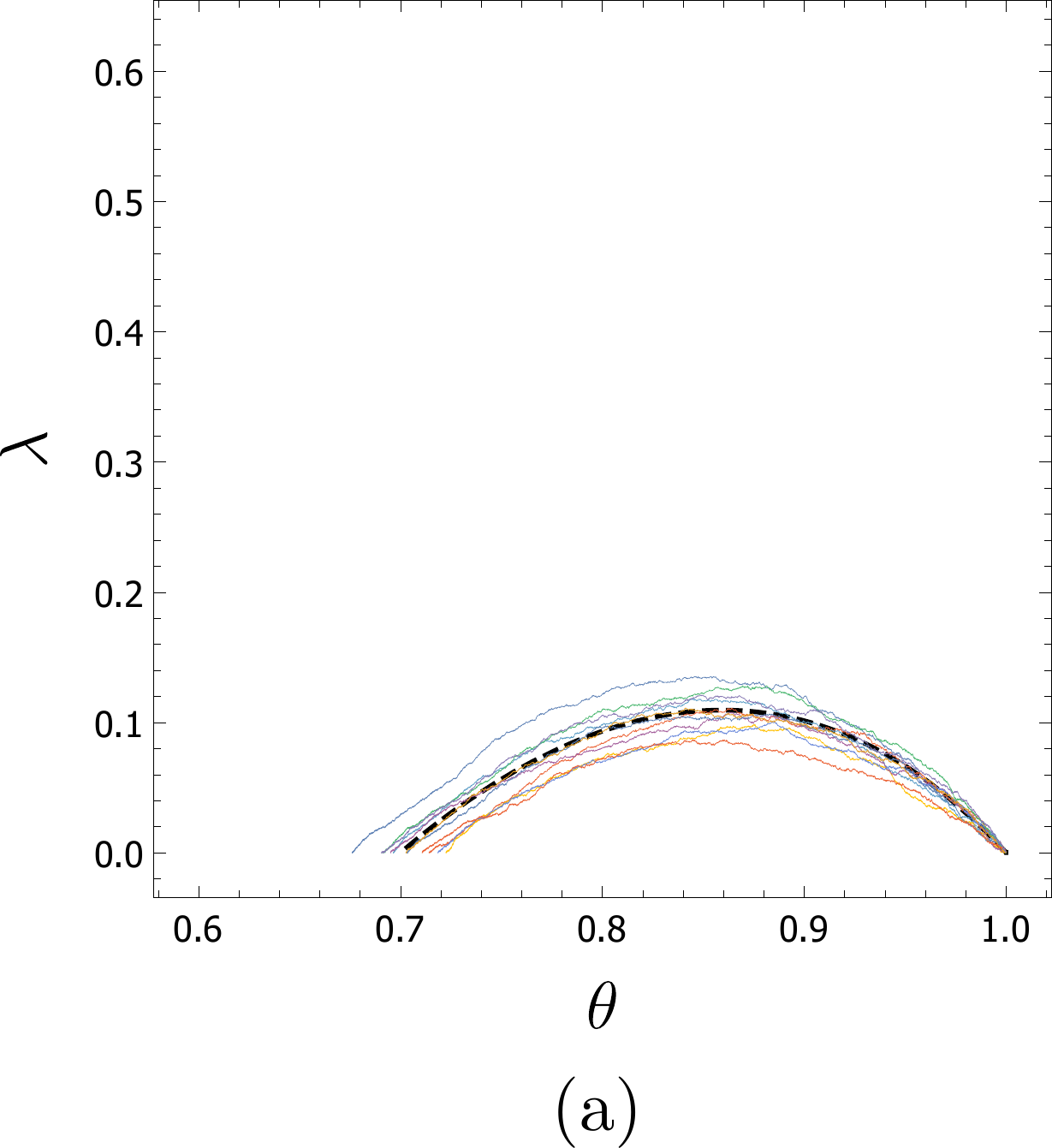}\quad \includegraphics[scale=0.42]{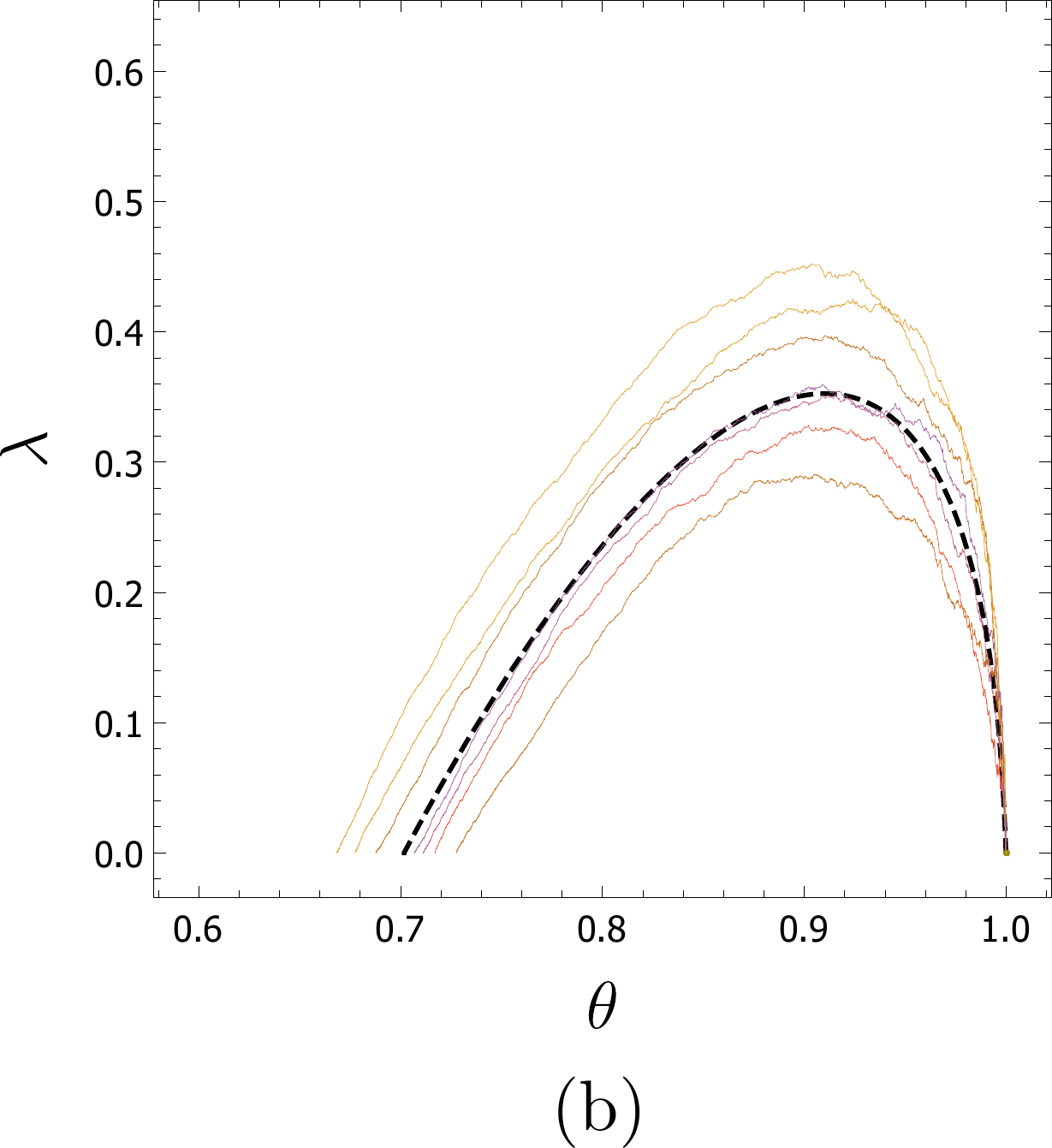}\quad \includegraphics[scale=0.42]{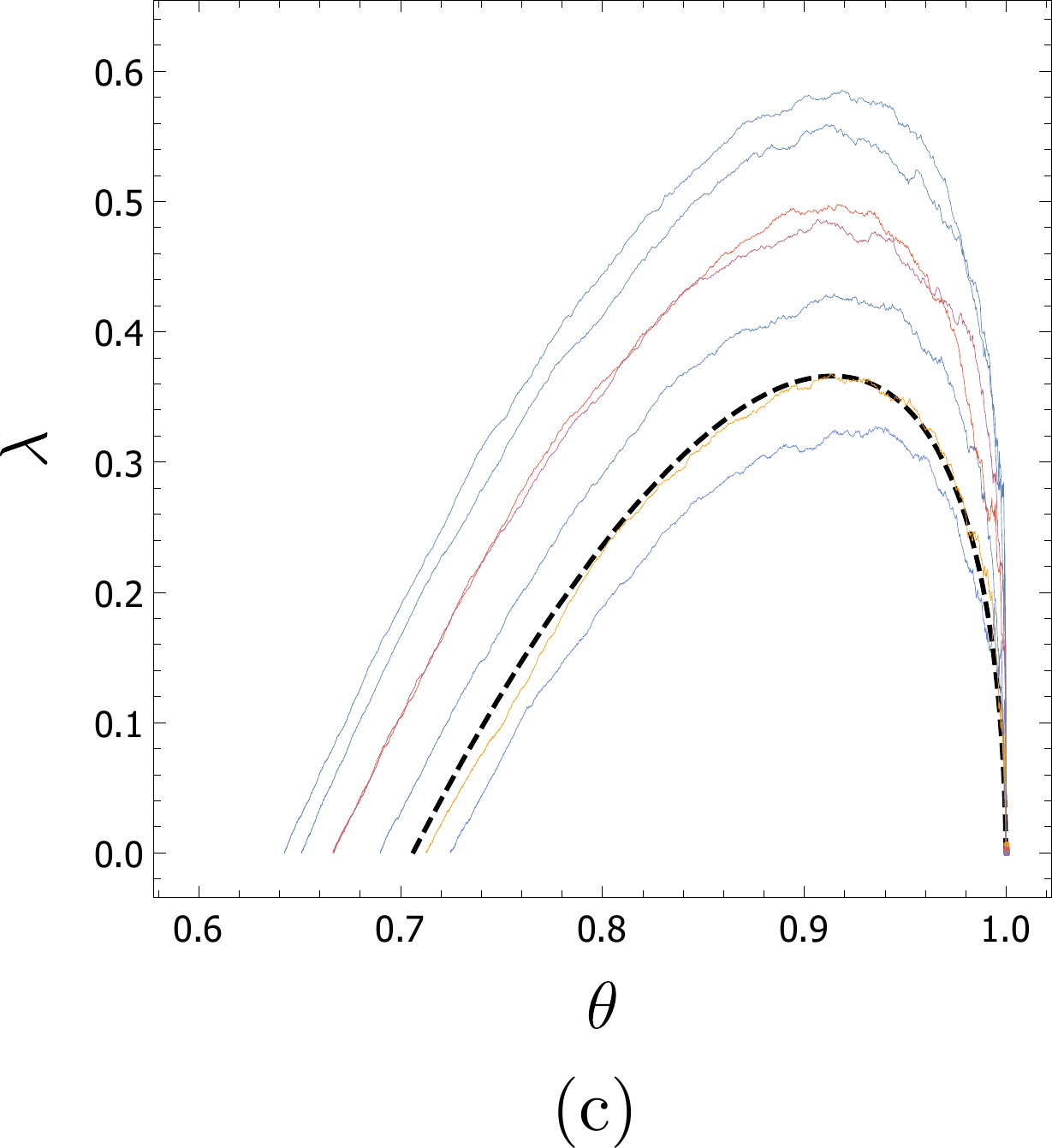}
\caption{Phase diagrams from the semi-deterministic mesoscopic model with $N=20000$,
$\gamma=1$, and $\theta^*=0.7$. Colored lines correspond to realizations of
Eqs.~\eqref{eqn:theta_simple}--\eqref{eqn:lambda_simple}, started with a total
number of 5 infectives with degrees sampled from a truncated Zipf distribution
with $\alpha=-2.5$ and \textbf{(a)} $K=10$, \textbf{(b)} $K=1000$, and \textbf{(c)} $K=19999$.
The black, dashed line corresponds to the deterministic solution.}
\label{fig:phase_cir}
\end{figure}

Therefore, if we assume that $\theta$ evolves discretely, staying constant from
time $\tau$ to time $\tau+\Delta \tau$ for reasonably small $\Delta \tau$, we
can use the result above to simulate
Eqs.~\eqref{eqn:theta_simple}--\eqref{eqn:lambda_simple} by sampling the values
of $\lambda$ directly from the distribution in Eq.~\eqref{eqn:CIR_pdf} with fixed
parameters $a(\theta(\tau))$, $\sigma(\theta(\tau))$ at every time-step. The
case $b=0$ corresponds to the special case of a non-central chi-square
distribution with zero degrees of freedom; thus, in order to simulate
Eq.~\eqref{eqn:lambda_simple} we follow the method described in
Refs.~\cite{glasserman_03,andersen_10} and sample from a central chi-square
distribution with Poisson-distributed degrees of freedom, with the convention
that the central chi-square distribution with zero degrees of freedom is
identically zero~\cite{siegel_79}. Figure~\ref{fig:phase_cir} shows the
dynamics obtained in this way. While we note an additional over-representation
of early extinctions, the other results are in very good agreement with the
reduced mesoscopic model, Eqs.~\eqref{eqn:slow_theta}--\eqref{eqn:slow_lambda}.

\section{Distribution of epidemic sizes}\label{sec:size}

In the previous sections, we have derived a two-dimensional reduction of the
$2K$-dimensional SIR dynamics described by Eqs.~\eqref{eqn:sk}--\eqref{eqn:ik},
by exploiting a separation of time-scales in the deterministic dynamics of the
system. We have found reasonably good agreement between the temporal behaviors
of the full and reduced models. Furthermore, we have also discussed an extra
simplification, consisting in neglecting the noise acting on the susceptible
variable $\theta$, and describing the system in terms of one ODE and one SDE
instead of two SDEs. Now we explore how accurately these reductions capture the
distribution of epidemic sizes, given by the number of recovered individuals
present in the system at the end of the epidemic, $r_\infty$, which can be
obtained as $r_\infty=1-\sum_k s_{k,\infty}$ and $r_\infty=1-G(\theta_\infty)$
in the full and reduced models, respectively, where $s_{k,\infty}$ and
$\theta_\infty$ are the final values of $s_k$ and $\theta$---not to be confused
with the fixed points in the deterministic limit, $s_k^*$ and $\theta^*$.
Formal mathematical work has considered central limit theorems for models that
generalize ours~\cite{Neal:2007}; we are interested here in explicit
calculation and simulation approaches.

We obtain the distribution of $r_\infty$ from the full model \emph{via} a
Sellke construction~\cite{house2014non}, and note that it is characterized by
two regions: one to the left of the interval, corresponding to early
extinctions, and one to the right corresponding to the epidemic taking off. The
first thing we note when comparing this to results from the two versions of the
reduced model---Eqs.\eqref{eqn:slow_theta}--\eqref{eqn:slow_lambda}, and
Eqs.~\eqref{eqn:theta_simple}--\eqref{eqn:lambda_simple}---is that, as
discussed previously, there is an over-representation of very early extinctions
in the case of the latter, and therefore the relative sizes of the left- and
right-most regions of the distribution are not well captured. We would not
expect our diffusion approximation to capture extinction probabilities
accurately, given that its derivation implicitly assumes that the infective
population is large and extinctions by definition involve small numbers of
infectives. We can, however, derive accurate extinction probabilities from
other results we have obtained.

First, we make an argument related to `susceptibility sets' in
probability~\cite{Ball:2000}. For each pair of individuals in the population,
we pick whether infection will be spread between them if one becomes infective
and draw a link between them if it will.  This forms a static graph, and if we
pick a node of the graph to be initially infective, then the size of the
epidemic will be the size of the network component that the initial pick is
part of.  This means that the size of the giant component of the graph is equal
to the final size of the epidemic if it is large, and the probability of early
extinction is equal to the proportion of the population in the small components
of this graph. Therefore, we can conclude that we have already calculated the
probability of early extinction of the epidemic starting with an individual
picked uniformly at random from a very large population; it will be equal to
$\sum_k s_k^* = G(\theta^*)$.

\begin{figure}
\centering
\subfigure{\includegraphics[scale=0.42]{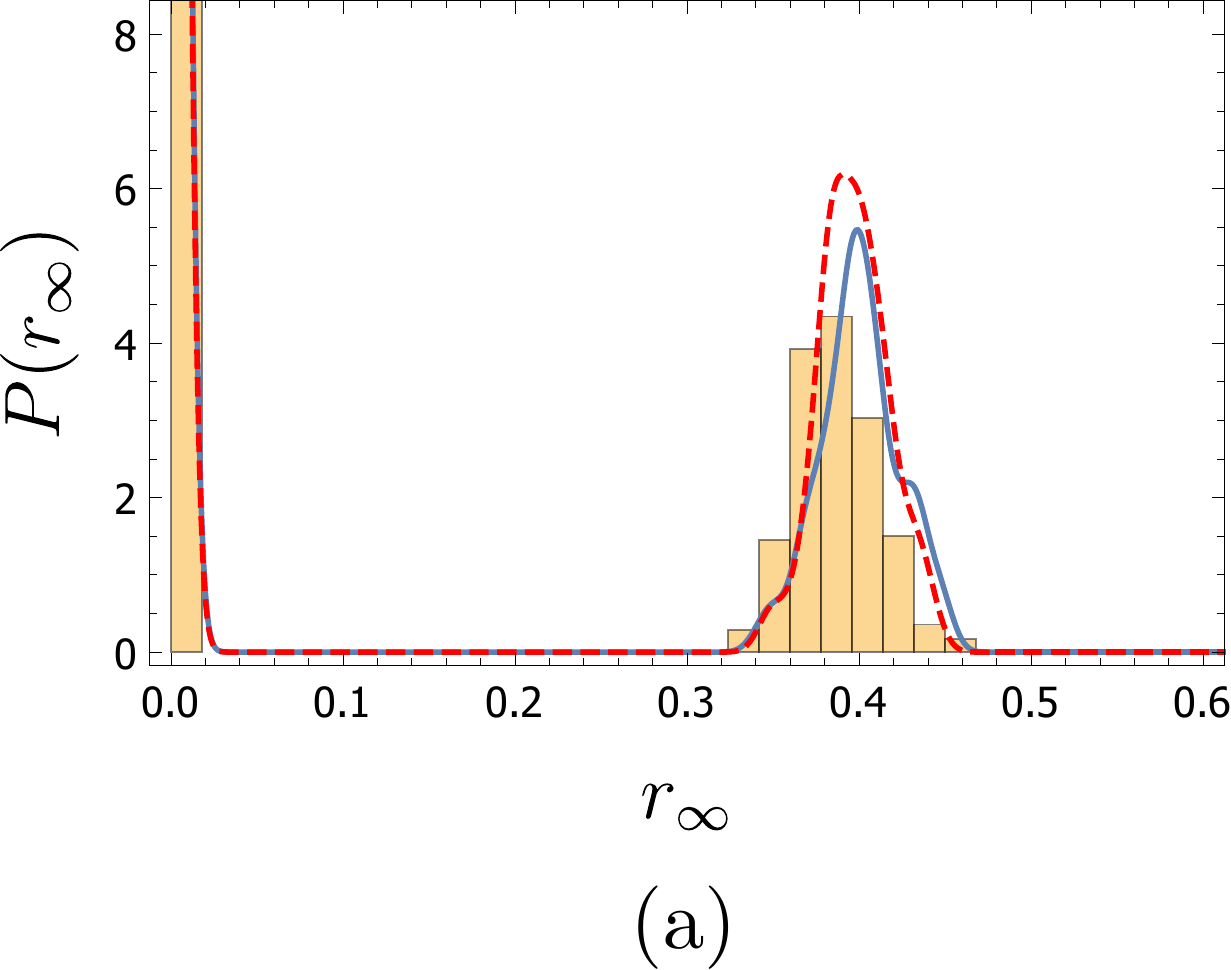}\quad \includegraphics[scale=0.42]{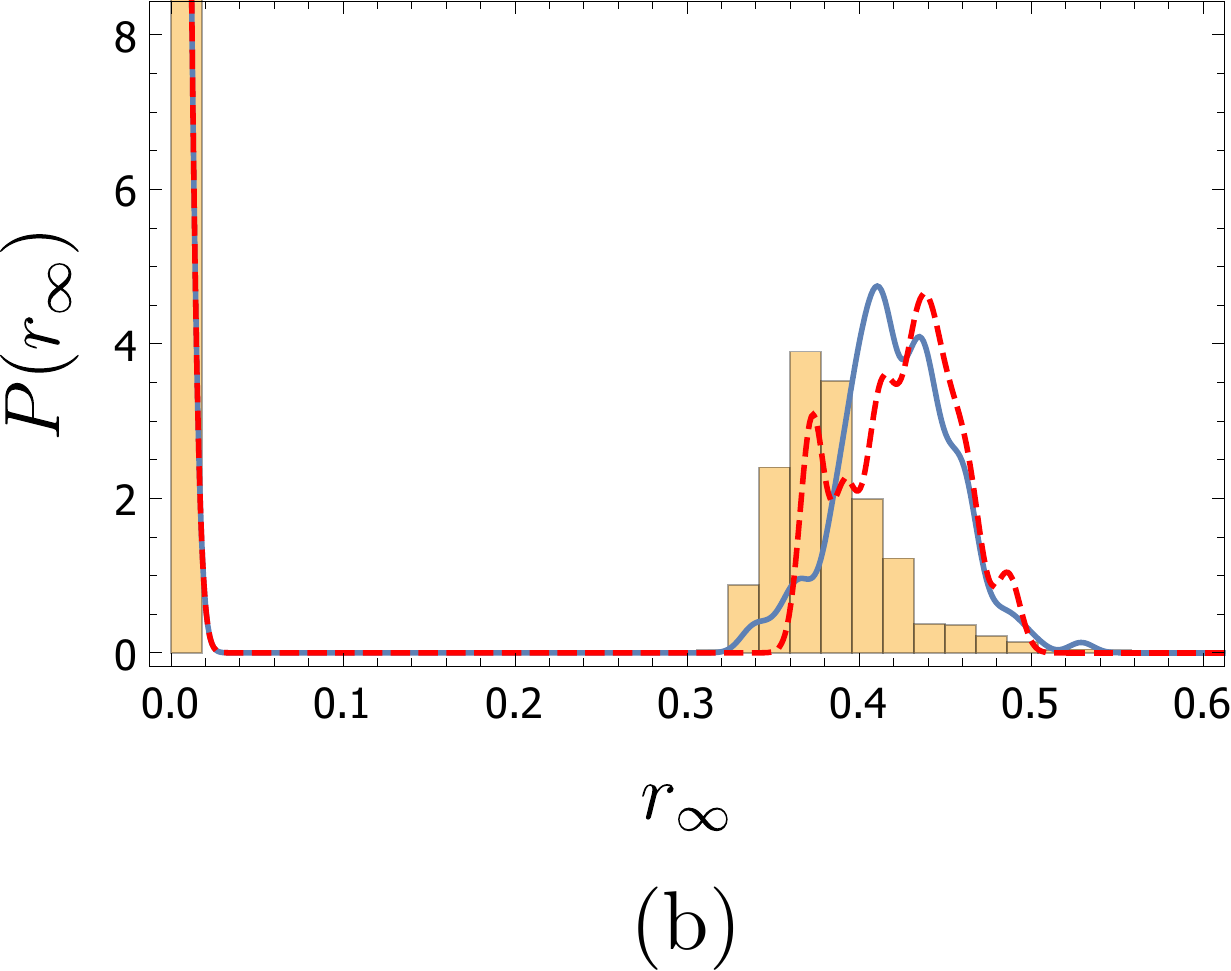}}\\
\ \\
\subfigure{\includegraphics[scale=0.42]{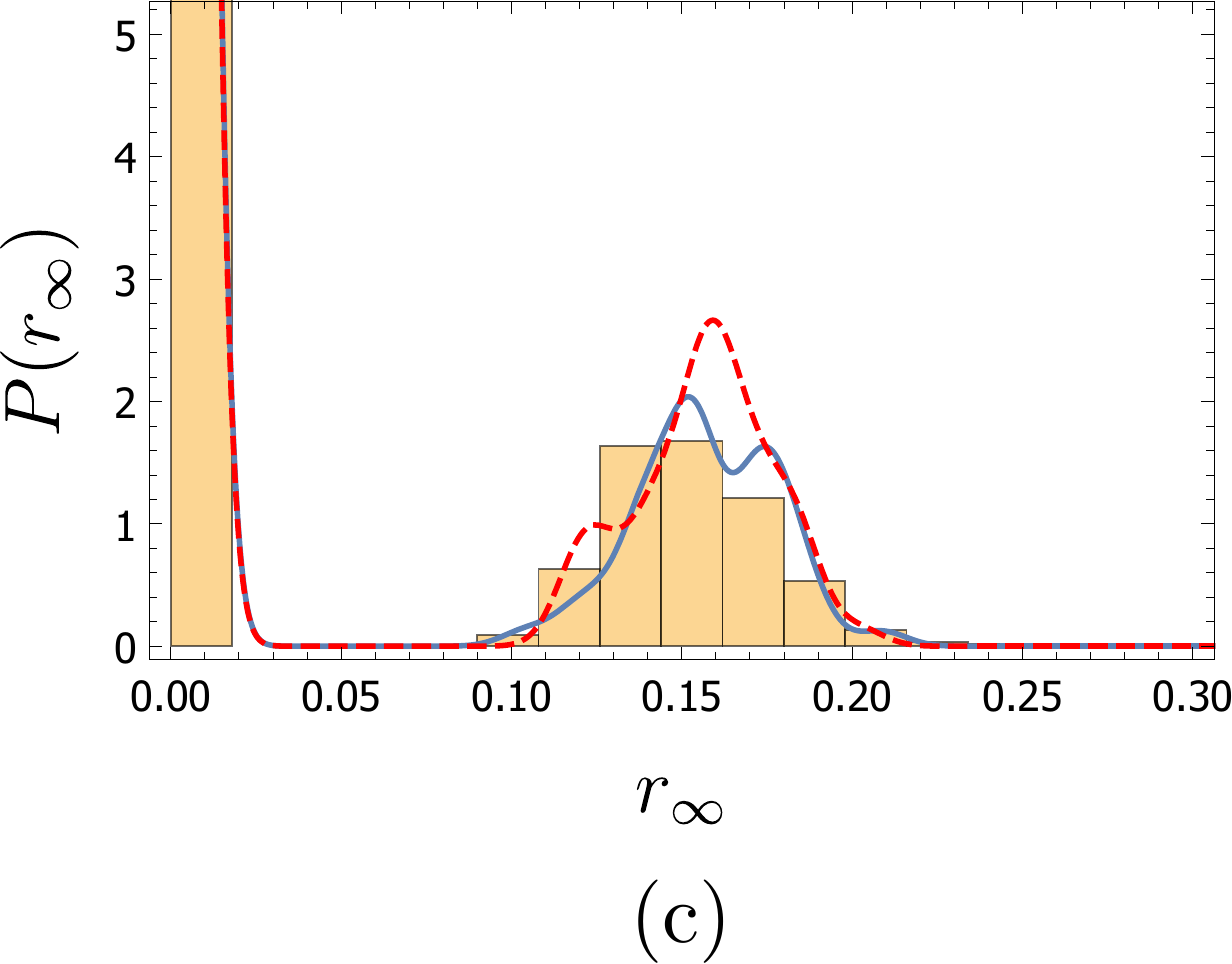}\quad \includegraphics[scale=0.42]{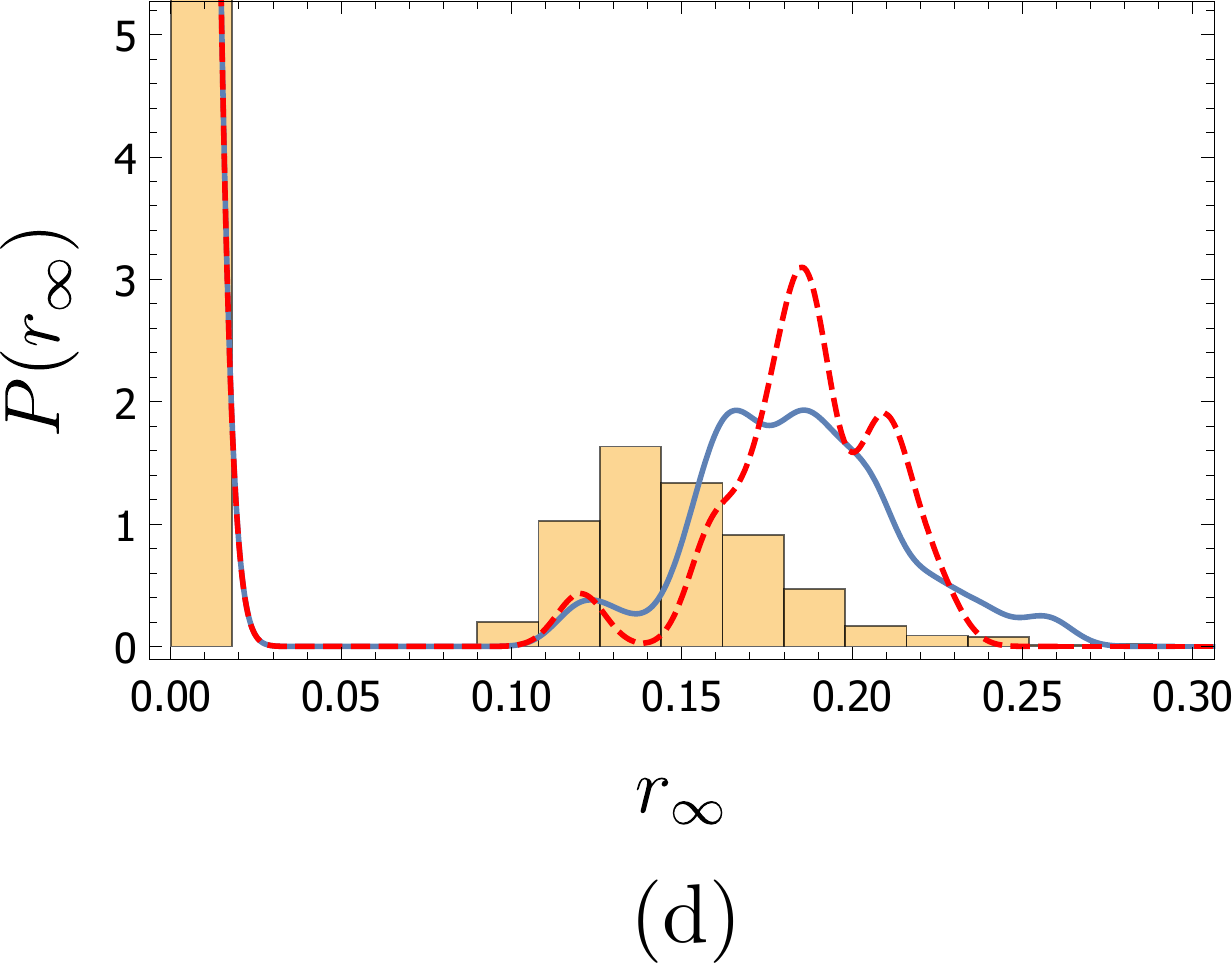}}
\caption{Distribution of epidemic sizes for $N=20000$, $\alpha=-2.5$ and
$\gamma=1$, for \textbf{(a)--(b)} $\theta^*=0.7$, and \textbf{(c)--(d)} $\theta^*=0.9$, with $K=1000$ (left panels) and $K=19999$ (right panels). Histogram: full model;
blue, solid line: reduced mesoscopic model
Eqs.~\eqref{eqn:slow_theta}--\eqref{eqn:slow_lambda}; red, dashed line:
semi-deterministic mesoscopic model
Eqs.~\eqref{eqn:theta_simple}--\eqref{eqn:lambda_simple}.}
\label{fig:r_inf}
\end{figure}

Figure~\ref{fig:r_inf} shows the distribution of final sizes from the reduced
and semi-deterministic mesoscopic models, adjusted to account for the
inaccuracy in early extinctions by removing these and assuming that a fraction
$G(\theta^*)$ of epidemics stop early, and a fraction $1-G(\theta^*)$
take the values from the mesoscopic models.  These results are compared to the
distribution from the full model and show that the argument we have introduced
provides a simple method to adjust for the inaccuracies in early extinctions,
yielding an accurate approximation overall.

\section{Discussion}\label{sec:discussion}

Stochastic epidemics on heterogeneous networks involve two potentially large
parameters.  The first of these is the size of the population, $N$, which is
proportional to the dimensionality of the microscopic description. For large
populations, this motivates the derivation of a mesoscopic model based on a
diffusion approximation whose errors are controlled by inverse powers of $N$,
and which we have formulated in this paper. The second of these is the
maximum degree $K$, which is proportional to the dimensionality of the full
mesoscopic model. We have been able to use fast-mode elimination techniques to
derive two-dimensional approximations to this scheme: the reduced mesoscopic
model, which involves three white-noise processes; and the semi-deterministic
mesoscopic model, which involves only one white-noise process.

Our numerical results back up our theoretical understanding that these
low-dimensional systems are accurate approximations to the microscopic model,
except in the case where $K$ approaches $N$, although this is to be expected since our derivation of the
diffusion limit implicitly assumes that $K\ll N$. Nevertheless, the reduction can be applied to cases with extremely large values of $K$, as long as they are not too close to $N$, thus significantly simplifying the original model.

We have observed that the elimination of fast variables in the system provides a good approximation to the full dynamics even for relatively large values of $\epsilon$, corresponding to a small separation between time-scales. We have also shown that the main inaccuracy of the reduction, relating to very early extinction probabilities of the
epidemic, can be dealt with using a simple argument. In
this way, we hope to have provided a useful tool in the study of both
infectious diseases and spreading processes on heterogeneous networks.

\begin{acknowledgments}
We thank Frank Ball for helpful comments on this manuscript, as well as the anonymous reviewers. C.P.-R. was funded by CONICYT Becas Chile Scholarship No. 72140425. T.H. was supported by the EPSRC (UK) under Grant No. EP/N033701/1.
\end{acknowledgments}

\appendix
\section{Eigenvectors of the deterministic system}\label{sec:eve}

\subsection*{Full system}

A set of right-eigenvectors corresponding to the eigenvalues from
Eq.~\eqref{eqn:eva_full} are given by
\begin{align}
\begin{split}
\bm{v}^{\{k\}} &= \bm{e}_{k}, \qquad k=1,\ldots, K,\\
\bm{v}^{\{k\}} &= \bm{e}_{3K+1-k}-(2K+1-k)\bm{e}_{K+1}, \qquad k=K+1,\ldots, 2K-1,\\
\bm{v}^{\{2K\}} &= -X \sum_{k=1}^K k s^*_k \bm{e}_{k}+ \sum_{k=1}^K k s^*_k \bm{e}_{K+k},
\end{split}\label{eqn:eve_full}
\end{align}
where $\bm{e}_i$ is the $i$-th unit vector and
\begin{align}
X = \frac{\beta \sum_{k=1}^{K} k^2 s^*_k}{\beta \sum_{k=1}^{K} k^2 s^*_k-\gamma}.
\end{align}

\subsection*{Totally reduced system}

A set of right-eigenvectors corresponding to the eigenvalues from
Eq.~\eqref{eqn:eva_total} are given by
\begin{align}
\begin{split}
\bm{v}^{\{1\}} &= \left( 1,0\right) ^\top,\\
\bm{v}^{\{2\}} &= \left( -\frac{\beta \theta^*}{\beta\phi(\theta^*)-\gamma},1\right) ^\top.
\end{split}\label{eqn:eve_total}
\end{align}

\subsection*{Partially reduced system}

The partially reduced system with eigenvalues given by Eq.~\eqref{eqn:eva_semi}
has a set of right-eigenvectors given by
\begin{align}
\begin{split}
\bm{v}^{\{1\}} &= \bm{e}_1,\\
\bm{v}^{\{k\}} &= \bm{e}_{K+3-k}-(K+2-k)\bm{e}_2,\qquad k=2,\ldots,K,\\
\bm{v}^{\{K+1\}} &= \frac{1}{Y_{K}}\left(\sum_{k=1}^{K}Y_k \bm{e}_{k+1} - X \bm{e}_1\right),
\end{split}\label{eqn:eve_semi}
\end{align}
where now
\begin{align}
\begin{split}
X &= \frac{\beta \theta^*}{\beta \phi(\theta^*)-\gamma},\\
Y_k &= \frac{k d_{k} {\theta^*}^{k}}{\phi(\theta^*)}.
\end{split}\label{XandY}
\end{align}

In Eqs.~(\ref{eqn:eve_full}) and (\ref{eqn:eve_total}) we did not normalize the
eigenvectors. However, in this case we do need the eigenvectors to be normalized
to simplify the application of the reduction technique described in
Sec.~\ref{sec:fastvarelim}. This requires finding the corresponding
left-eigenvectors. When finding these, we have a complication given by the fact
that there is one eigenvalue which is $(K-1)$-fold degenerate, and the
condition~\eqref{eqn:normal} will not necessarily hold in general. However, if
we construct a matrix $\bm{V}$ in such a way that its $i$-th column corresponds
to $\bm{v}^{\{i\}}$, we can choose $\bm{u}^{\{j\}}$ to be the $j$-th row of the
inverse matrix $\bm{V}^{-1}$, if such an inverse exists. We find that indeed
this is the case, and the left-eigenvectors in the partially reduced model are
then given by
\begin{align}
\begin{split}
\bm{u}^{\{1\}} &= \bm{e}_1 + X\sum_{k=1}^K k \bm{e}_{k+1},\\
\bm{u}^{\{k\}} &= \bm{e}_{K+3-k}-Y_{K+2-k}\sum_{l=1}^K l \bm{e}_{l+1},\qquad k=2,\ldots,K, \\
\bm{u}^{\{K+1\}} &= Y_{K}\sum_{k=1}^K k \bm{e}_{k+1}.
\end{split}\label{eqn:leve_semi}
\end{align}

We note that
\begin{align}
\sum_{k=1}^{K}k Y_k &= \frac{\sum_{k=1}^K k^2 d_k {\theta^*}^k}{\phi(\theta^*)}=\frac{\phi(\theta^*)}{\phi(\theta^*)}=1.
\end{align}

Using this property, we can verify that the orthogonality
condition~\eqref{eqn:normal} is satisfied:
\begin{align}
\begin{split}
\bm{u}^{\{1\}} \cdot \bm{v}^{\{1\}} &= 1,\\
\bm{u}^{\{1\}} \cdot \bm{v}^{\{j\}} &= (K+2-j)X-(K+2-j)X=0, \qquad j=2,\ldots,K,\\
\bm{u}^{\{1\}} \cdot \bm{v}^{\{K+1\}} &= -\frac{X}{Y_K}+\frac{X}{Y_K}\sum_{k=1}^K kY_k = 0,\\
\bm{u}^{\{i\}} \cdot \bm{v}^{\{1\}} &= 0, \qquad i=2,\ldots,K,\\
\bm{u}^{\{i\}} \cdot \bm{v}^{\{j\}} &= \delta_{ij}+Y_{K+2-i}(K+2-j)-Y_{K+2-i}(K+2-j)=\delta_{ij}, \qquad i,j=2,\ldots,K,\\
\bm{u}^{\{i\}} \cdot \bm{v}^{\{K+1\}} &= \frac{Y_{K+2-i}}{Y_K}-\frac{Y_{K+2-i}}{Y_K}\sum_{k=1}^K kY_k = 0, \qquad i=2,\ldots,K,\\
\bm{u}^{\{K+1\}} \cdot \bm{v}^{\{1\}} &= 0,\\
\bm{u}^{\{K+1\}} \cdot \bm{v}^{\{j\}} &= Y_k(K+2-j)-Y_k(K+2-j)=0,\qquad j=2,\ldots,K,\\
\bm{u}^{\{K+1\}} \cdot \bm{v}^{\{K+1\}} &= \sum_{k=1}^K kY_k = 1.
\end{split}
\end{align}

\section{Fast-variable elimination}\label{sec:projection}
In this Appendix we derive the two-dimensional reduced mesoscopic model, by
projecting the dynamics of the ${(K+1)}$-dimensional partially reduced
mesoscopic model onto the slow degrees of freedom of its corresponding
deterministic limit.

Using the projector defined in Eq.~(\ref{eqn:projector}) on $\bm{x}=(\theta,
i_1,\ldots, i_K)$ gives the slow variables defined by Eqs.~(\ref{eqn:slow_z1})
and (\ref{eqn:slow_z2}). Later, the inverse transformation, back to the more
familiar variables $\theta$ and $\sum_k k i_k (=\lambda)$, will be required. It
is
\begin{align}
\theta(\bm{z})&=z_1-\frac{\beta  z_2 \phi(\theta^*)}{K {\theta^*}^{K-1} d_{K} \left[-\gamma + \beta  \phi(\theta^*)\right]},\label{eqn:theta_slow}\\
\lambda(\bm{z}) &= \frac{\phi(\theta^*)}{K d_K{\theta^*}^{K}}\,z_2,
\label{eqn:lambda_slow}
\end{align}
however for the moment we will use the $z_1$ and $z_2$ variables, and will
denote the right-hand side of Eq.~(\ref{eqn:theta_slow}) as $\chi(\bm{z})$.

The deterministic slow dynamics, obtained by applying the projector to $\bm
A(\bm x)$, will be given by
\begin{align}
\dot{z}_1 &= \frac{\phi(\theta^*)}{K d_{K}{\theta^*}^{K}} \beta z_2\left[-\chi(\bm{z})+\frac{\theta^*\left(\beta \phi( \chi(\bm{z}))-\gamma \right)}{-\gamma+\beta\phi(\theta^*)}\right],\label{eqn:slow_dz1}\\
\dot{z}_2 &= z_2\left(\beta \phi( \chi(\bm{z}))-\gamma \right),\label{eqn:slow_dz2}
\end{align}
where the dot represents the time derivative.

Applying now the projector to the noise terms in the finite-$N$ case we find,
for the component along $\bm{v}^{\{1\}}$,
\begin{align}
\eta_{z_1}(\tau) &= \frac{1}{\sqrt{N}} \bm{u}_1 \cdot \left(\frac{1}{G^\prime (\theta)}\sum_k \sigma_1^{(k)}\rho_1^{(k)}(\tau),-\sigma_1^{(1)}\rho_1^{(1)}(\tau)+\sigma_2^{(1)}\rho_2^{(1)}(\tau),\ldots,-\sigma_1^{(K)}\rho_1^{(K)}(\tau)+\sigma_2^{(K)}\rho_2^{(K)}(\tau)\right)\nonumber\\
&= \frac{1}{\sqrt{N}}\left[\xi_1(\tau) + \frac{\beta \theta^*}{-\gamma+\beta\phi(\theta^*)}\left(\xi_2(\tau)+\xi_3(\tau)\right)\right]\label{RHS_z2dot} \\
&= \frac{1}{\sqrt{N}}\left[\left(1-\frac{\phi(\chi(\bm z))}{\chi(\bm z)}\frac{\beta \theta^*}{-\gamma +\beta \phi(\theta^*)}\right)\bar{\sigma}_1(\bm z)\zeta_1(\tau) + \frac{\beta \theta^*}{-\gamma +\beta \phi(\theta^*)}\left(\bar{\sigma}_2(\bm z)\zeta_2(\tau)+\bar{\sigma}_3(\bm z)\zeta_3(\tau)\right)\right],
\label{eqn:eta_z1}
\end{align}
where we obtain, from Eqs.~\eqref{eqn:slow_ik}, \eqref{eqn:slow_z1} and
\eqref{eqn:slow_z2},
\begin{align}
\bar{\sigma}_1(\bm z) &= \left[\frac{\beta \phi(\theta^*) z_2 \chi(\bm z)}{K d_K {\theta^*}^K G^\prime (\chi(\bm z))}\right]^{1/2},\\
\bar{\sigma}_2(\bm{z})&= \left[\frac{\beta \phi(\theta^*)z_2}{Kd_K{\theta^*}^K}\left(\phi(\chi(\bm z))+\psi(\chi(\bm z))-\frac{\phi(\chi(\bm z))^2}{\chi(\bm z)G^\prime (\chi(\bm z))}\right)\right]^{1/2},\\
\bar{\sigma}_3(\bm z) &= \left[\frac{z_2}{K d_{K}{\theta^*}^{K}}\left(\gamma \left[\phi(\theta^*)+\psi(\theta^*)\right]+\beta \left[\phi(\theta^*)\psi(\chi(\bm{z}))-\phi(\chi(\bm{z}))\psi(\theta^*)\right]\right)\right]^{1/2}.
\label{eqn:sigma_3_bar_determined}
\end{align}
Similarly, for the component of the fluctuations along $\bm{v}^{\{K+1\}}$:
\begin{align}
\eta_{z_2}(\tau)&= \frac{1}{\sqrt{N}}\frac{Kd_K {\theta^*}^K}{\phi(\theta^*)}\left(-\frac{\phi(\chi(\bm z))}{\chi(\bm z)}\bar{\sigma}_1(\bm z)\zeta_1(\tau)+\bar{\sigma}_2(\bm z)\zeta_2(\tau)+\bar{\sigma}_3(\bm z)\zeta_3(\tau)\right).
\label{eqn:eta_z2}
\end{align}

We now form $\dot{\theta}$ and $\dot{\lambda}$ as linear combinations of
$\dot{z}_1$ and $\dot{z}_2$, using Eqs.~(\ref{eqn:theta_slow}) and
(\ref{eqn:lambda_slow}). These are equal to the appropriate linear combinations
of Eqs.~(\ref{eqn:slow_dz1}), (\ref{eqn:slow_dz2}), (\ref{eqn:eta_z1}), and
(\ref{eqn:eta_z2}). One finds that Eqs.~(\ref{eqn:theta_stoch}) and
(\ref{eqn:lambda_stoch}) are exactly recovered, except that now
$\bar{\sigma}_3\equiv \sqrt{\bar{B}_{33}}$ is determined, given by
Eq.~(\ref{eqn:sigma_3_bar_determined}).

\end{document}